%% file: paper.tex
\documentclass[10pt,journal,compsoc]{IEEEtran}
\usepackage{graphicx}
\usepackage{xspace}
\usepackage{booktabs}
\usepackage{multirow}
\usepackage{subcaption}
\usepackage{url}
\usepackage{times}
\usepackage{slashbox}
\usepackage{xcolor}
\newcommand{\subparagraph}{}
\usepackage{titlesec}
\usepackage{fancyhdr}

\titlespacing\section{0pt}{0pt plus 6pt minus 2pt}{0pt plus 6pt minus 2pt}
\titlespacing\subsection{0pt}{0pt plus 6pt minus 2pt}{0pt plus 6pt minus 2pt}
\newcommand{\bliss}{BLISS\xspace}
\def\httilde{\mbox{\tt\raisebox{-.5ex}{\symbol{126}}}}

\newcommand{\squishlist}{
   \begin{list}{$\bullet$}
    { \setlength{\itemsep}{0pt}      \setlength{\parsep}{0pt}
      \setlength{\topsep}{0pt}       \setlength{\partopsep}{0pt}
      \setlength{\leftmargin}{1em} \setlength{\labelwidth}{1em}
      \setlength{\labelsep}{0.5em} } }

\newcommand{\squishend}{
    \end{list}  }

\fancypagestyle{plain}{
\fancyhf{}  %clear all header and footer fields
\chead{SAFARI Technical Report No. 2015-004 (March, 2015)}
\cfoot{\thepage}
}

\begin{document}

\title{
The Blacklisting Memory Scheduler: Balancing Performance, Fairness and Complexity
}
\author{
Lavanya Subramanian,
Donghyuk Lee,
Vivek Seshadri,
Harsha Rastogi,
Onur Mutlu\\
%Carnegie Mellon University, 5000, Forbes Avenue, Pittsburgh, PA 15217\\
Carnegie Mellon University\\
\{lsubrama,donghyu1,visesh,harshar,onur\}@cmu.edu\\
SAFARI Technical Report No. 2015-004\\
\vspace{-5mm}
}
\maketitle

\thispagestyle{plain}
\input{sections/abstract}
\input{sections/introduction}

\input{sections/background}

\input{sections/observations}

\input{sections/mechanism1}
\input{sections/implementation}
\input{sections/methodology}

\input{sections/evaluation}

\input{sections/related_work}

\input{sections/conclusion}

\bstctlcite{IEEEtrans:BSTcontrol}
\bstctlcite{bstctl:nodash, bstctl:simpurl}
\linespread{0.93}
\bibliographystyle{IEEEtranS}
\bibliography{references}
\end{document}

%% file: sections/abstract.tex
\begin{abstract}

In a multicore system, applications running on different cores
interfere at main memory. This inter-application interference degrades
overall system performance and unfairly slows down applications. Prior
works have developed application-aware memory request schedulers to
tackle this problem. State-of-the-art application-aware memory request
schedulers prioritize memory requests of applications that are
vulnerable to interference, by ranking individual applications based
on their memory access characteristics and enforcing a total rank
order.

In this paper, we observe that state-of-the-art application-aware
memory schedulers have two major shortcomings. First, such
schedulers trade off hardware complexity in order to achieve high
performance or fairness, since ranking applications individually
with a total order based on memory access characteristics leads to
high hardware cost and complexity. Such complexity could prevent
the scheduler from meeting the stringent timing requirements of
state-of-the-art DDR protocols. Second, ranking can unfairly slow
down applications that are at the bottom of the ranking stack,
thereby sometimes leading to high slowdowns and low overall system
performance. To overcome these shortcomings, we propose \emph{the
Blacklisting Memory Scheduler (\bliss)}, which achieves high
system performance and fairness {\em while} incurring low hardware
cost and complexity. \bliss design is based on two new
observations. First, we find that, to mitigate interference, it is
sufficient to separate applications into only two groups, one
containing applications that are vulnerable to interference and
another containing applications that cause interference, instead
of ranking individual applications with a total order.
Vulnerable-to-interference group is prioritized over the
interference-causing group. Second, we show that this grouping can
be efficiently performed by simply counting the number of
consecutive requests served from each application -- an
application that has a large number of consecutive requests served
is dynamically classified as interference-causing.

We evaluate \bliss across a wide variety of workloads and system
configurations and compare its performance and hardware complexity
(via RTL implementations), with five state-of-the-art memory
schedulers. Our evaluations show that \bliss achieves 5\% better
system performance and 25\% better fairness than the
best-performing previous memory scheduler while greatly reducing
critical path latency and hardware area cost of the memory
scheduler (by 79\% and 43\%, respectively), thereby achieving
a good trade-off between performance, fairness and hardware
complexity.
\end{abstract}

%% file: sections/introduction.tex
\section{Introduction}

In modern systems, the high latency of accessing large-capacity
off-chip memory and limited memory bandwidth have made main memory a
critical performance bottleneck. In a multicore system, main memory is
typically shared by applications running on different cores (or,
hardware contexts). Requests from such applications contend for the
off-chip memory bandwidth, resulting in interference. Several prior
works~\cite{fqm,mph,stfm,parbs} demonstrated that this
inter-application interference can severely degrade overall system
performance and fairness. This problem will likely get worse as the
number of cores on a multicore chip increases~\cite{mph}.

Prior works proposed different solution approaches to mitigate
inter-application interference, with the goal of improving system
performance and
fairness (e.g.,~\cite{stfm,parbs,podc-08,atlas,tcm,crit-scheduling-cornell,mcp,mop,fst,mise,hyoseung-rtas14,tang-thread-scheduling,zhuravlev-thread-scheduling}).
A prevalent solution direction is application-aware memory request
scheduling (e.g.,~\cite{stfm,parbs,podc-08,atlas,tcm,mise}).  The basic
idea of application-aware memory scheduling is to prioritize
requests of different applications differently, based on the
applications' memory access characteristics.  State-of-the-art
application-aware memory schedulers typically i) \emph{monitor}
applications' memory access characteristics, ii) \emph{rank
applications individually} based on these characteristics such
that applications that are vulnerable to interference are ranked
higher and iii) \emph{prioritize} requests based on the computed
ranking.

We observe that there are two major problems with past
ranking-based schedulers. First, such schedulers trade off
hardware complexity in order to improve performance or fairness.
They incur high hardware complexity (logic and storage overhead as
well as critical path latency) to schedule requests based on a
scheme that ranks individual applications with a total order. As a
result, the critical path latency and chip area cost of such
schedulers are significantly higher compared to
application-unaware schedulers.  For example, as we demonstrate in
Section~\ref{sec:complexity}, based on our RTL designs,
TCM~\cite{tcm}, a state-of-the-art application-aware scheduler is
8x slower and 1.8x larger than a commonly-employed
application-unaware scheduler, FRFCFS~\cite{frfcfs}. Second,
such schedulers not only increase hardware complexity, but
also cause unfair slowdowns. When a total order based ranking is
employed, applications that are at the bottom of the ranking stack
get heavily deprioritized and unfairly slowed down. This greatly
degrades system fairness.

\textbf{Our goal}, in this work, is to design a new memory
scheduler that does not suffer from these two problems: one that
achieves high system performance \emph{and} fairness {\em while}
incurring low hardware cost and low scheduling latency. To this
end, we propose the {\em Blacklisting memory scheduler} ({\em
\bliss}). Our \bliss design is based on two new observations.

\textbf{Observation 1.} In contrast to forming a total rank order
of all applications (as done in prior works), we find that, to
mitigate interference, it is sufficient to i) separate
applications into \emph{only two} groups, one group containing
applications that are vulnerable to interference and another
containing applications that cause interference, and ii)
prioritize the requests of the {\em vulnerable-to-interference}
group over the requests of the {\em interference-causing} group.
Although one prior work, TCM~\cite{tcm}, proposed to group
applications based on memory intensity, TCM ranks applications
individually within each group and enforces the total rank order
during scheduling. Our approach overcomes the two major problems
with such schedulers that employ per-application ranking. First,
separating applications into only two groups, as opposed to
employing ranking based on a total order of applications,
significantly reduces hardware complexity
(Section~\ref{sec:complexity}). Second, since our approach
prioritizes only one dynamically-determined group of applications
over another dynamically-determined group, no single application
is heavily deprioritized, improving overall system fairness
(Section~\ref{sec:eval}).

\textbf{Observation 2.} We observe that applications can be
efficiently classified as either {\em vulnerable-to-interference}
or {\em interference-causing} by simply counting the number of
consecutive requests served from an application in a short time
interval. Applications with a large number of consecutively-served
requests are classified as interference-causing. The rationale
behind this approach is that when a large number of consecutive
requests are served from the same application, requests of other
applications are more likely to be delayed, causing those
applications to stall. On the other hand, applications with very
few consecutive requests will likely not delay other applications
and are in fact vulnerable to interference from other
applications that have a large number of requests generated and
served. Our approach to classifying applications is simpler to
implement than prior approaches~(e.g.,~\cite{parbs,atlas,tcm})
that use more complicated metrics such as memory intensity,
row-buffer locality, bank-level parallelism or long-term memory
service as proxies for vulnerability to interference
(Section~\ref{sec:complexity}).

{\bf Mechanism Overview.} Based on these two observations, our
mechanism, the Blacklisting Memory Scheduler (\bliss), counts the
number of consecutive requests served from the same application within
a short time interval. When this count exceeds a threshold, BLISS
places the application in the interference-causing group, which we
also call the {\em blacklisted} group. In other words, BLISS {\em
  blacklists} the application such that it is deprioritized. During
scheduling, non-blacklisted (vulnerable-to-interference) applications'
requests are given higher priority over requests of blacklisted
(interference-causing) applications. No per-application ranking is
employed. Prioritization is based solely on two groups as opposed to a
total order of applications.

This paper makes the following contributions:

\squishlist
\item
We present two new observations on how a simple grouping scheme
that avoids per-application ranking can mitigate interference,
based on our analyses and studies of previous memory schedulers.
These observations can enable simple and effective memory
interference mitigation techniques including and beyond the ones
we propose in this work.
\item 
We propose the Blacklisting memory scheduler (\bliss), which
achieves high system performance and fairness while incurring low
hardware cost and complexity. The key idea is to separate
applications into \emph{only two} groups, {\em
vulnerable-to-interference} and {\em interference-causing}, and
deprioritize the latter during scheduling, rather than ranking
individual applications with a total order based on their access
characteristics (like prior work did).
\item
We provide a comprehensive complexity analysis of five previously
proposed memory schedulers, comparing their critical path latency
and area via RTL implementations (Section~\ref{sec:eval}). Our
results show that \bliss reduces critical path latency/area of the
memory scheduler by 79\%/43\% respectively, compared to the
best-performing ranking-based scheduler, TCM~\cite{tcm}.
\item
We evaluate \bliss against five previously-proposed memory
schedulers in terms of system performance and fairness across a
wide range of workloads (Section~\ref{sec:complexity}). Our
results show that \bliss achieves 5\% better system performance
and 25\% better fairness than the best-performing previous scheduler,
TCM~\cite{tcm}.

\item
We evaluate the trade-off space between performance, fairness and
hardware complexity for five previously-proposed memory schedulers
and \bliss
(Section~\ref{sec:tradeoff-performance-fairness-complexity}). We
demonstrate that \bliss achieves the best trade-off between
performance, fairness and complexity, compared to previous memory
schedulers.
\squishend

%% file: sections/background.tex
\section{Background and Motivation}
\label{sec:background}

In this section, we first provide a brief background on the
organization of a DRAM main memory system. We then describe
previous memory scheduling proposals and their
shortcomings that motivate the need for a new memory scheduler -
our Blacklisting memory scheduler.

\subsection{DRAM Background}
The DRAM main memory system is organized hierarchically as
channels, ranks and banks. Channels are independent and can
operate in parallel. Each channel consists of ranks (typically 1 -
4) that share the command, address and data buses of the channel.
A rank consists of multiple banks that can operate in parallel.
However, all banks within a channel share the command, address and
data buses of the channel. Each bank is organized as a
two-dimensional array of rows and columns. On a data access, the
entire row containing the data is brought into an internal
structure called the row buffer. Therefore, a subsequent access
to the same row can be served from the row buffer itself and need
not access the array. Such an access is called a row hit. On an
access to a different row, however, the array itself needs to be
accessed. Such an access is called a row miss/conflict. A row hit
is served {\raise.17ex\hbox{$\scriptstyle\mathtt{\sim}$}}2-3x
faster than a row miss/conflict~\cite{jedec-ddr3}. For more detail
on DRAM operation, we refer the reader
to~\cite{salp,tldram,aldram,rowclone}.

\subsection{Memory Scheduling}
Commonly employed memory controllers employ a memory scheduling
policy called First Ready First Come First Served
(FRFCFS)~\cite{frfcfs-patent,frfcfs} that leverages the row buffer
by prioritizing row hits over row misses/conflicts. Older requests
are then prioritized over newer requests. FRFCFS aims to maximize
DRAM throughput by prioritizing row hits. However, it unfairly
prioritizes requests of applications that generate a large number
of requests to the same row (high-row-buffer-locality) and access
memory frequently (high-memory-intensity)~\cite{mph,stfm}.
Previous work (e.g.,~\cite{stfm,parbs,podc-08,atlas,tcm}) proposed
application-aware memory scheduling techniques that take into
account the memory access characteristics of applications and
schedule requests appropriately in order to mitigate
inter-application interference and improve system performance and
fairness. We will focus on four state-of-the-art schedulers, which
we evaluate quantitatively in Section~\ref{sec:eval}.

Mutlu and Moscibroda propose PARBS~\cite{parbs}, an application-aware
memory scheduler that batches the oldest requests from applications
and prioritizes the batched requests, with the goals of preventing
starvation and improving fairness. Within each batch, PARBS ranks individual applications
based on the number of outstanding requests of each application and,
using this total rank order, prioritizes requests of applications that
have low-memory-intensity to improve system throughput.

Kim et al.~\cite{atlas} observe that applications that receive low
memory service tend to experience interference from applications that
receive high memory service. Based on this observation, they propose
ATLAS, an application-aware memory scheduling policy that ranks
individual applications based on the amount of long-term memory
service each receives and prioritizes applications that receive low
memory service, with the goal of improving overall system throughput.

Thread cluster memory scheduling (TCM)~\cite{tcm} ranks individual
applications by memory intensity such that low-memory-intensity
applications are prioritized over high-memory-intensity
applications (to improve system throughput). Kim et al.~\cite{tcm}
also observed that ranking all applications based on memory
intensity and prioritizing low-memory-intensity applications could
slow down the deprioritized high-memory-intensity applications
significantly and unfairly. With the goal of mitigating this
unfairness, TCM clusters applications into low and high
memory-intensity clusters and employs a different ranking scheme
in each cluster. In the low-memory-intensity cluster, applications
are ranked by memory intensity, whereas, in the
high-memory-intensity cluster, applications' ranks are shuffled to
provide fairness. Both clusters employ a total rank order among
applications at any given time.

More recently, Ghose et al.~\cite{crit-scheduling-cornell} propose
a memory scheduler that aims to prioritize \emph{critical} memory
requests that stall the instruction window for long lengths of
time. The scheduler predicts the criticality of a load instruction
based on how long it has stalled the instruction window in the
past (using the instruction address (PC)) and prioritizes requests
from load instructions that have large total and maximum stall
times measured over a period of time. Although this scheduler is
not application-aware, we compare to it as it is the most recent
scheduler that aims to maximize performance by mitigating memory
interference.
\subsection{Shortcomings of Previous Schedulers}
\label{sec:background-shortcomings}
These state-of-the-art schedulers attempt to achieve two main goals -
high system performance and high fairness. However, previous
schedulers have two major shortcomings. First, these schedulers
increase hardware complexity in order to achieve high system
performance and fairness. Specifically, most of these schedulers rank
individual applications with a total order, based on their memory
access characteristics (e.g.,~\cite{parbs,podc-08,atlas,tcm}). Scheduling requests
based on a total rank order incurs high hardware complexity, as we
demonstrate in Section~\ref{sec:complexity}, slowing down the memory
scheduler significantly (by 8x for TCM compared to FRFCFS), while also
increasing its area (by 1.8x). Such high critical path delays in the
scheduler directly increase the time it takes to schedule a request,
potentially making the memory controller latency a bottleneck. Second,
a total-order ranking is unfair to applications at the bottom of the ranking stack.
Even shuffling the ranks periodically (like TCM does) does not fully
mitigate the unfairness and slowdowns experienced by an application
when it is at the bottom of the ranking stack, as we show in
Section~\ref{sec:observations}.

\begin{figure}[h]
    \vspace{-4mm}
    \centering
    \includegraphics[scale=0.25, angle=0]{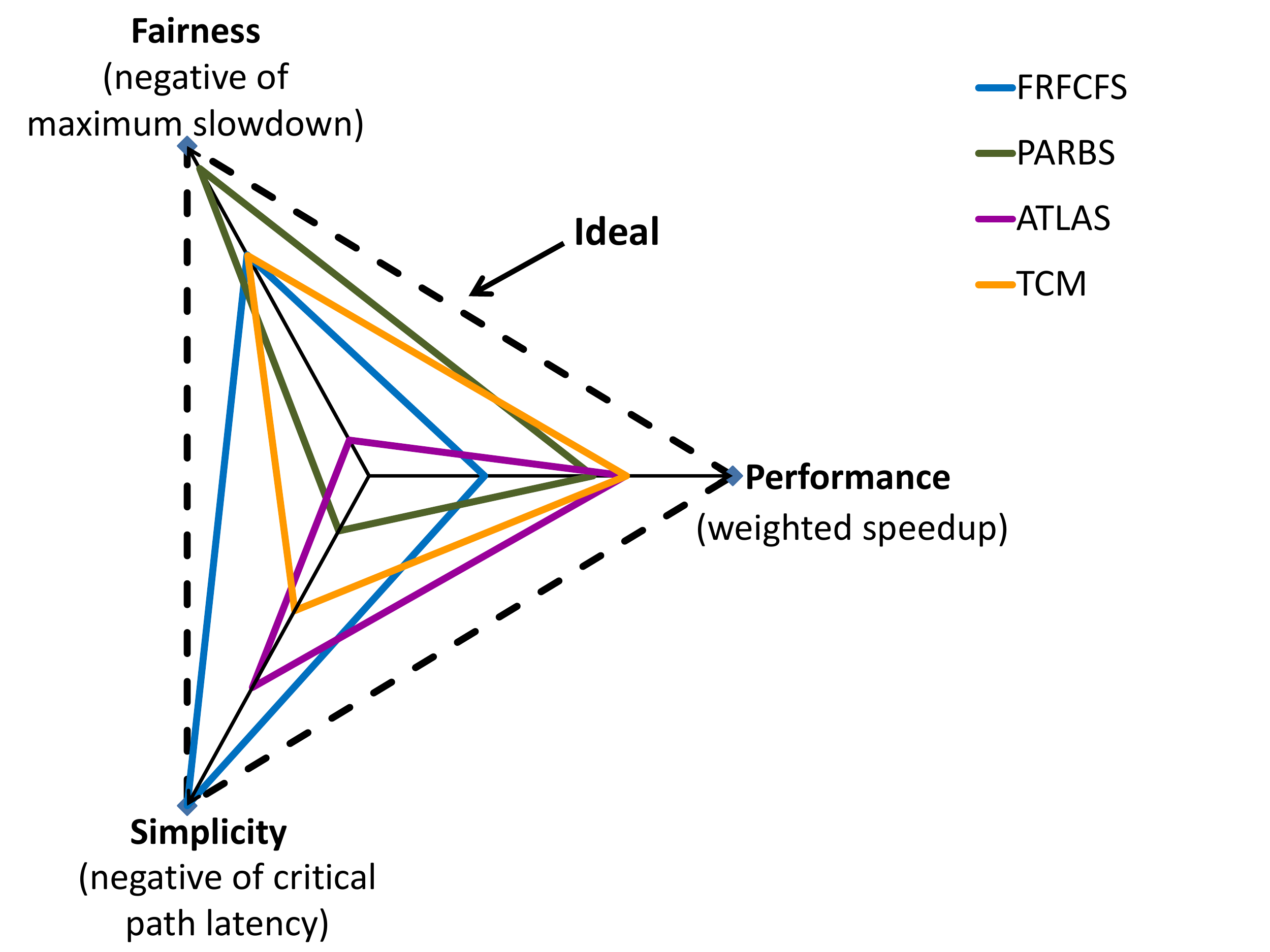}
    \vspace{-2mm}
    \caption{Performance vs. fairness vs. simplicity}
    \label{fig:perf-fairness-simplicity-initial}
    \vspace{-3mm}
\end{figure}

Figure~\ref{fig:perf-fairness-simplicity-initial} compares four
major previous schedulers using a three-dimensional plot with
performance, fairness and simplicity on three different
axes.\footnote{Results across 80 simulated workloads on a 24-core,
4-channel system. Section~\ref{sec:methodology} describes our
methodology and metrics.} On the fairness axis, we plot the negative of
maximum slowdown, and on the simplicity axis, we plot the negative
of critical path latency. Hence, the ideal scheduler would have
high performance, fairness and simplicity, as indicated by the
black triangle.  As can be seen, previous ranking-based
schedulers, PARBS, ATLAS and TCM, increase complexity
significantly, compared to the currently employed FRFCFS
scheduler, in order to achieve high performance and/or fairness.

\textbf{Our goal}, in this work, is to design a new memory scheduler
that does not suffer from these shortcomings: one that achieves high
system performance and fairness {\em while} incurring low hardware
cost and complexity. To this end, we propose the {\em Blacklisting
  memory scheduler} ({\em \bliss}) based on two new observations
described in the next section.

%% file: sections/observations.tex
\section{Key Observations}
\label{sec:observations}

\begin{sloppypar}
As we described in the previous section, several major
state-of-the-art memory schedulers rank individual applications
with a total order, to mitigate inter-application interference.
While such ranking is one way to mitigate interference, it has
shortcomings, as described in
Section~\ref{sec:background-shortcomings}. We seek to overcome
these shortcomings by exploring an alternative means to protecting
vulnerable applications from interference. We make two key
observations on which we build our new memory scheduling
mechanism.
\end{sloppypar}

\begin{figure*} [ht!]
\vspace{-2mm}
  \centering 
  \begin{subfigure}{0.31\textwidth} 
    \centering
    \includegraphics[scale=0.16, angle=270]{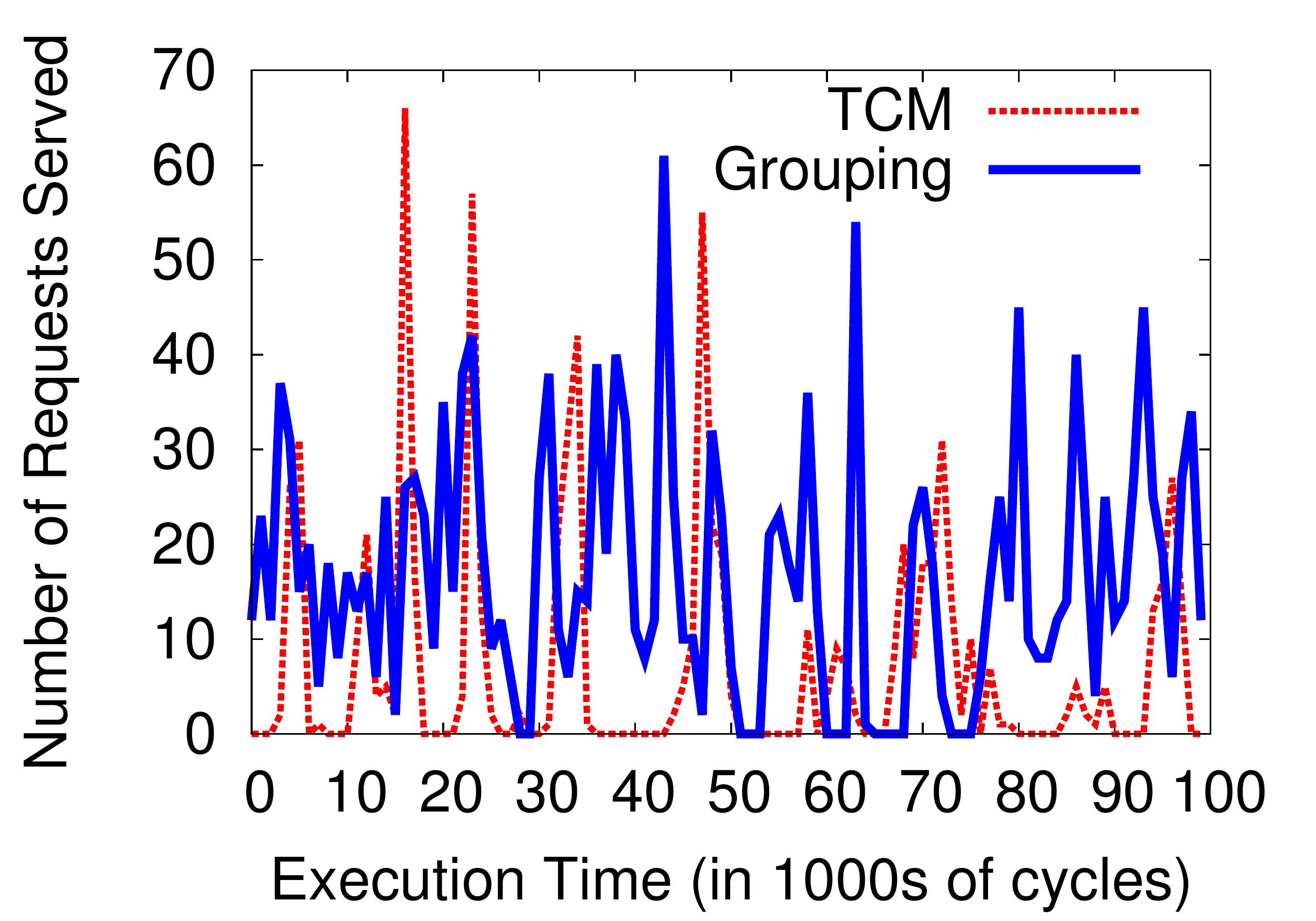}
     \vspace{-1mm}
    \caption{astar}
  \end{subfigure} 
  \begin{subfigure}{0.31\textwidth}
    \centering 
    \includegraphics[scale=0.16, angle=270]{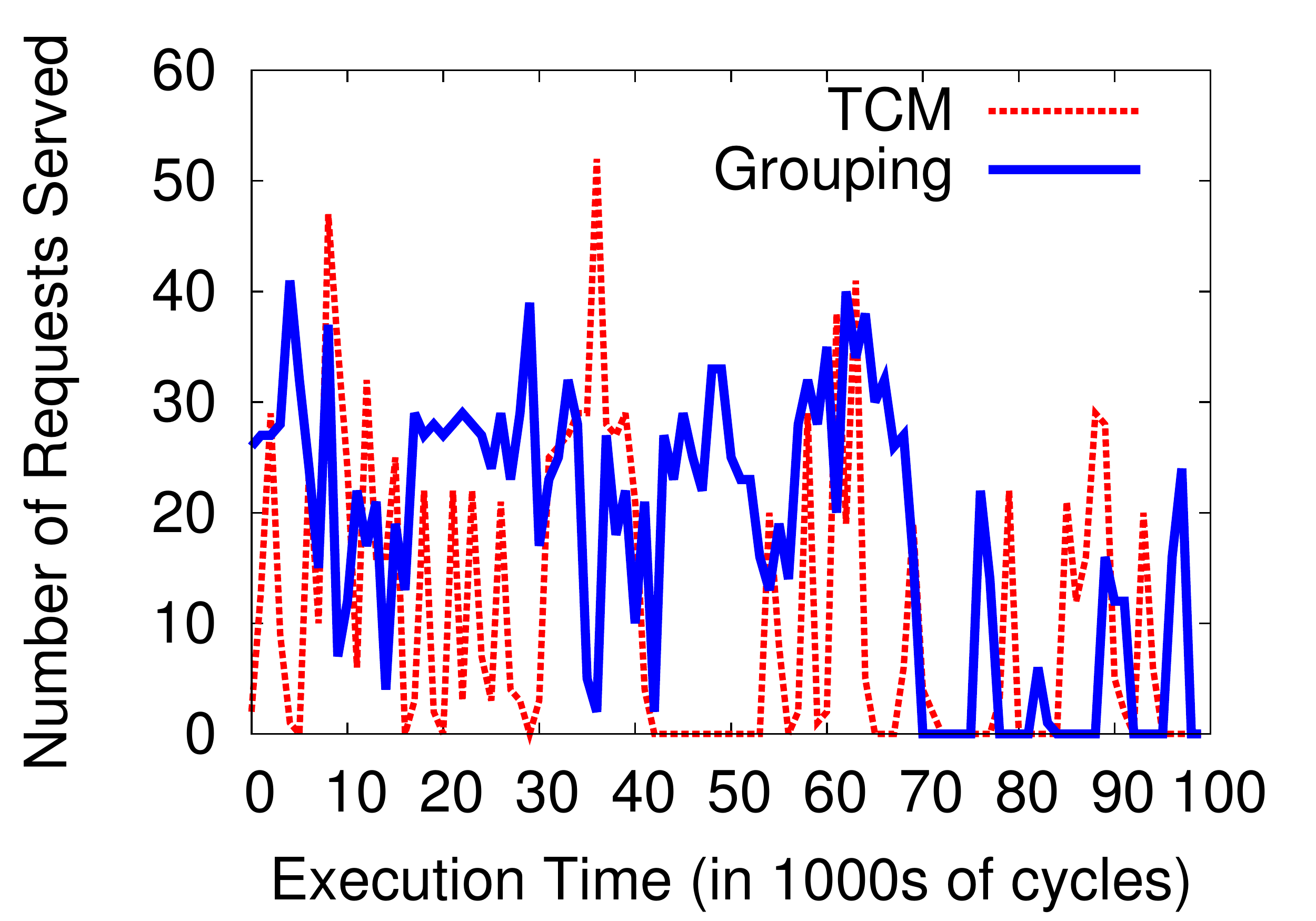}
     \vspace{-1mm}
    \caption{hmmer}
  \end{subfigure} 
  \begin{subfigure}{0.31\textwidth}
    \centering 
    \includegraphics[scale=0.16, angle=270]{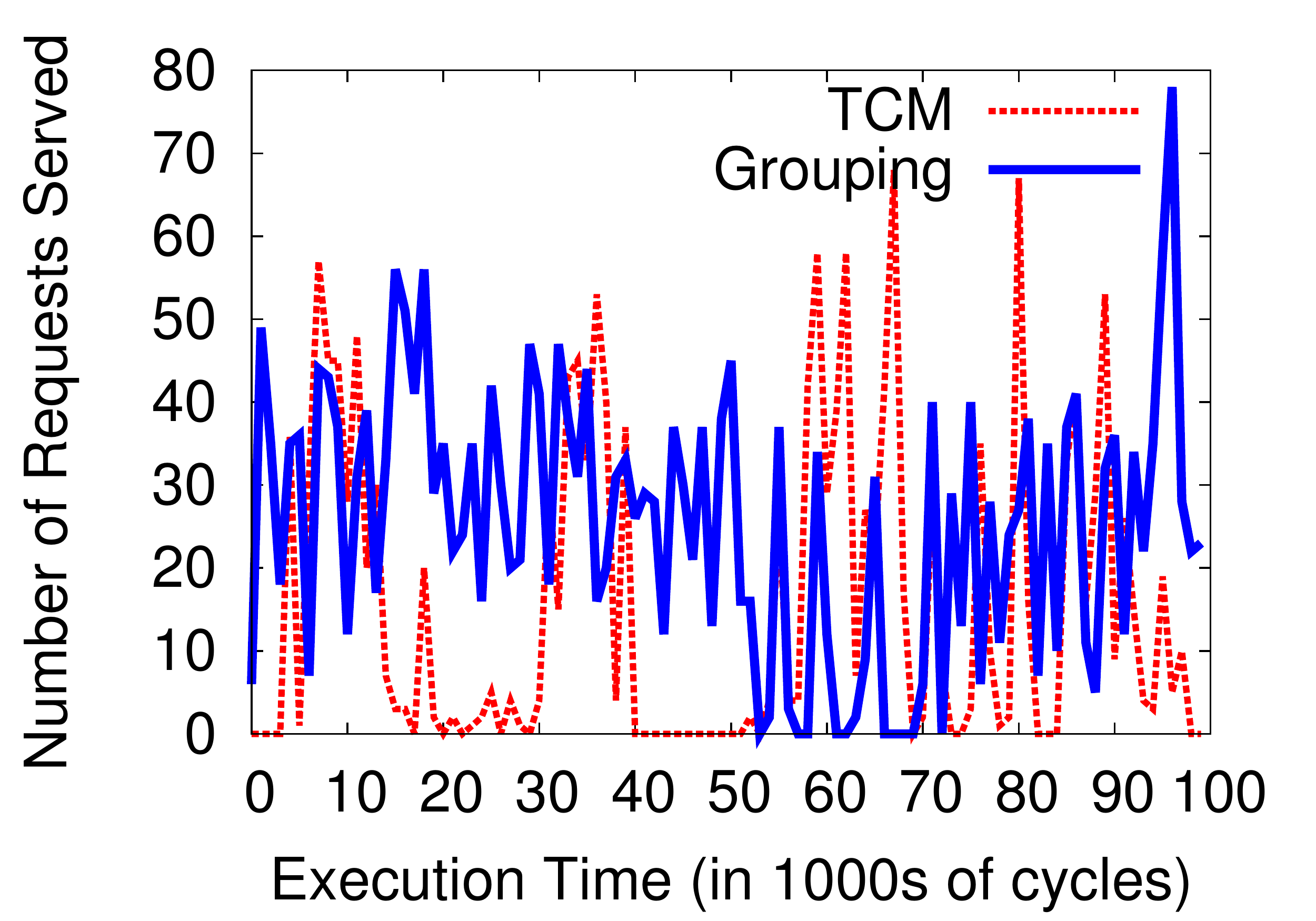}
     \vspace{-1mm}
    \caption{lbm}
  \end{subfigure} 
  \vspace{-2mm}
  \caption{Request service distribution over time with TCM and Grouping schedulers}
  \vspace{-1mm}
  \label{fig:request-dist}
\vspace{-4mm}
\end{figure*}
 
\textbf{Observation 1. } {\em Separating applications into only
two groups (interference-causing and vulnerable-to-interference),
without ranking individual applications using a total order, is sufficient to mitigate
inter-application interference. This leads to higher performance,
fairness and lower complexity, all at the same time.}

We observe that applications that are vulnerable to interference
can be protected from interference-causing applications by simply
separating them into two groups, one containing
interference-causing applications and another containing
vulnerable-to-interference applications, rather than ranking
individual applications with a total order as many
state-of-the-art schedulers do. To motivate this, we contrast
TCM~\cite{tcm}, which clusters applications into two groups and
employs a total rank order within each cluster, with a simple
scheduling mechanism ({\em Grouping}) that simply groups
applications only into two groups, based on memory intensity (as
TCM does), and prioritizes the low-intensity group {\em without}
employing ranking in each group. {\em Grouping} uses the FRFCFS
policy within each group. Figure~\ref{fig:request-dist} shows the
number of requests served during a 100,000 cycle period at
intervals of 1,000 cycles, for three representative applications,
astar, hmmer and lbm from the SPEC CPU2006 benchmark
suite~\cite{spec2006}, using these two schedulers.\footnote{All
these three applications are in the high-memory-intensity group.
We found very similar behavior in all other such applications we
examined.} These three applications are executed with other
applications in a simulated 24-core 4-channel system.\footnote{See
Section~\ref{sec:methodology} for our methodology.}

Figure~\ref{fig:request-dist} shows that TCM has high variance in
the number of requests served across time, with very few requests
being served during several intervals and many requests being
served during a few intervals. This behavior is seen in most
applications in the high-memory-intensity cluster since TCM ranks
individual applications with a total order. This ranking causes
some high-memory-intensity applications' requests to be
prioritized over \emph{other} high-memory-intensity applications'
requests, at any point in time, resulting in high interference.
Although TCM periodically shuffles this total-order ranking, we observe that
an application benefits from ranking \emph{only} during those periods
when it is ranked very high. These very highly ranked periods
correspond to the spikes in the number of requests served (for
TCM) in Figure~\ref{fig:request-dist} for that application. During
the other periods of time when an application is ranked lower
(i.e., most of the {\em shuffling intervals}), only a small number
of its requests are served, resulting in very slow progress.
Therefore, most high-memory-intensity applications experience high
slowdowns due to the total-order ranking employed by TCM.

\begin{figure} [ht!]
  \vspace{-2mm} 
  \centering 
  \begin{minipage}{0.24\textwidth}
    \centering 
    \includegraphics[scale=0.15, angle=270]{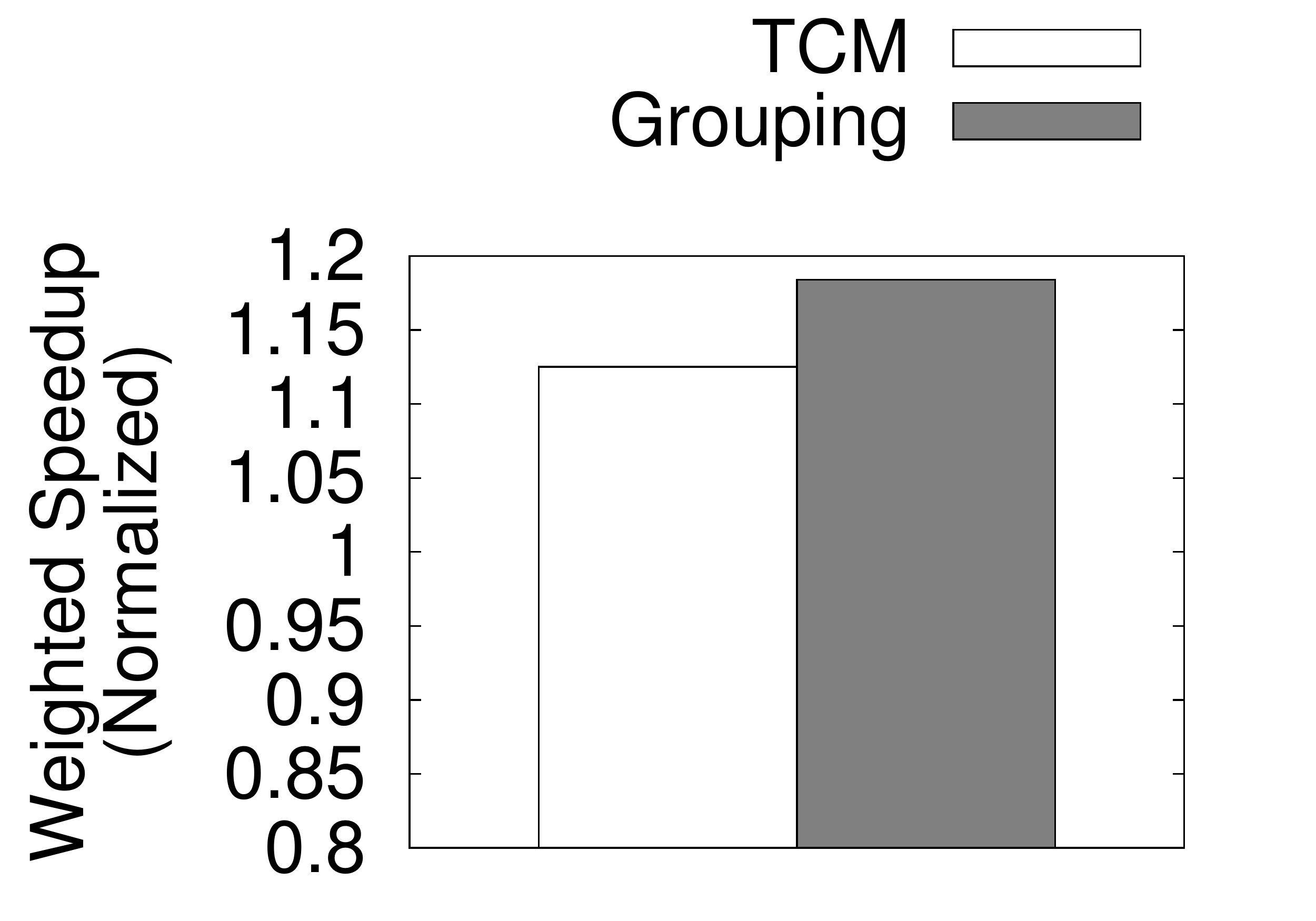}
  \end{minipage} 
%  \hspace{10pt}
  \begin{minipage}{0.24\textwidth} 
    \centering
    \includegraphics[scale=0.15, angle=270]{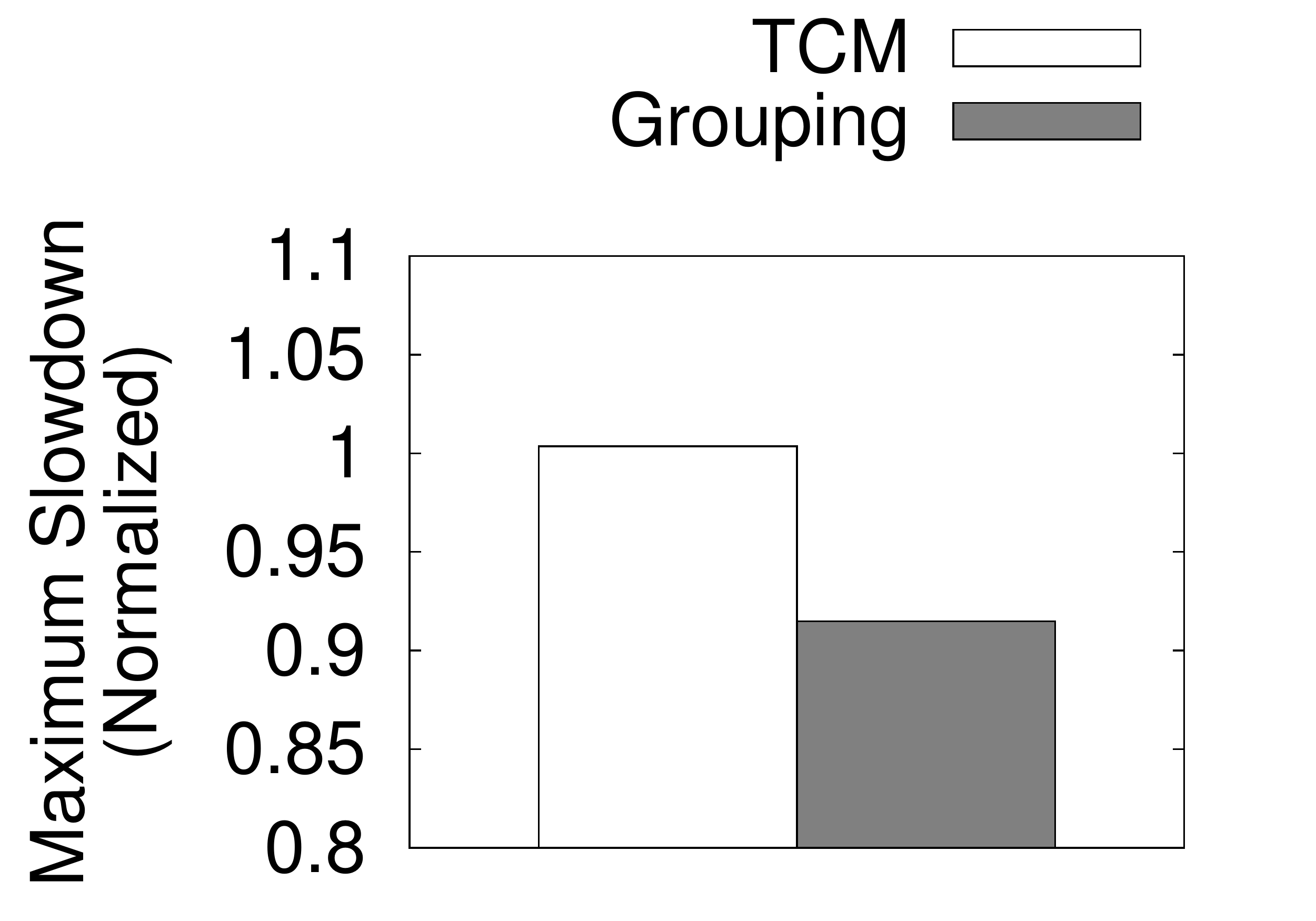}
  \end{minipage} 
  \vspace{-2mm} 
  \caption{Performance and fairness of Grouping vs. TCM}
  \label{fig:tcm-clustering}
  \vspace{-6mm} 
\end{figure}

On the other hand, when applications are separated into only two
groups based on memory intensity and no per-application ranking is
employed within a group, some interference exists among
applications within each group (due to the application-unaware
FRFCFS scheduling in each group). In the high-memory-intensity
group, this interference contributes to the few
low-request-service periods seen for {\em Grouping} in
Figure~\ref{fig:request-dist}. However, the request service
behavior of {\em Grouping} is less spiky than that of TCM, resulting
in lower memory stall times and a more steady and overall higher
progress rate for high-memory-intensity applications, as compared
to when applications are ranked in a total order. In the
low-memory-intensity group, there is not much of a difference
between TCM and \emph{Grouping}, since applications anyway have low
memory intensities and hence, do not cause significant
interference to each other. Therefore, {\em Grouping} results in
higher system performance and significantly higher fairness than
TCM, as shown in Figure~\ref{fig:tcm-clustering} (across 80
24-core workloads on a simulated 4-channel system).

Solely grouping applications into two also requires much lower hardware
overhead than ranking-based schedulers that incur high overhead for
computing and enforcing a total rank order for all
applications. Therefore, grouping can not only achieve better system
performance and fairness than ranking, but it also can do so while
incurring lower hardware cost. However, classifying applications into
two groups at coarse time granularities, on the order of a few million
cycles, like TCM's clustering mechanism does (and like what we have
evaluated in Figure~\ref{fig:tcm-clustering}), can still cause unfair
application slowdowns. This is because applications in one group would
be deprioritized for a long time interval, which is especially
dangerous if application behavior changes during the interval. Our
second observation, which we describe next, minimizes such unfairness
and at the same time reduces the complexity of grouping even further.

\textbf{Observation 2. }{\em Applications can be classified into
\emph{interference-causing} and \emph{vulnerable-to-interference}
groups by monitoring the number of consecutive requests served
from each application at the memory controller. This leads to
higher fairness and lower complexity, at the same time, than
grouping schemes that rely on coarse-grained memory intensity
measurement.}

Previous work actually attempted to perform grouping, along with ranking, to
mitigate interference. Specifically, TCM~\cite{tcm} ranks applications
by memory intensity and classifies applications that make up a certain
fraction of the total memory bandwidth usage into a \emph{group} called the
\emph{low-memory-intensity cluster} and the remaining applications
into a second group called the \emph{high-memory-intensity cluster}. While employing such a grouping
scheme, without ranking individual applications, reduces hardware
complexity and unfairness compared to a total order based ranking
scheme (as we show in Figure~\ref{fig:tcm-clustering}), it i) \emph
{can still cause unfair slowdowns due to classifying applications into
  groups at coarse time granularities, which is especially dangerous
  if application behavior changes during an interval}, and ii)
\emph{incurs additional hardware overhead and scheduling latency to
  compute and rank applications by long-term memory intensity and total memory
  bandwidth usage}.

We propose to perform application grouping using a significantly
simpler, novel scheme: simply by counting the number of requests served from each
application in a short time interval. Applications that have a
large number (i.e., above a threshold value) of consecutive
requests served are classified as interference-causing (this
classification is periodically reset). The rationale behind this
scheme is that when an application has a large number of
consecutive requests served within a short time period, which is
typical of applications with high memory intensity or
high row-buffer locality, it delays other applications' requests,
thereby stalling their progress. Hence, identifying and
essentially \emph{blacklisting} such interference-causing
applications by placing them in a separate group and
deprioritizing requests of this blacklisted group can prevent such
applications from hogging the memory bandwidth. As a result, the
interference experienced by vulnerable applications is mitigated.
The blacklisting classification is cleared periodically, at short
time intervals (on the order of 1000s of cycles) in order not to
deprioritize an application for too long of a time period to cause
unfairness or starvation. Such clearing and re-evaluation of application
classification at short time intervals significantly reduces unfair
application slowdowns (as we quantitatively show in
Section~\ref{sec:tcm-clustering}), while reducing complexity
compared to tracking per-application metrics such as memory
intensity.

\textbf{Summary of Key Observations.} In summary, we make two key
novel observations that lead to our design in
Section~\ref{sec:mechanism}. First, separating applications into
only two groups can lead to a less complex and more fair and
higher performance scheduler. Second, the two application groups
can be formed seamlessly by monitoring the number of consecutive
requests served from an application and deprioritizing the ones
that have too many requests served in a short time interval.

%% file: sections/mechanism1.tex
\section{Mechanism}
\label{sec:mechanism}

In this section, we present the details of our Blacklisting memory
scheduler (\bliss) that employs a simple grouping scheme motivated by
our key observations from Section~\ref{sec:observations}. The basic
idea behind \bliss is to observe the number of consecutive requests
served from an application over a short time interval and blacklist
applications that have a relatively large number of consecutive
requests served. The blacklisted (interference-causing) and
non-blacklisted (vulnerable-to-interference) applications are thus
separated into two different groups. The memory scheduler then
prioritizes the non-blacklisted group over the blacklisted group. The
two main components of \bliss are i) the blacklisting mechanism and
ii) the memory scheduling mechanism that schedules requests based on
the blacklisting mechanism. We describe each in turn.

\subsection{The Blacklisting Mechanism}
\label{sec:blacklist-mechanism}

The blacklisting mechanism needs to keep track of three quantities:
1) the application (i.e., hardware context) ID of the last scheduled
request (\textit{Application ID})\footnote{An application here denotes
  a hardware context. There can be as many applications executing
  actively as there are hardware contexts. Multiple hardware contexts
  belonging to the same application are considered separate
  applications by our mechanism, but our mechanism can be extended to
  deal with such multithreaded applications.}, 2) the number of
requests served from an application (\textit{\#Requests Served}), and 3)
the blacklist status of each application.

When the memory controller is about to issue a request, it
compares the application ID of the request with the
\textit{Application ID} of the \emph{last scheduled request}.
\squishlist
\item 
If the application IDs of the two requests are the same, the
\textit{\#Requests Served} counter is incremented.
\item
If the application IDs of the two requests are not the same, the
\textit{\#Requests Served} counter is reset to zero and the
\textit{Application ID} register is updated with the application ID
of the request that is being issued.  
\squishend

If the \textit{\#Requests Served} exceeds a \textit{Blacklisting
  Threshold} (set to 4 in most of our evaluations): 
\squishlist
\item
The application with ID \textit{Application ID} is blacklisted
(classified as interference-causing).
\item
The \textit{\#Requests Served} counter is reset to zero.
\squishend

The blacklist information is cleared periodically after every
\textit{Clearing Interval} (set to 10000 cycles in our major
evaluations).

\subsection{Blacklist-Based Memory Scheduling}

Once the blacklist information is computed, it is used to determine
the scheduling priority of a request. Memory requests are prioritized
in the following order:

\begin{enumerate}\itemsep0em
\item Non-blacklisted applications' requests
\item Row-buffer hit requests
\item Older requests
\end{enumerate}

Prioritizing requests of non-blacklisted applications over
requests of blacklisted applications mitigates interference.
Row-buffer hits are then prioritized to optimize DRAM bandwidth
utilization. Finally, older requests are prioritized over younger
requests for forward progress.

%% file: sections/implementation.tex
\section{Implementation}

The Blacklisting memory scheduler requires additional storage (flip
flops) and logic over an FRFCFS scheduler to 1) perform blacklisting
and 2) prioritize non-blacklisted applications' requests. We
analyze the storage and logic cost of it. 

\subsection{Storage Cost}
In order to perform blacklisting, the memory scheduler needs the
following storage components:
\squishlist
\item one register to store \textit{Application ID} 
%(5 bits for 24 applications)
\item one counter for \textit{\#Requests Served} 
%(8 bits is more
%than sufficient for the values of request count threshold
%\textrm{N} that we observe achieves high performance and fairness.)
\item one register to store the \textit{Blacklisting Threshold} that
determines when an application should be blacklisted
\item a blacklist bit vector to indicate the blacklist
status of each application (one bit for each hardware context)
%(24 bits for 24 applications)
\squishend

%\noindent
In order to prioritize non-blacklisted applications' requests, the
memory controller needs to store the application ID (hardware context
ID) of each request so it can determine the blacklist status of the
application and appropriately schedule the request.

\subsection{Logic Cost}
The memory scheduler requires comparison logic to
\squishlist
\item
determine when an application's \textit{\#Requests Served} exceeds
the \textit{Blacklisting Threshold} and set the bit corresponding to the
application in the \textit{Blacklist} bit vector.
\item
prioritize non-blacklisted applications' requests.
\squishend

%\noindent
We provide a detailed quantitative evaluation of the hardware area
cost and logic latency of implementing \bliss and previously
proposed memory schedulers, in Section~\ref{sec:complexity}.

%%% ONUR-8-24: Not a huge deal but I am not sure if noindent is a good
%%% idea in all cases above and in the mechanism description.

%% file: sections/methodology.tex
\newcommand{\aloneipc}{\textrm{IPC}_{i}^{\scriptstyle{alone}}\xspace}
\newcommand{\sharedipc}{\textrm{IPC}_{i}^{\scriptstyle{shared}}\xspace}

\section{Methodology}
\label{sec:methodology}

\subsection{System Configuration}
\begin{sloppypar}
We model the DRAM memory system using a cycle-level in-house
DDR3-SDRAM simulator. The simulator was validated against Micron's
behavioral Verilog model~\cite{ddr3verilog} and
DRAMSim2~\cite{dramsim2}. This DDR3 simulator is integrated with a
cycle-level in-house simulator that models out-of-order execution
cores, driven by a Pin~\cite{pin} tool at the frontend, Each core
has a private cache of 512 KB size. We present most of our results
on a system with the DRAM main memory as the only shared resource
in order to isolate the effects of memory bandwidth interference
on application performance. We also present results with shared
caches in Section~\ref{sec:sensitivity-system}. Table~\ref{tab:meth}
provides more details of our simulated system. We perform most of
our studies on a system with 24 cores and 4 channels. We provide a
sensitivity analysis for a wide range of core and channel counts,
in Section~\ref{sec:sensitivity-system}. Each channel has one
rank and each rank has eight banks. We stripe data across channels
and banks at the granularity of a row.
\end{sloppypar}

%%% ONUR-8-24: what if you employ some other granularity of
%%% interleaving? Did you evaluate this and are you providing results?
%%% Can we have more results in a tech report and cite it from the
%%% paper?

%%% I evaluated other interleaving schemes briefly. But, did not
%%% evaluate them thoroughly. I plan to include them in the journal
%%% version

%\vspace{-2mm}
\input{tables/methodology}

%\vspace{10mm}
\subsection{Workloads}
We perform our main studies using 24-core multiprogrammed
workloads made of applications from the SPEC CPU2006
suite~\cite{spec2006}, TPC-C, Matlab and the NAS parallel
benchmark suite~\cite{nas}.\footnote{Each benchmark is single
threaded.} We classify a benchmark as memory-intensive if it has a
Misses Per Kilo Instruction (MPKI) greater than 5 and
memory-non-intensive otherwise. We construct four categories of
workloads (with 20 workloads in each category), with 25, 50, 75
and 100 percent of memory-intensive applications. This makes up a
total of 80 workloads with a range of memory intensities,
constructed using random combinations of benchmarks, modeling a
cloud computing like scenario where workloads of various types are
consolidated on the same node to improve efficiency. We also
evaluate 16-, 32- and 64- core workloads, with different memory
intensities, created using a similar methodology as described
above for the 24-core workloads. We simulate each workload for 100
million representative cycles, as done by previous studies in
memory scheduling~\cite{parbs,atlas,tcm}.

\subsection{Metrics}
We quantitatively compare \bliss with previous memory schedulers in terms of
system performance, fairness and complexity. We use the weighted
speedup~\cite{stc,weighted-speedup,symbjobscheduling} metric to measure system
performance.  We use the maximum slowdown
metric~\cite{stc,atlas,tcm,max-slowdown} to measure unfairness. We report the
harmonic speedup metric~\cite{harmonic-speedup} as another measure of system
performance. The harmonic speedup metric also serves as a measure of balance
between system performance and fairness~\cite{harmonic-speedup}. We report area
in $micro meter^2$ ($um^2$) and scheduler critical path latency in nanoseconds
(ns) as measures of complexity.

%\vspace{-6mm}
%
%\begin{small}
%  \begin{eqnarray*}
%    \textrm{Weighted Speedup} & = & \Sigma_i \frac{\sharedipc}{\aloneipc}\\
%    \textrm{Maximum Slowdown} & = & \textrm{max}\left(\frac{\aloneipc}{\sharedipc}\right)\\
%    \textrm{Harmonic Speedup} & = & N/\left(\Sigma_i\frac{\aloneipc}{\sharedipc}\right)\\
%  \end{eqnarray*}
%\end{small}
%
%\vspace{-8mm}

%%% ONUR-8-24: You can omit the above equations, especially if you
%%% need space to explain stuff related to BLISS.

\subsection{RTL Synthesis Methodology}

In order to obtain timing/area results for \bliss and previous
schedulers, we implement them in Register Transfer Level (RTL), using
Verilog. We synthesize the RTL implementations with a commercial 32 nm
standard cell library, using the Design Compiler tool from Synopsys.

\subsection{Mechanism Parameters}

%%% ONUR-8-24: If you name the thresholds and intervals well, you do
%%% not need to redefine what they are here. Can you?
For \bliss, we use a value of four for \textit{Blacklisting Threshold}, and a
value of 10000 cycles for \textit{Clearing Interval}. These values provide a
good balance between performance and fairness, as we observe from our
sensitivity studies in Section~\ref{sec:sensitivity-algorithm}. For the other
schedulers, we tuned their parameters to achieve high performance and fairness
on our system configurations and workloads. We use a \textit{Marking-Cap} of 5 for
PARBS, cap of 4 for FRFCFS-Cap, \textit{HistoryWeight} of 0.875 for ATLAS,
\textit{ClusterThresh} of 0.2 and \textit{ShuffleInterval} of 1000 cycles for
TCM.

%%% ONUR-8-24: Please use the names provided by the respective papers,
%%% above.

%%% ONUR-8-24: I was mistaken. Ignore the below.
%%% ONUR-8-24: Figure placement. Fig 5 is not references on Page 4,
%%% but it is on top of Page 4. Makes life harder for the reader. I am
%%% assuming you have not fixed the figure placement yet, so you
%%% probably will address this anyway.

%% file: tables/methodology.tex
\begin{table}[h!]
\begin{scriptsize}
  \vspace{-2mm}
  \centering
    \begin{tabular}{ll}
        \toprule
Processor           &  16-64 cores, 5.3GHz, 3-wide issue,\\ & 8 MSHRs, 128-entry instruction window     \\
        \cmidrule(rl){1-2}
Last-level cache    &  64B cache-line, 16-way associative,\\ & 512KB private cache-slice per core     \\
        \cmidrule(rl){1-2} Memory controller   &  128-entry read/write request queue per controller   \\
        \cmidrule(rl){1-2}
\multirow{2}[2]{*}{\centering Memory}              &  Timing: DDR3-1066 (8-8-8)~\cite{micron} \\
 & Organization: 1-8 channels, 1
 rank-per-channel,\\ & 8 banks-per-rank, 8 KB row-buffer \\ 
        \bottomrule
    \end{tabular}%
  \vspace{-1mm}
  \caption{Configuration of the simulated system}
  \label{tab:meth}%
  \vspace{-2mm}
\end{scriptsize}%
\end{table}%

%% file: sections/evaluation.tex
\section{Evaluation}
\label{sec:eval}

\begin{figure*}[ht!]
  \vspace{-2mm}
  \centering
  \begin{minipage}{0.31\textwidth}
    \centering
    \includegraphics[scale=0.175, angle=270]{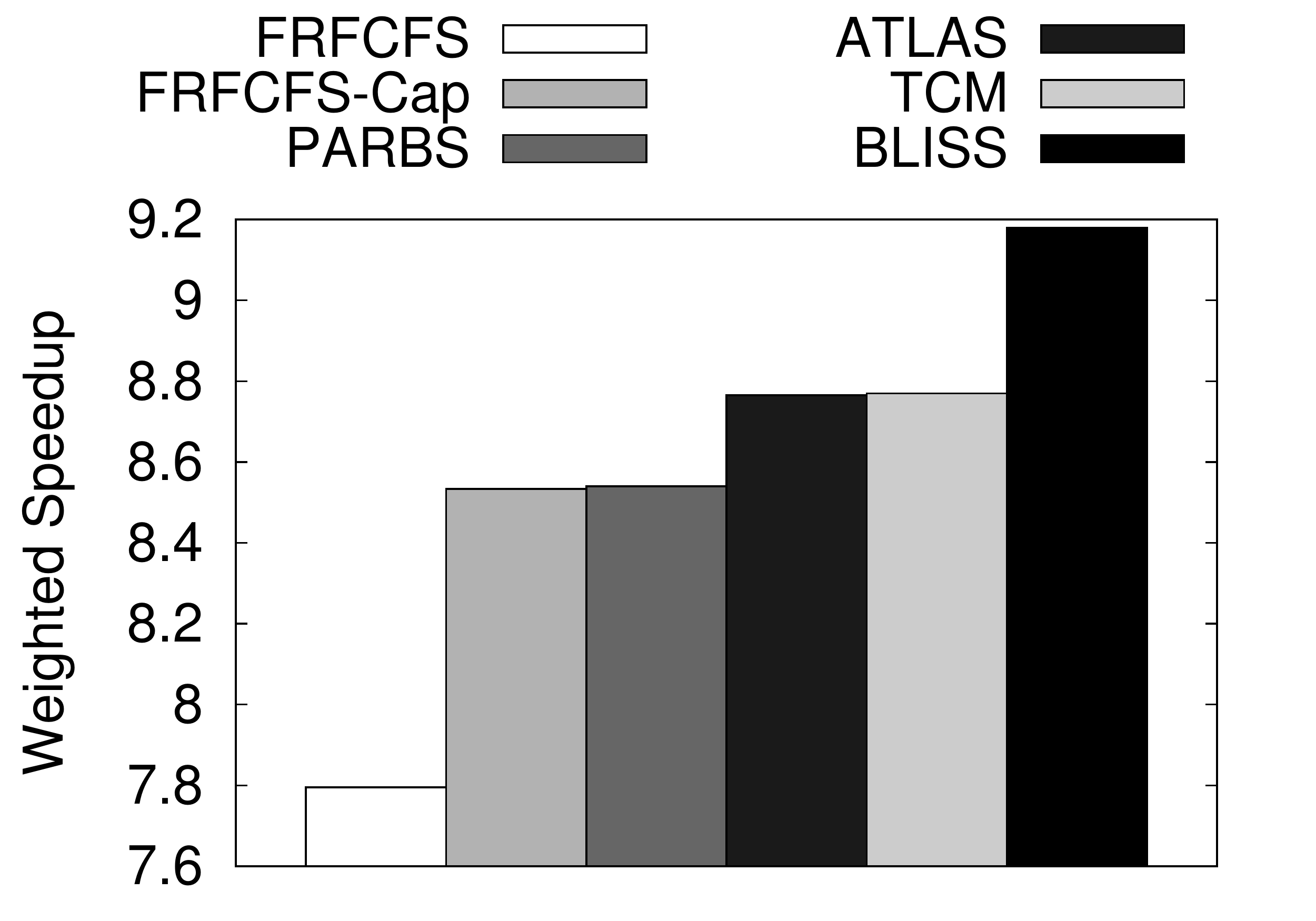}
  \end{minipage}
  \begin{minipage}{0.31\textwidth}
    \centering
    \includegraphics[scale=0.175, angle=270]{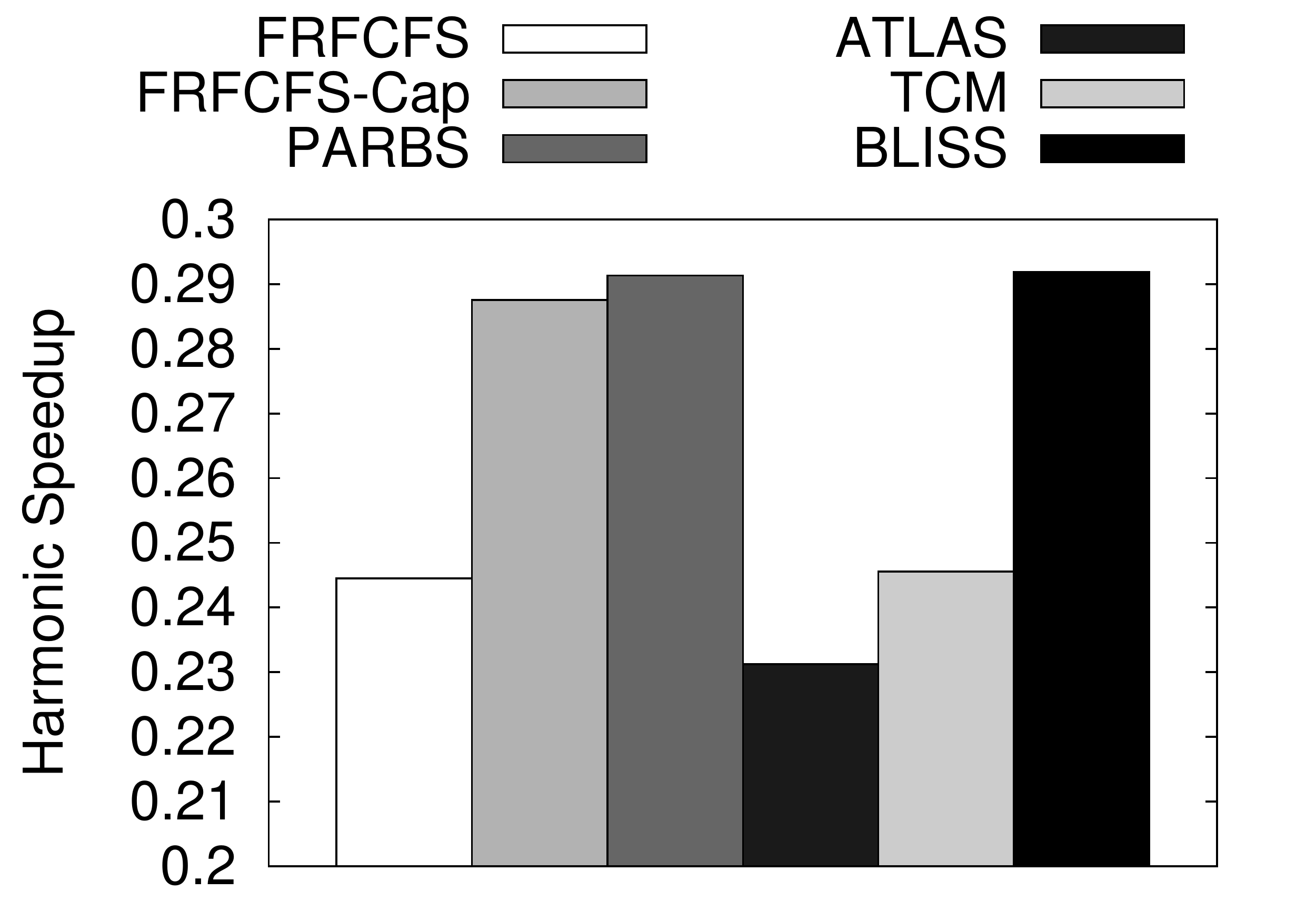}
  \end{minipage}
  \begin{minipage}{0.31\textwidth}
    \centering
    \includegraphics[scale=0.175, angle=270]{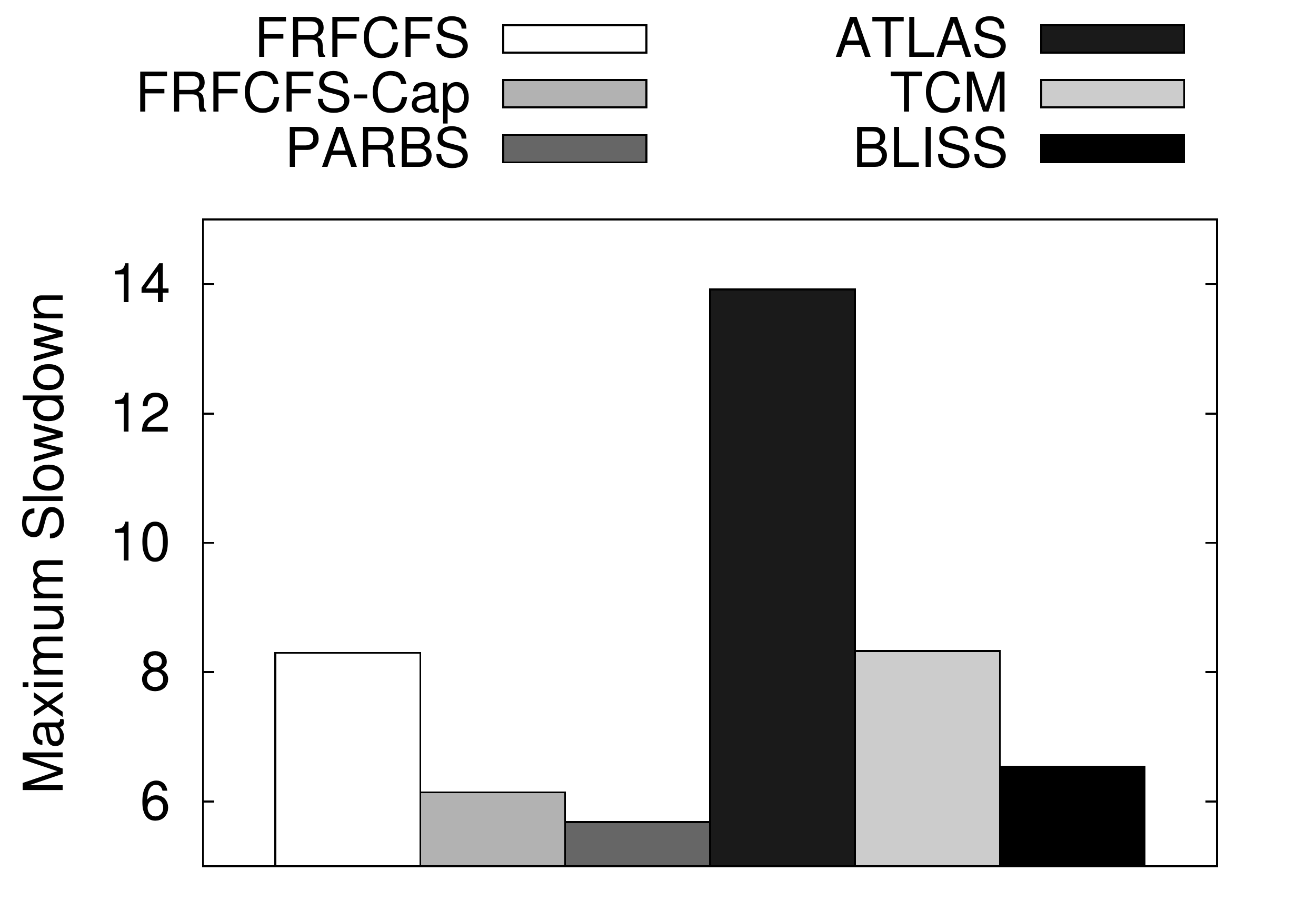}
  \end{minipage}
  \vspace{-2mm}
  \caption{System performance and fairness of \bliss compared to previous schedulers}
  \label{fig:main-results}
  \vspace{-5mm}
\end{figure*}

\begin{figure}[h]
  \vspace{-2mm}
    \centering
    \includegraphics[scale=0.20, angle=0]{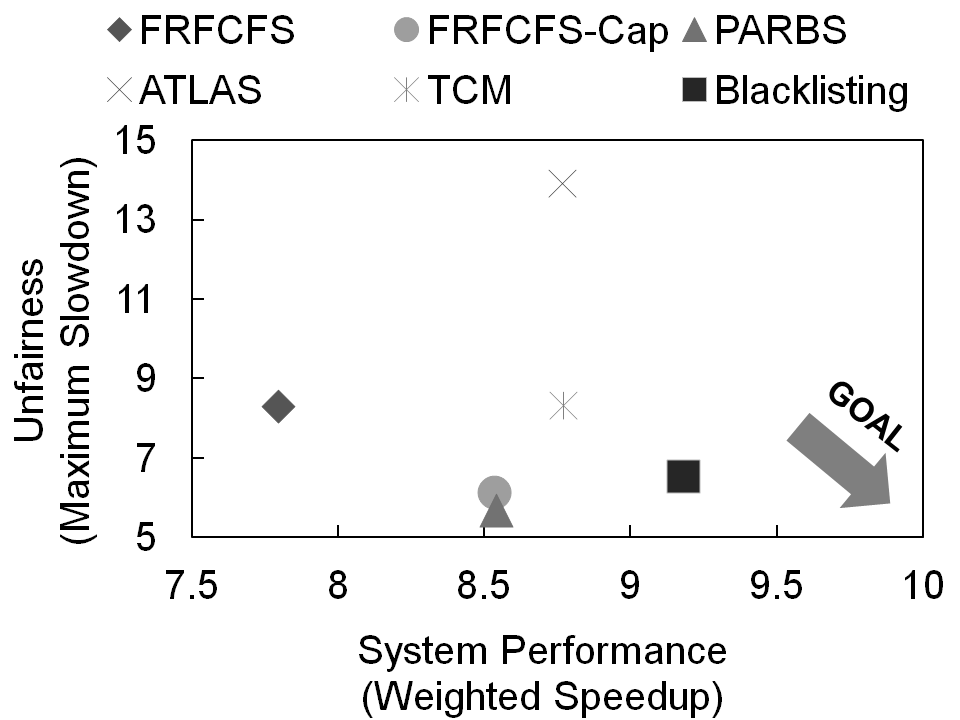}
    \vspace{-1mm}
    \caption{Pareto plot of system performance and fairness}
    \label{fig:pareto}
  \vspace{-6mm}
\end{figure}

We compare \bliss with five previously proposed memory schedulers,
FRFCFS, FRFCFS with a cap (FRFCFS-Cap)~\cite{stfm}, PARBS, ATLAS
and TCM.  FRFCFS-Cap is a modified version of FRFCFS that caps the
number of consecutive row-buffer hitting requests that can be
served from an application~\cite{stfm}.
Figure~\ref{fig:main-results} shows the average system performance
(weighted speedup and harmonic speedup) and unfairness (maximum
slowdown) across all our workloads.  Figure~\ref{fig:pareto} shows
a Pareto plot of weighted speedup and maximum slowdown. We make
three major observations. First, \bliss achieves 5\% better
weighted speedup, 25\% lower maximum slowdown and 19\% better
harmonic speedup than the best performing previous scheduler
(in terms of weighted speedup), TCM, while reducing the critical path
and area by 79\% and 43\% respectively (as we will show in
Section~\ref{sec:complexity}). Therefore, we conclude that \bliss
achieves both high system performance and fairness, at low
hardware cost and complexity.

Second, \bliss significantly outperforms all these five previous
schedulers in terms of system performance, however, it has 10\%
higher unfairness than PARBS, the previous scheduler with the
least unfairness. PARBS creates request batches containing the
oldest requests from each application. Older batches are
prioritized over newer batches. However, within each batch,
individual applications' requests are ranked and prioritized based
on memory intensity. PARBS aims to preserve fairness by batching
older requests, while still employing ranking within a batch to
prioritize low-memory-intensity applications. We observe that the
batching aspect of PARBS is quite effective in mitigating
unfairness, although it increases complexity. This unfairness
reduction also contributes to the high harmonic speedup of PARBS.
However, batching restricts the amount of request reordering that
can be achieved through ranking. Hence, low-memory-intensity
applications that would benefit from prioritization via aggressive
request reordering have lower performance. As a result, PARBS has
8\% lower weighted speedup than \bliss. Furthermore, PARBS has a
6.5x longer critical path and \httilde2x greater area than \bliss,
as we will show in Section~\ref{sec:complexity}. Therefore, we
conclude that \bliss achieves better system performance than
PARBS, at much lower hardware cost, while slightly trading off
fairness.

Third, \bliss has 4\% higher unfairness than FRFCFS-Cap, but it
also 8\% higher performance than FRFCFS-Cap. FRFCFS-Cap has
higher fairness than \bliss since it restricts the length of
{\em only} the {\em ongoing} row hit streak, whereas blacklisting
an application can deprioritize the application {\em for a longer
time}, until the next clearing interval. As a result, FRFCFS-Cap
slows down high-row-buffer-locality applications to a lower degree
than \bliss. However, restricting \emph{only} the on-going streak
rather than blacklisting an interfering application for a longer
time causes more interference to other applications, degrading
system performance compared to \bliss. Furthermore, FRFCFS-Cap is
unable to mitigate interference due to applications with high
memory intensity yet low-row-buffer-locality, whereas \bliss is
effective in mitigating interference due to such applications as
well. Hence, we conclude that \bliss achieves higher performance
(weighted speedup) than FRFCFS-Cap, while slightly trading off
fairness.

\subsection{Analysis of Individual Workloads}
\label{sec:scurve-analysis}

\begin{figure*}[t!]
  \vspace{-3mm}
  \centering
  \begin{minipage}{0.47\textwidth}
    \centering
    \includegraphics[scale=0.26, angle=270]{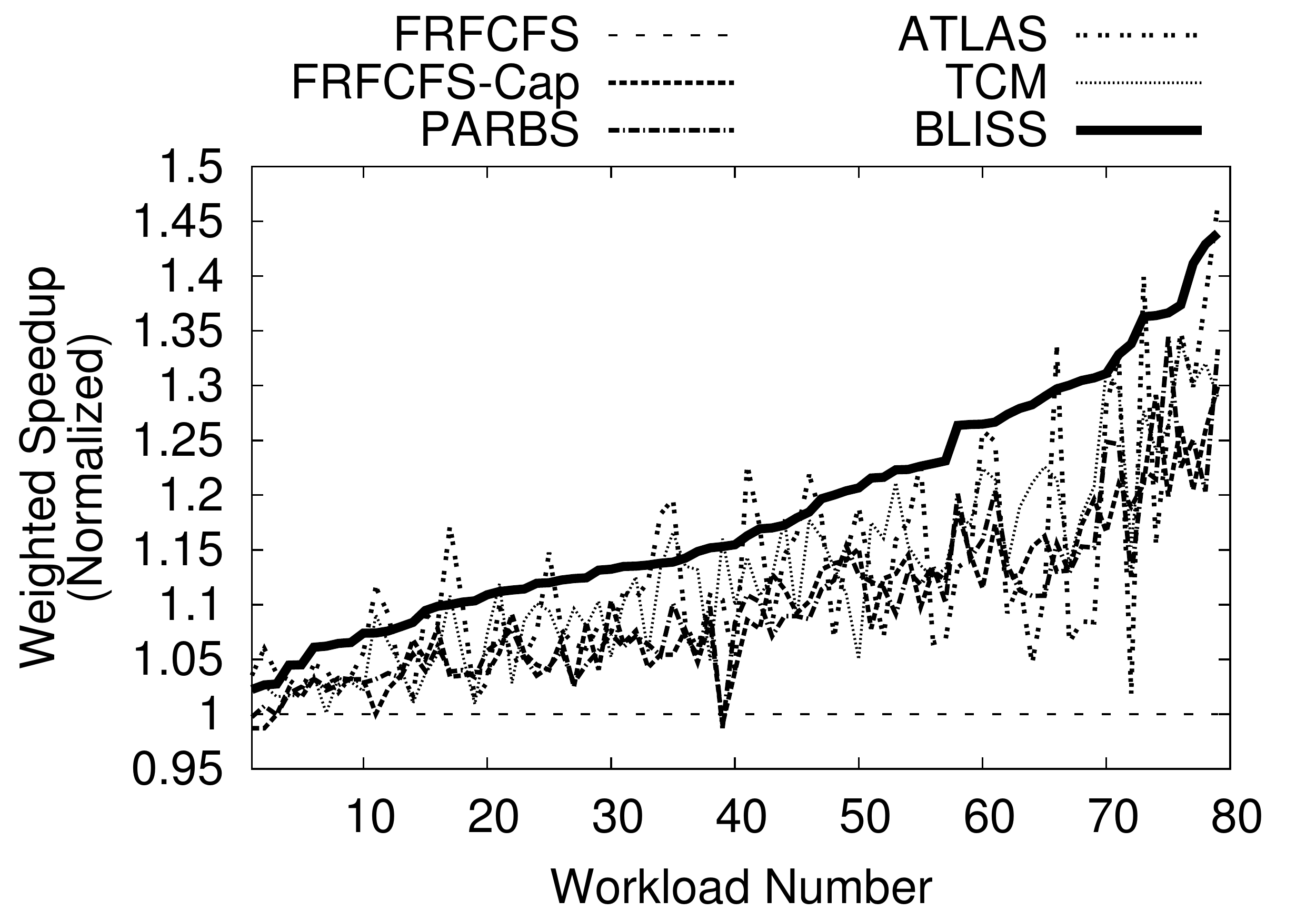}
  \end{minipage}
  \begin{minipage}{0.47\textwidth}
    \centering
    \includegraphics[scale=0.26, angle=270]{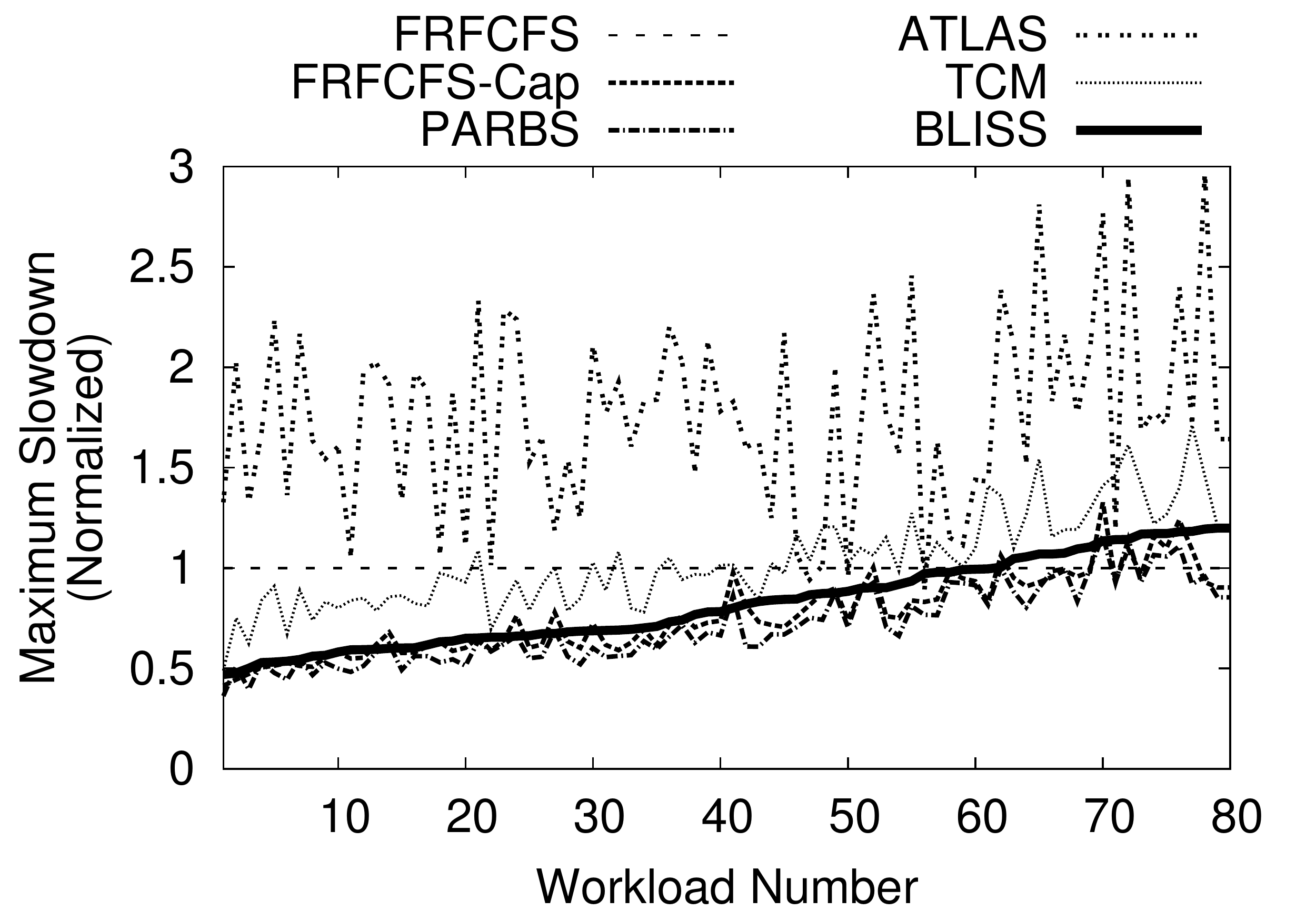}
  \end{minipage}
  \vspace{-1.5mm}
  \caption{System performance and fairness for all workloads}
  \label{fig:scurve}
  \vspace{-5mm}
\end{figure*}

In this section, we analyze the performance and fairness for
individual workloads, when employing different schedulers.
Figure~\ref{fig:scurve} shows the performance and fairness
normalized to the baseline FRFCFS scheduler for all our 80
workloads, for BLISS and previous schedulers, in the form of
S-curves~\cite{eaf}. The workloads are sorted based on the
performance improvement of BLISS. We draw three major
observations. First, BLISS achieves the best performance among all
previous schedulers for most of our workloads. For a few
workloads, ATLAS achieves higher performance, by virtue of always
prioritizing applications that receive low memory service.
However, always prioritizing applications that receive low memory
service can unfairly slow down applications with high memory
intensities, thereby degrading fairness significantly (as shown in
the maximum slowdown plot, Figure~\ref{fig:scurve} bottom).
Second, BLISS achieves significantly higher fairness than ATLAS
and TCM, the best-performing previous schedulers, while also
achieving higher performance than them and approaches the fairness
of the fairest previous schedulers, PARBS and FRFCFS-Cap. As
described in the analysis of average performance and fairness
results above, PARBS, by virtue of request batching and
FRFCFS-Cap, by virtue of restricting only the current row hit
streak achieve higher fairness (lower maximum slowdown) than BLISS
for a number of workloads. However, these schedulers achieve
higher fairness at the cost of lower system performance, as shown
in Figure~\ref{fig:scurve}. Third, for some workloads with very
high memory intensities, the default FRFCFS scheduler achieves the
best fairness. This is because memory bandwidth becomes a very
scarce resource when the memory intensity of a workload is very
high.  Hence, prioritizing row hits utilizes memory bandwidth
efficiently for such workloads, thereby resulting in higher
fairness. Based on these observations, we conclude that BLISS
achieves the best performance and a good trade-off between
fairness and performance for most of the workloads we examine.

\subsection{Hardware Complexity}
\label{sec:complexity}

\begin{figure}[ht]
  \vspace{-3mm}
  \centering
  \begin{minipage}{0.45\textwidth}
     \centering
     \includegraphics[scale=0.17, angle=270]{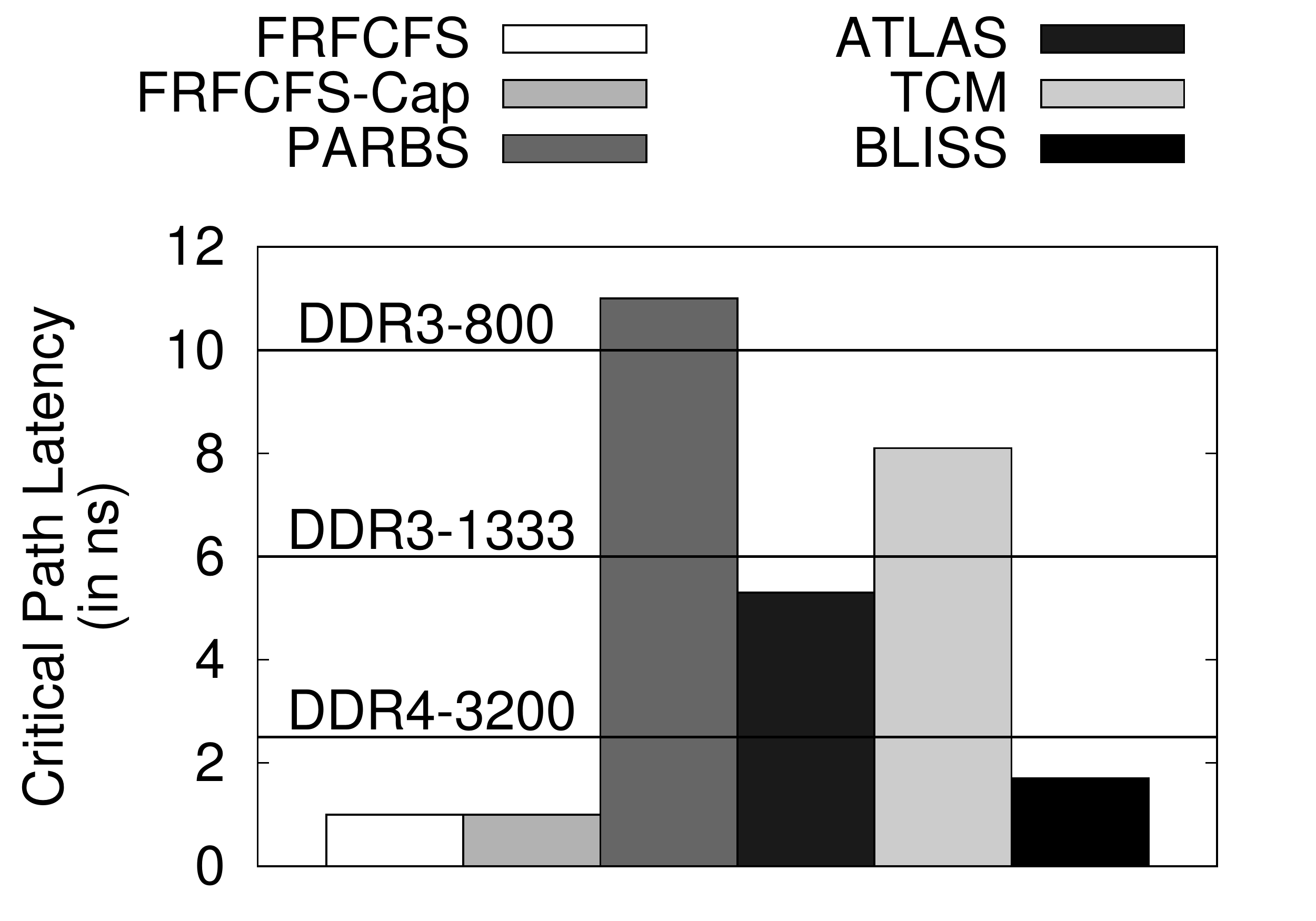}
     \vspace{-1mm}
     \caption{Critical path: \bliss vs. previous schedulers}
     \label{fig:crit-path-latency}
  \end{minipage}
  \begin{minipage}{0.45\textwidth}
%      \vspace{-1mm}
      \centering
      \includegraphics[scale=0.17, angle=270]{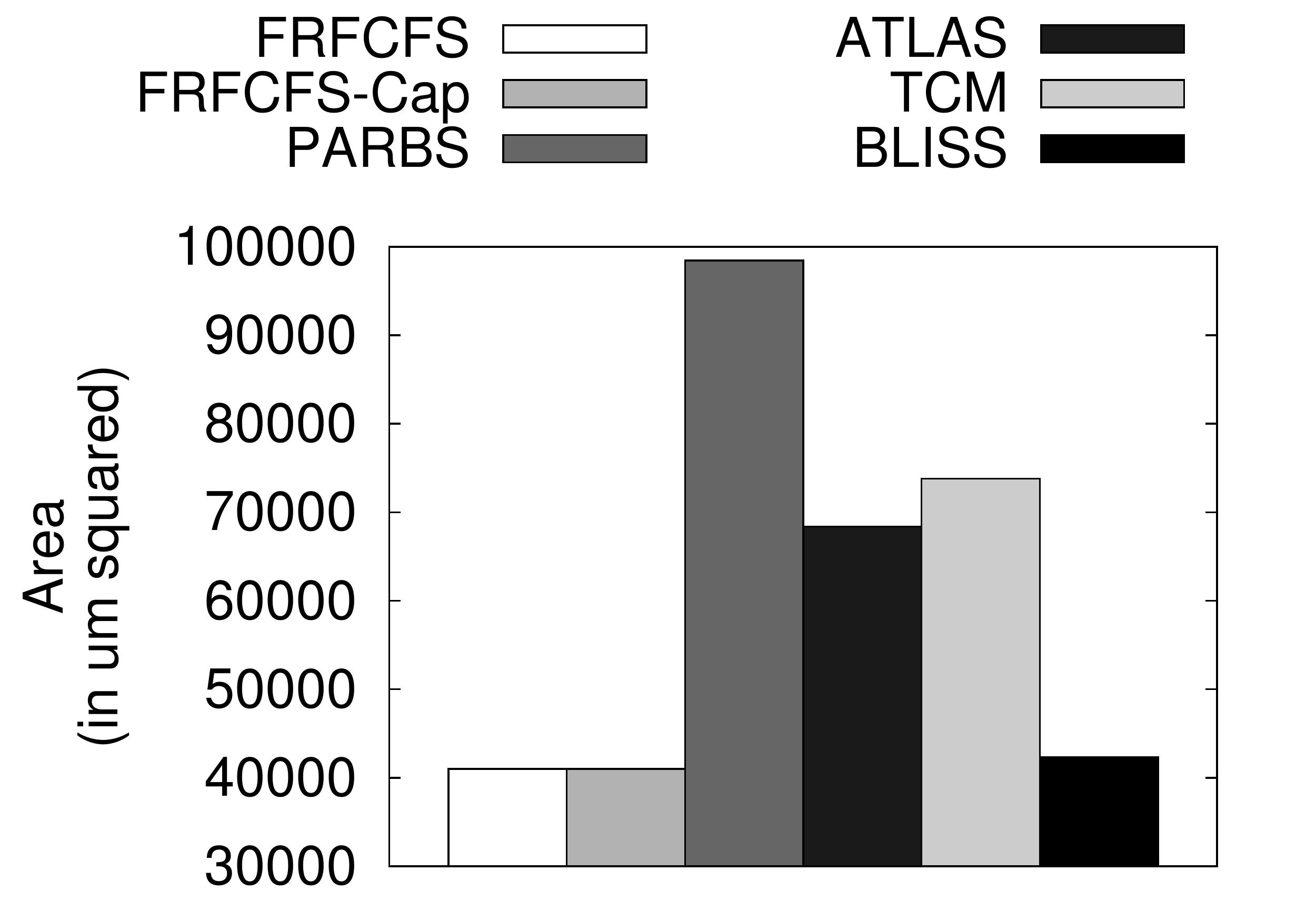}
      \vspace{-1mm}
      \caption{Area: \bliss vs. previous schedulers}
      \label{fig:area}
  \end{minipage}
  \vspace{-8mm}
\end{figure}

Figures~\ref{fig:crit-path-latency} and~\ref{fig:area} show the
critical path latency and area of five previous schedulers and
\bliss for a 24-core system for every memory channel. We draw two
major conclusions. First, previously proposed ranking-based
schedulers, PARBS/ATLAS/TCM, greatly increase the critical path
latency and area of the memory scheduler: by 11x/5.3x/8.1x and
2.4x/1.7x/1.8x respectively, compared to FRFCFS and FRFCFS-Cap,
whereas \bliss increases latency and area by only 1.7x and 3.2\%
over FRFCFS/FRFCFS-Cap.\footnote{The area numbers are for the
lowest value of critical path latency that the scheduler is able
to meet.} Second, PARBS, ATLAS and TCM cannot meet the stringent
worst-case timing requirements posed by the DDR3 and DDR4
standards~\cite{jedec-ddr3,jedec-ddr4}. In the case where every
request is a row-buffer hit, the memory controller would have to
schedule a request every read-to-read cycle time ($t_{CCD}$), the
minimum value of which is 4 cycles for both DDR3 and DDR4. TCM and
ATLAS can meet this worst-case timing only until DDR3-800
(read-to-read cycle time of 10 ns) and DDR3-1333 (read-to-read
cycle time of 6 ns) respectively, whereas \bliss can meet the
worst-case timing all the way down to the highest released
frequency for DDR4, DDR4-3200 (read-to-read time of 2.5 ns).
Hence, the high critical path latency of PARBS, ATLAS and TCM is a
serious impediment to their adoption in today's and future memory
interfaces. Techniques like pipelining could potentially be
employed to reduce the critical path latency of these previous
schedulers. However, the additional flops required for pipelining
would increase area, power and design effort significantly.
Therefore, we conclude that \bliss, with its greatly lower
complexity and cost as well as higher system performance and
competitive or better fairness, is a more effective alternative to
state-of-the-art application-aware memory schedulers.

\subsection{Analysis of Trade-offs Between Performance, Fairness and Complexity}
\label{sec:tradeoff-performance-fairness-complexity}

\begin{figure}[h]
    \vspace{-4mm}
    \centering
    \includegraphics[scale=0.25, angle=0]{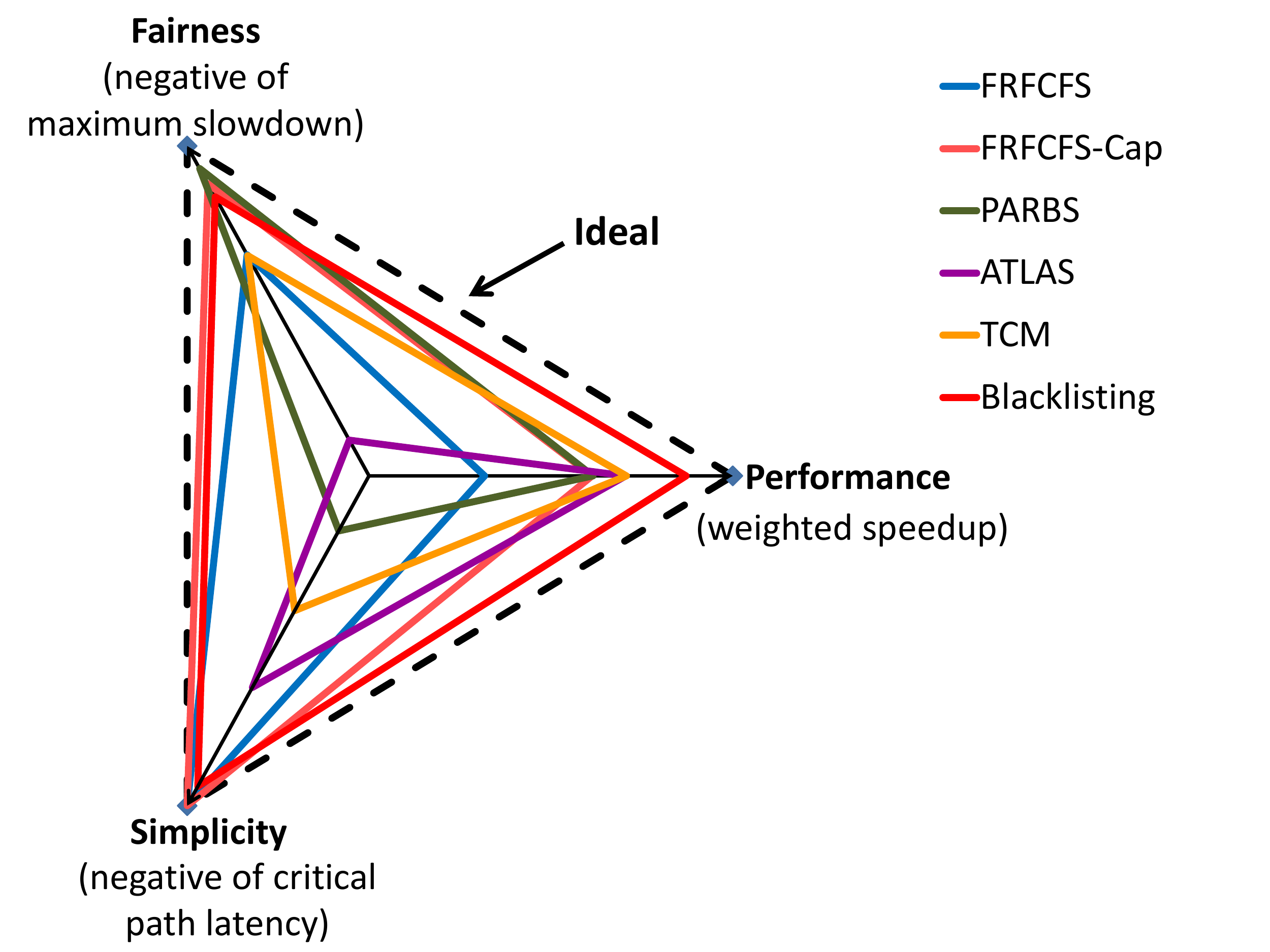}
    \vspace{-1mm}
    \caption{Performance, fairness and simplicity trade-offs}
    \label{fig:perf-fairness-simplicity}
    \vspace{-4mm}
\end{figure}

In the previous sections, we studied the performance, fairness and
complexity of different schedulers individually. In this section,
we will analyze the trade-offs between these metrics for
different schedulers. Figure~\ref{fig:perf-fairness-simplicity}
shows a three-dimensional radar plot with performance, fairness and
simplicity on three different axes. On the fairness axis, we plot
the negative of the maximum slowdown numbers, and on the simplicity axis, we
plot the negative of the critical path latency numbers. Hence, the ideal
scheduler would have high performance, fairness and simplicity, as
indicated by the encompassing, dashed black triangle. We draw three major conclusions
about the different schedulers we study. First,
application-unaware schedulers, such as FRFCFS and FRFCFS-Cap, are
simple. However, they have low performance and/or fairness. This
is because, as described in our performance analysis above, FRFCFS
allows long streaks of row hits from one application to cause
interference to other applications. FRFCFS-Cap attempts to tackle
this problem by restricting the length of current row hit streak.
While such a scheme improves fairness, it still does not improve
performance significantly. Second, application-aware schedulers,
such as PARBS, ATLAS and TCM, improve performance or fairness by
ranking based on applications' memory access characteristics.
However, they do so at the cost of increasing complexity (reducing
simplicity) significantly, since they employ a full ordered
ranking across all applications. Third, BLISS, achieves high
performance and fairness, while keeping the design simple, thereby
approaching the ideal scheduler design (i.e., leading to a
triangle that is closer to the ideal triangle). This is because BLISS
requires only simple hardware changes to the memory controller to
blacklist applications that have long streaks of requests served,
which effectively mitigates interference. Therefore, we conclude
that BLISS achieves the best trade-off between performance,
fairness and simplicity.

\subsection{Understanding the Benefits of \bliss}
\label{sec:bliss-benefits}

\begin{figure*}[ht]
  \centering
  \begin{subfigure}[t]{0.32\textwidth}
    \centering
    \includegraphics[scale=0.17, angle=270]{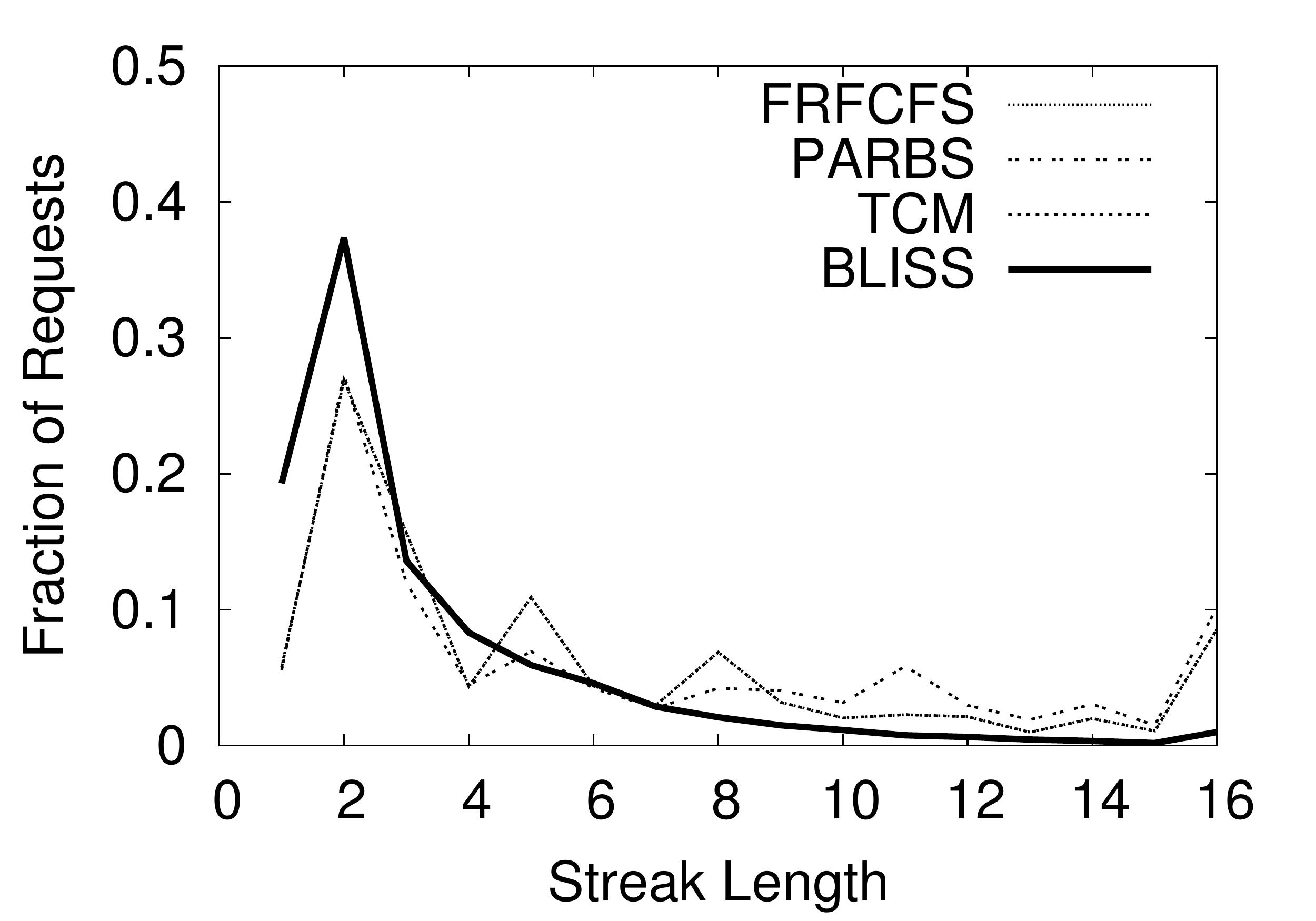}
    \vspace{-1mm}
    \caption{libquantum (MPKI: 52; RBH: 99\%)}
    \label{fig:libq}
  \end{subfigure}
  \begin{subfigure}[t]{0.32\textwidth}
    \centering
    \includegraphics[scale=0.17, angle=270]{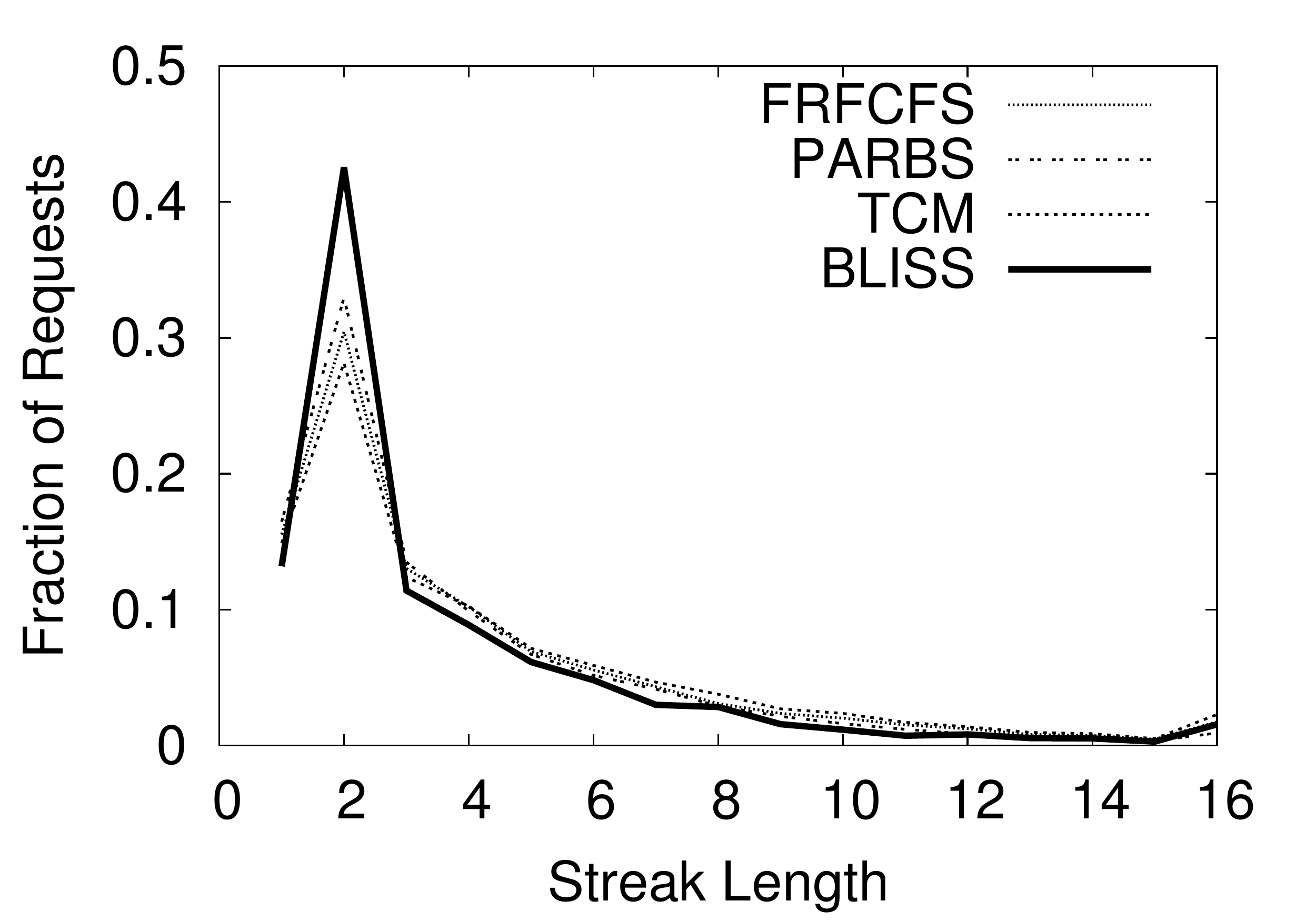}
    \vspace{-1mm}
    \caption{mcf (MPKI: 146; RBH: 40\%)}
    \label{fig:mcf}
  \end{subfigure}
  \begin{subfigure}[t]{0.32\textwidth}
    \centering
    \includegraphics[scale=0.17, angle=270]{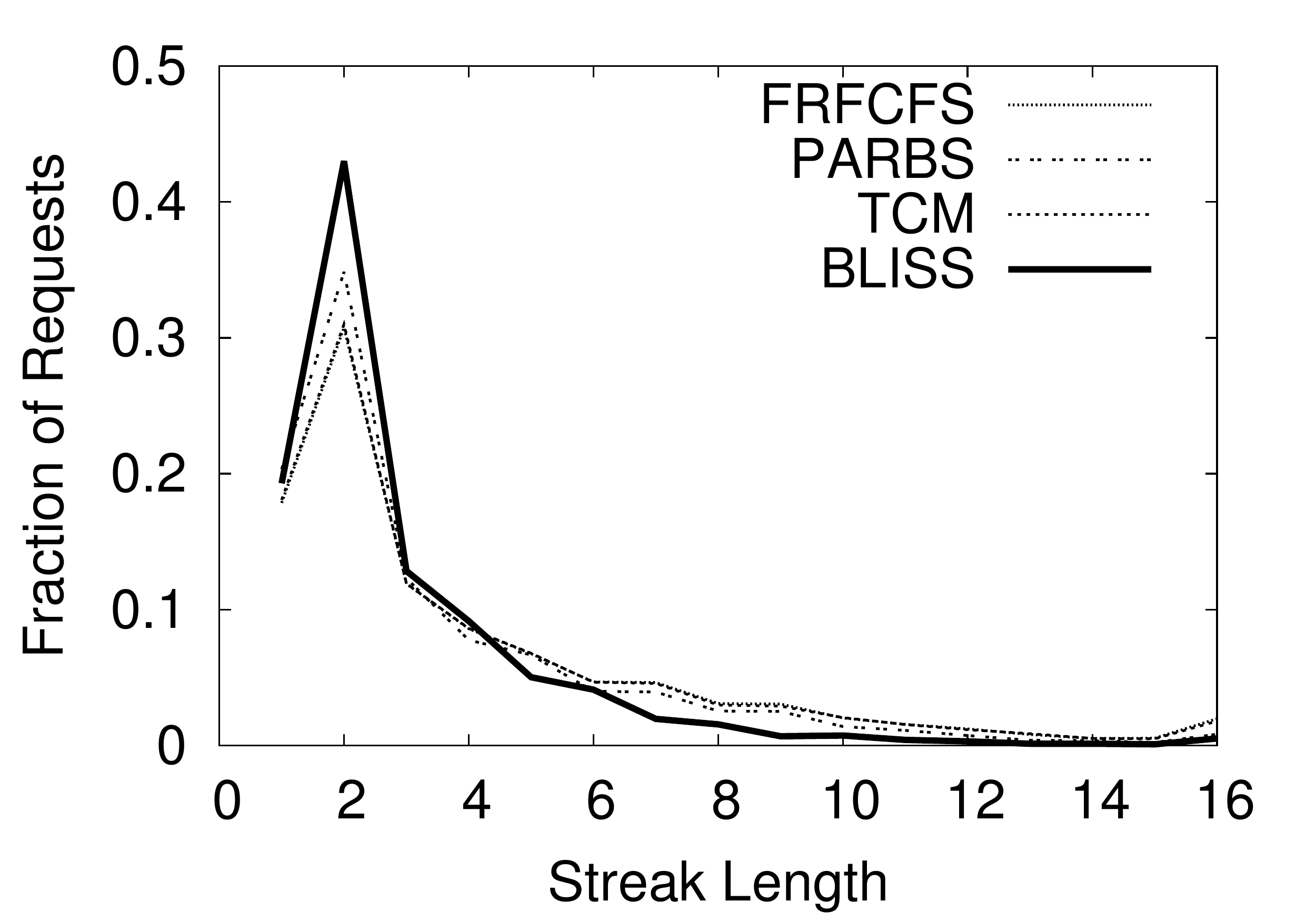}
    \vspace{-1mm}
    \caption{lbm (MPKI: 41; RBH: 89\%)}
    \label{fig:lbm}
  \end{subfigure}
  \begin{subfigure}[t]{0.32\textwidth}
    \centering
    \includegraphics[scale=0.17, angle=270]{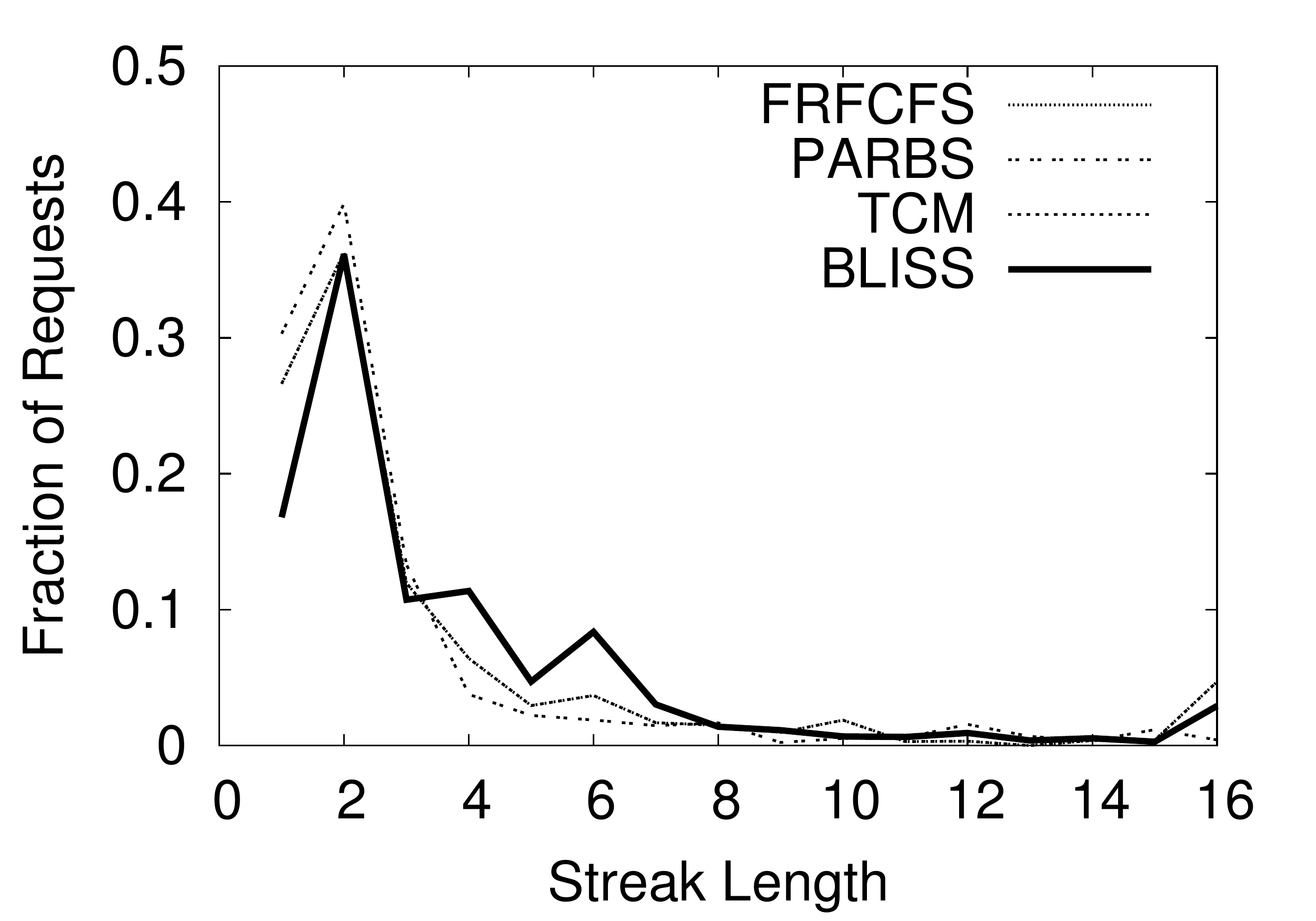}
    \vspace{-1mm}
    \caption{calculix (MPKI: 0.1; RBH: 85\%)}
    \label{fig:calculix}
  \end{subfigure}
  \begin{subfigure}[t]{0.32\textwidth}
    \centering
    \includegraphics[scale=0.17, angle=270]{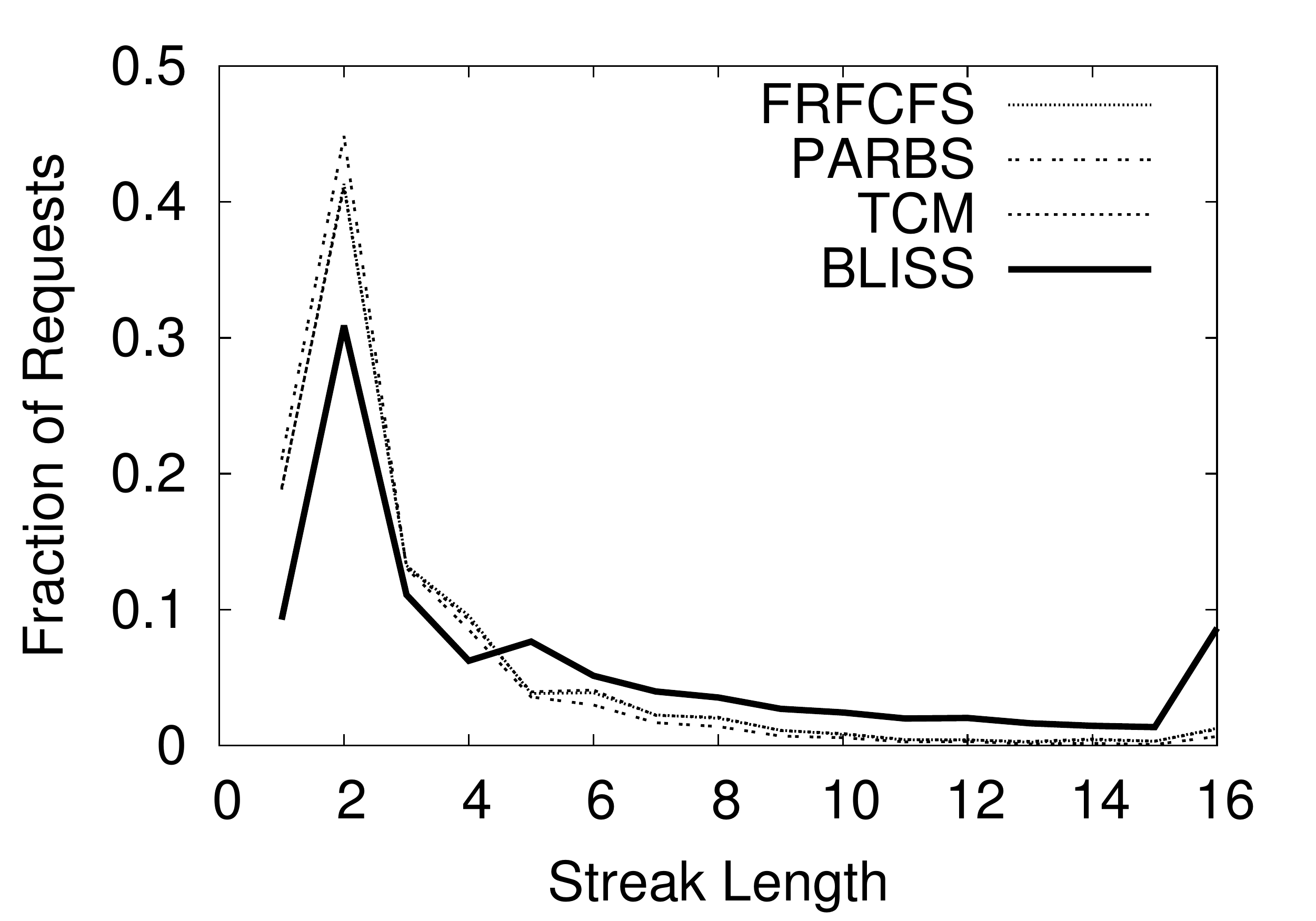}
    \vspace{-1mm}
    \caption{sphinx3 (MPKI: 24; RBH: 91\%)}
    \label{fig:sphinx3}
  \end{subfigure}
  \begin{subfigure}[t]{0.32\textwidth}
    \centering
    \includegraphics[scale=0.17, angle=270]{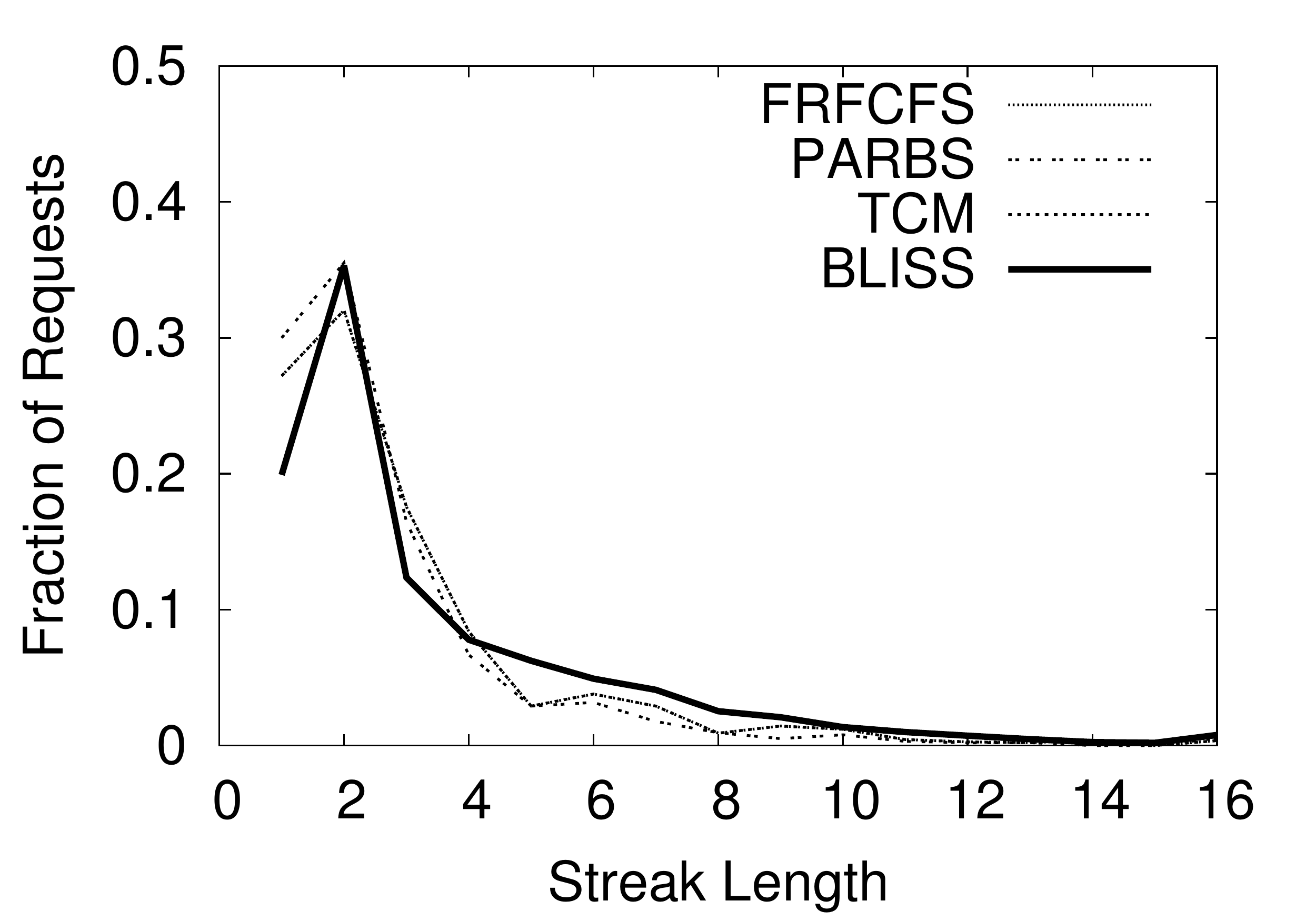}
    \vspace{-1mm}
    \caption{cactusADM (MPKI: 7; RBH: 49\%)}
    \label{fig:cactus}
  \end{subfigure}
  \vspace{-1mm}
  \caption{Distribution of streak lengths}
  \label{fig:streak_length}
\vspace{-6mm}
\end{figure*}

We present the distribution of the number of consecutive requests
served (streaks) from individual applications to better understand
why \bliss effectively mitigates interference.
Figure~\ref{fig:streak_length} shows the distribution of requests
served across different streak lengths ranging from 1 to
16 for FRFCFS,
PARBS, TCM and \bliss for \emph{six representative applications
from the same 24-core workload}.\footnote{A value of 16 captures streak
lengths 16 and above.} The figure captions indicate the
memory intensity, in misses per kilo instruction (MPKI) and
row-buffer hit rate (RBH) of each application when it is run
alone. Figures~\ref{fig:libq},~\ref{fig:mcf} and~\ref{fig:lbm}
show the streak length distributions of applications that have a
tendency to cause interference ({\em libquantum}, {\em mcf} and
{\em lbm}). All these applications have high memory intensity
and/or high row-buffer locality.
Figures~\ref{fig:calculix},~\ref{fig:sphinx3} and~\ref{fig:cactus}
show applications that are vulnerable to interference ({\em
calculix}, {\em cactusADM} and {\em sphinx3}). These applications
have lower memory intensities and row-buffer localities, compared
to the interference-causing applications. We observe that \bliss
shifts the distribution of streak lengths towards the left for the
interference-causing applications, while it shifts the streak
length distribution to the right for the interference-prone
applications. Hence, \bliss breaks long streaks of consecutive
requests for interference-causing applications, while enabling
longer streaks for vulnerable applications. This enables such
vulnerable applications to make faster progress, thereby resulting
in better system performance and fairness. We have observed
similar results for most of our workloads.

\subsection{Average Request Latency}
\label{sec:average-request-latency}

\begin{figure}[h!]
  \vspace{-4mm}
  \centering
  \includegraphics[scale=0.16, angle=270]{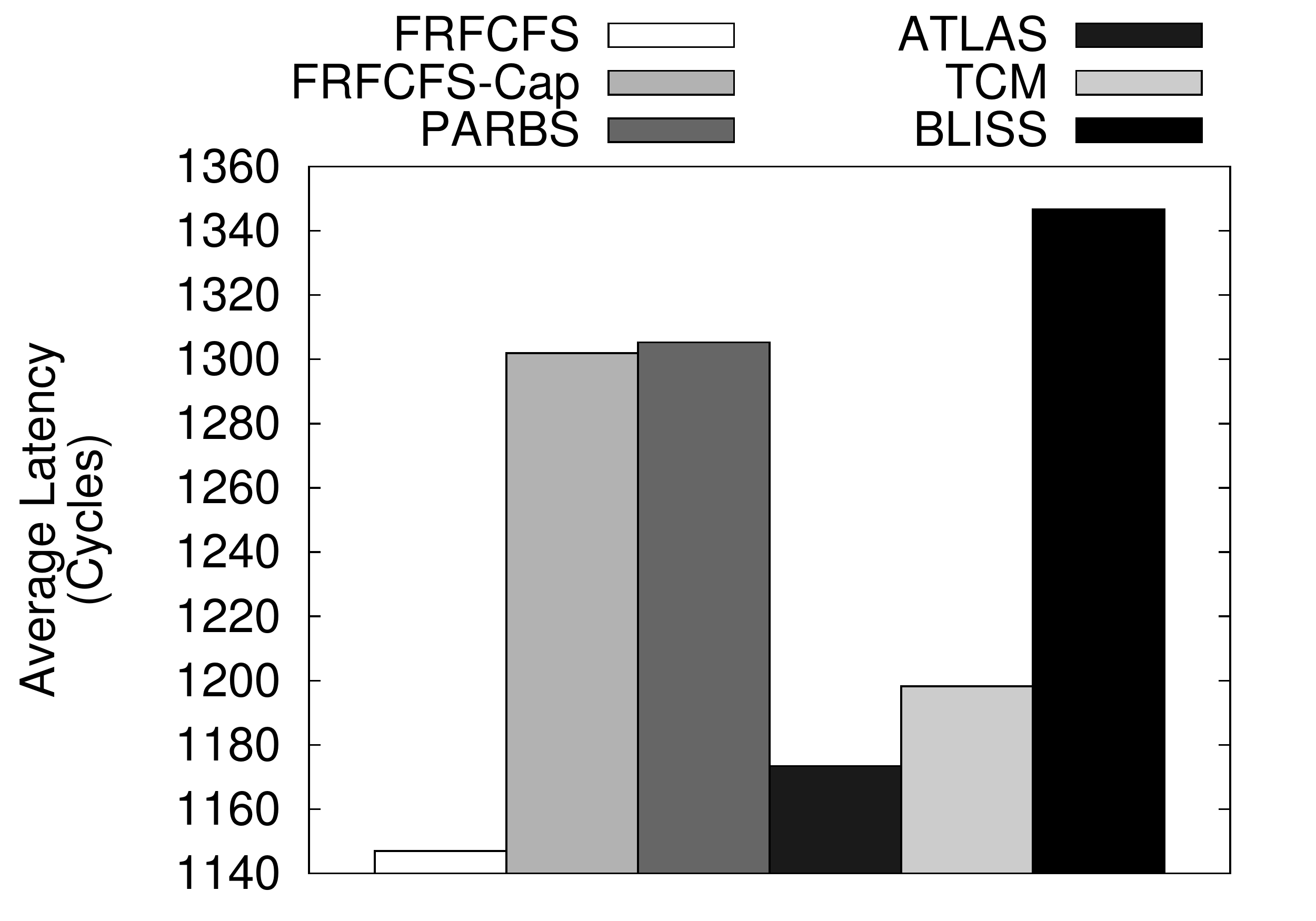}
  \vspace{-1mm}
  \caption{The Average Request Latency Metric}
  \label{fig:latency}
  \vspace{-3mm}
\end{figure}

In this section, we evaluate the average memory request latency
(from when a request is generated until when it is served)
metric and seek to understand its correlation with performance and
fairness. Figure~\ref{fig:latency} presents the average memory request
latency (from when the request is generated until when it is served)
for the five previously proposed memory schedulers and \bliss. Two
major observations are in order. First, FRFCFS has the lowest
average request latency among all the schedulers. This is
expected since FRFCFS maximizes DRAM throughput by prioritizing
row-buffer hits. Hence, the number of requests served is maximized
overall (across all applications). However, maximizing throughput
(i.e., minimizing overall average
request latency) degrades the performance of low-memory-intensity
applications, since these applications' requests are often delayed
behind row-buffer hits and older requests. This results in
degradation in system performance and fairness, as shown in
Figure~\ref{fig:main-results}. 

\begin{figure*}[hbt!]
  \vspace{-1mm}
  \centering
  \begin{minipage}{0.31\textwidth}
    \centering
    \includegraphics[scale=0.16, angle=270]{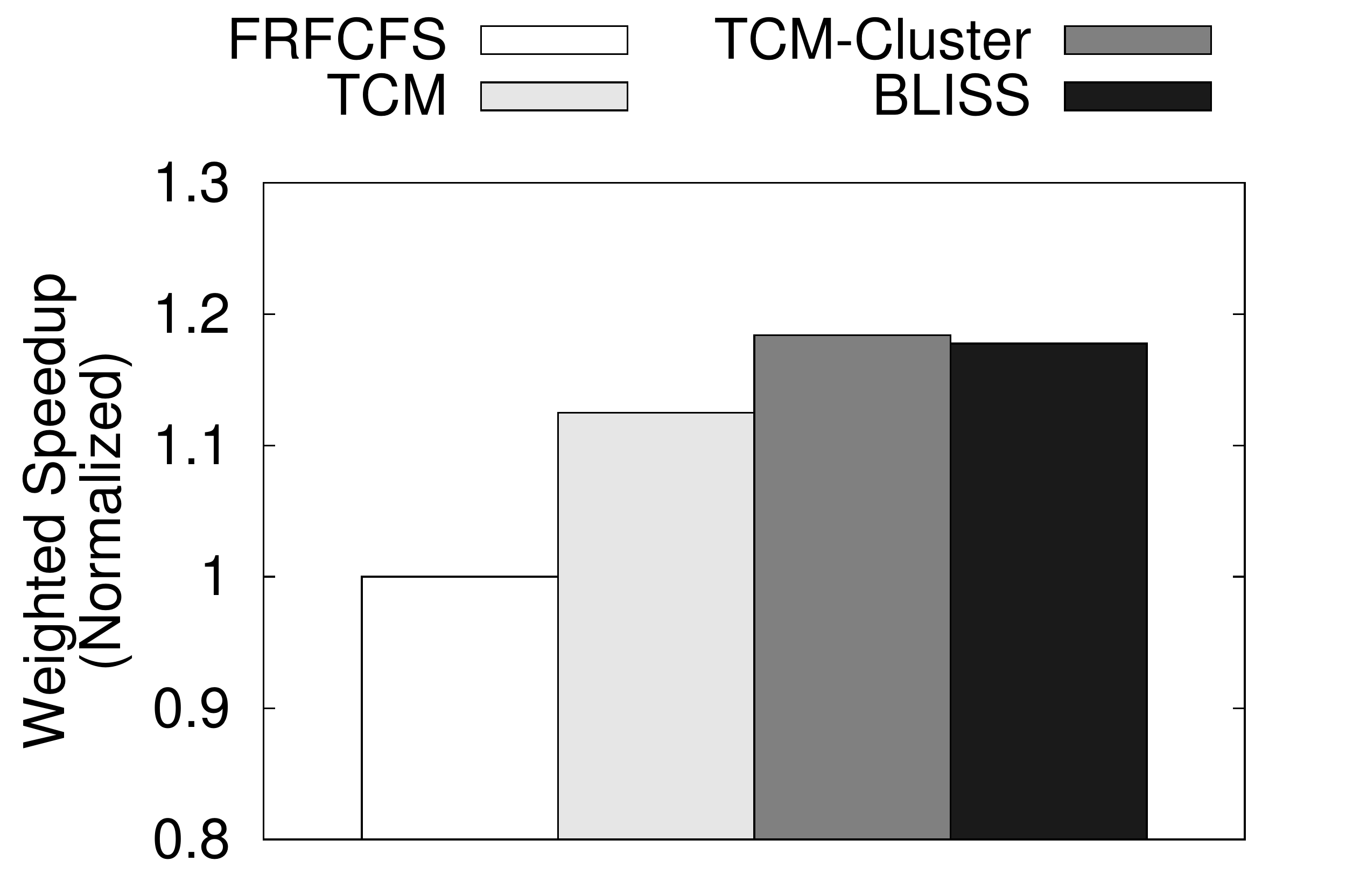}
  \end{minipage}
  \begin{minipage}{0.31\textwidth}
    \centering
    \includegraphics[scale=0.16, angle=270]{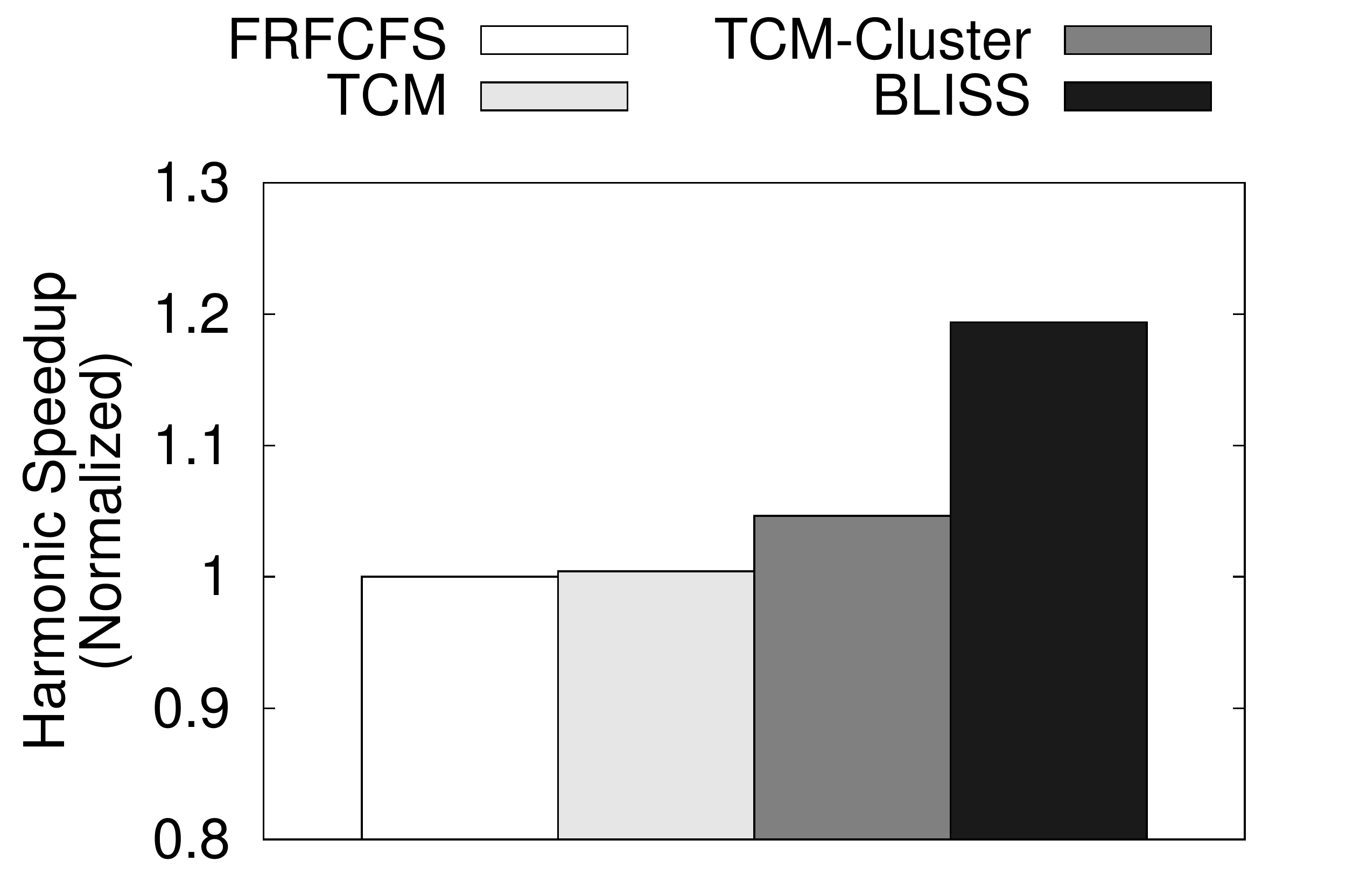}
  \end{minipage} 
  \begin{minipage}{0.31\textwidth}
    \centering
    \includegraphics[scale=0.16, angle=270]{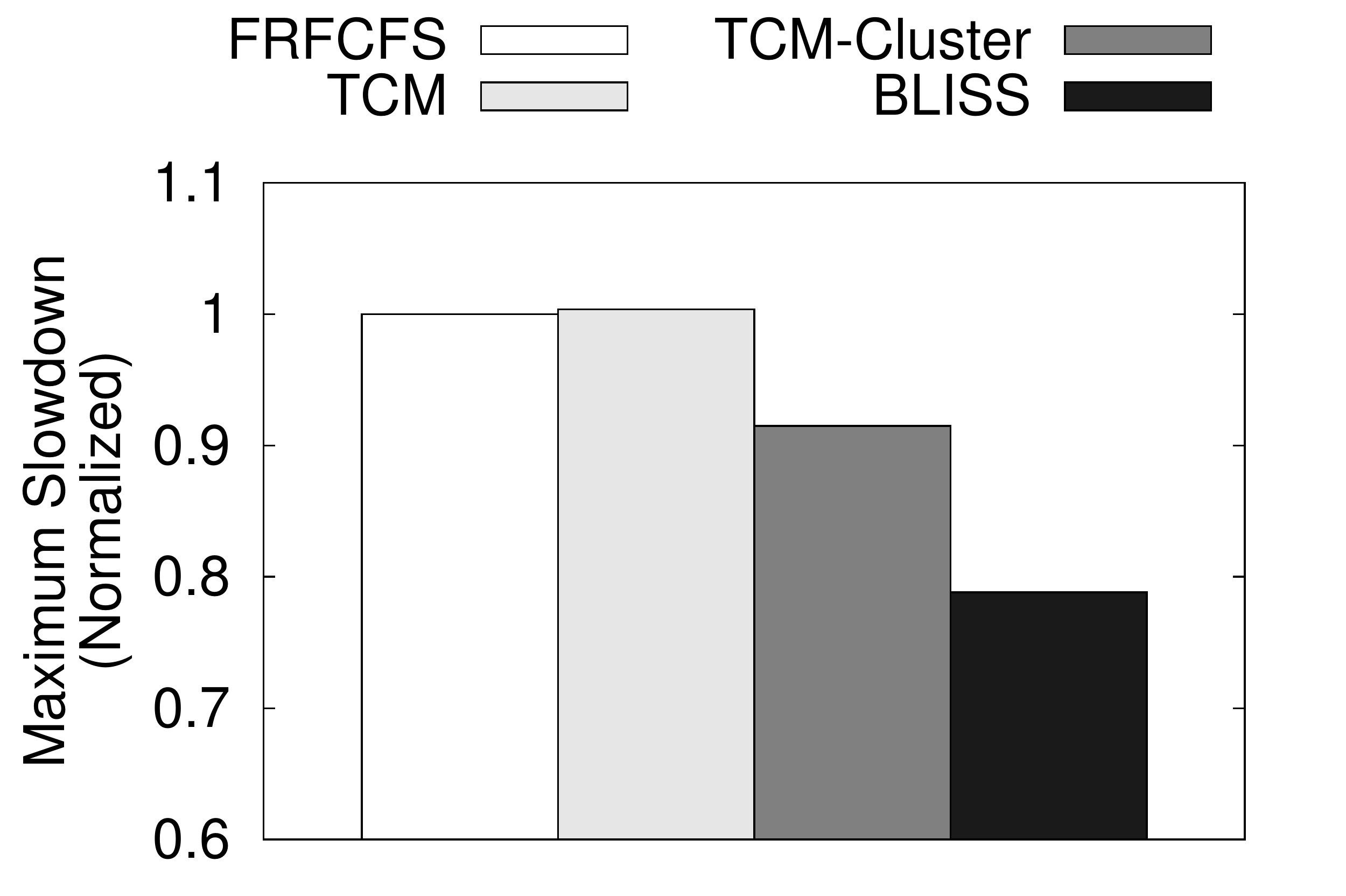}
  \end{minipage} 
  \vspace{-1mm}
  \caption{Comparison with TCM's clustering mechanism}
  \label{fig:tcm-no-ranking}
  \vspace{-2mm}
\end{figure*}

\begin{figure*}[ht!]
  \vspace{-3mm}
  \centering
  \begin{minipage}{0.31\textwidth}
    \centering
    \includegraphics[scale=0.16, angle=270]{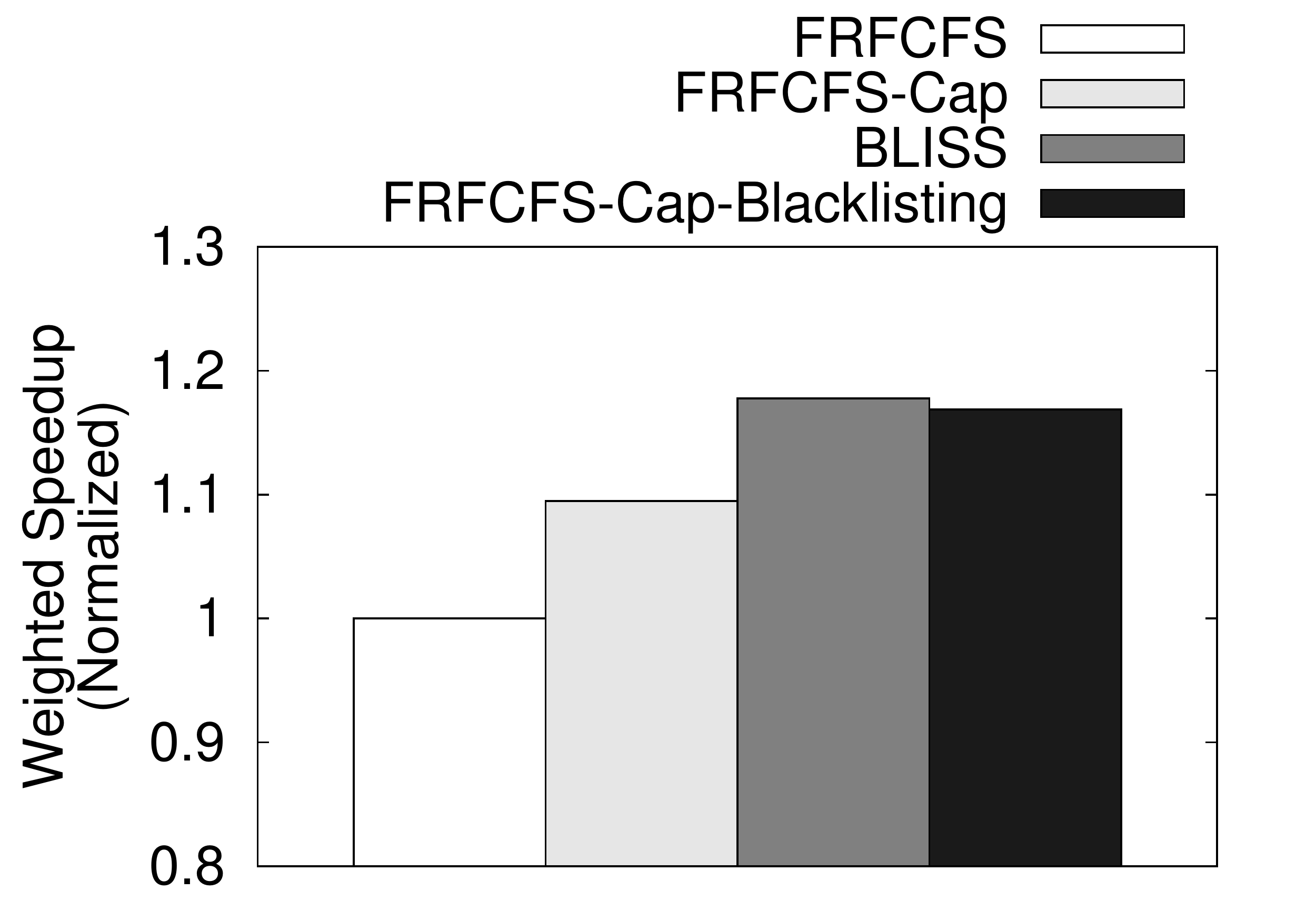}
  \end{minipage}
  \begin{minipage}{0.31\textwidth}
    \centering
    \includegraphics[scale=0.16, angle=270]{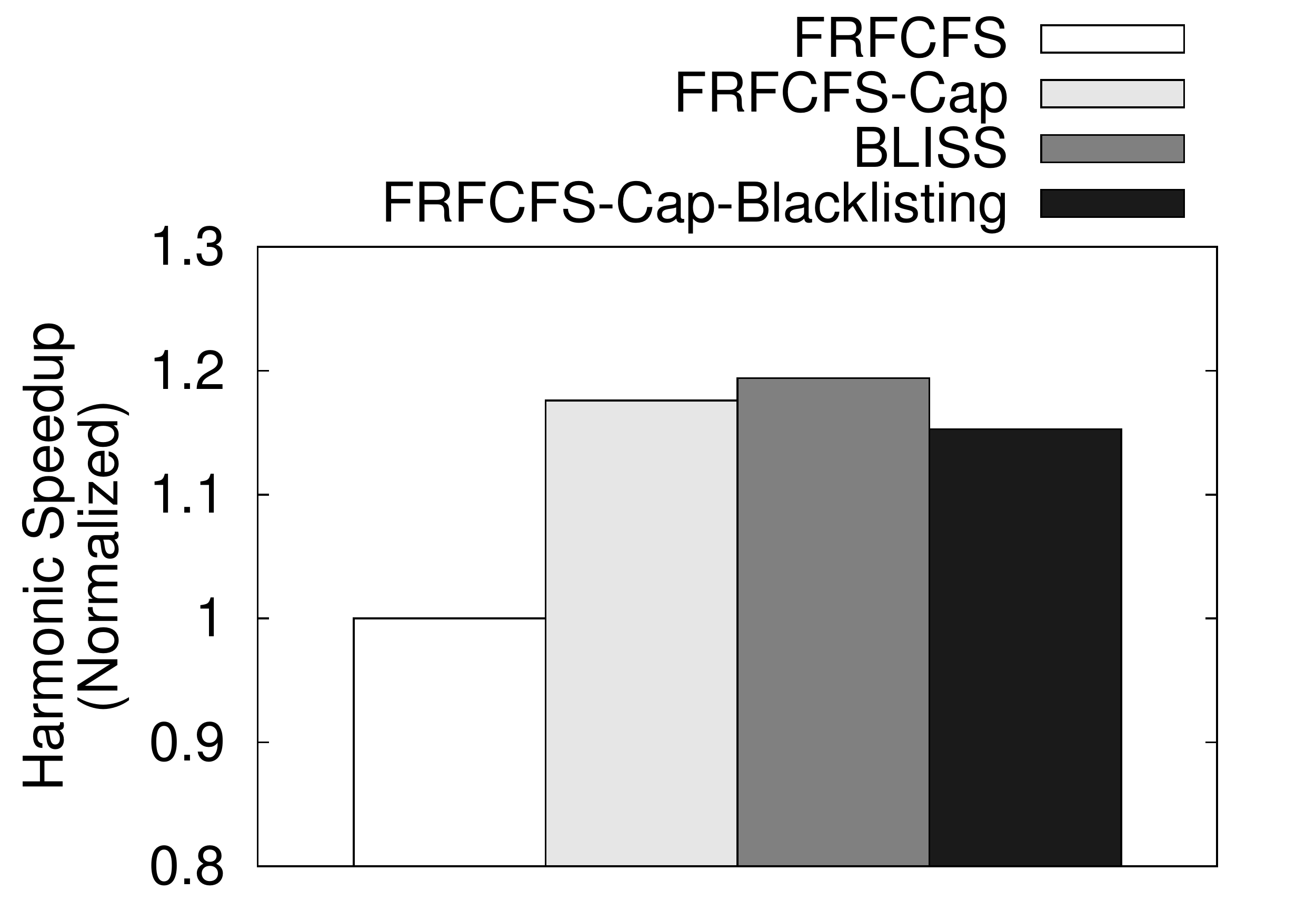}
  \end{minipage}
  \begin{minipage}{0.31\textwidth}
    \centering
    \includegraphics[scale=0.16, angle=270]{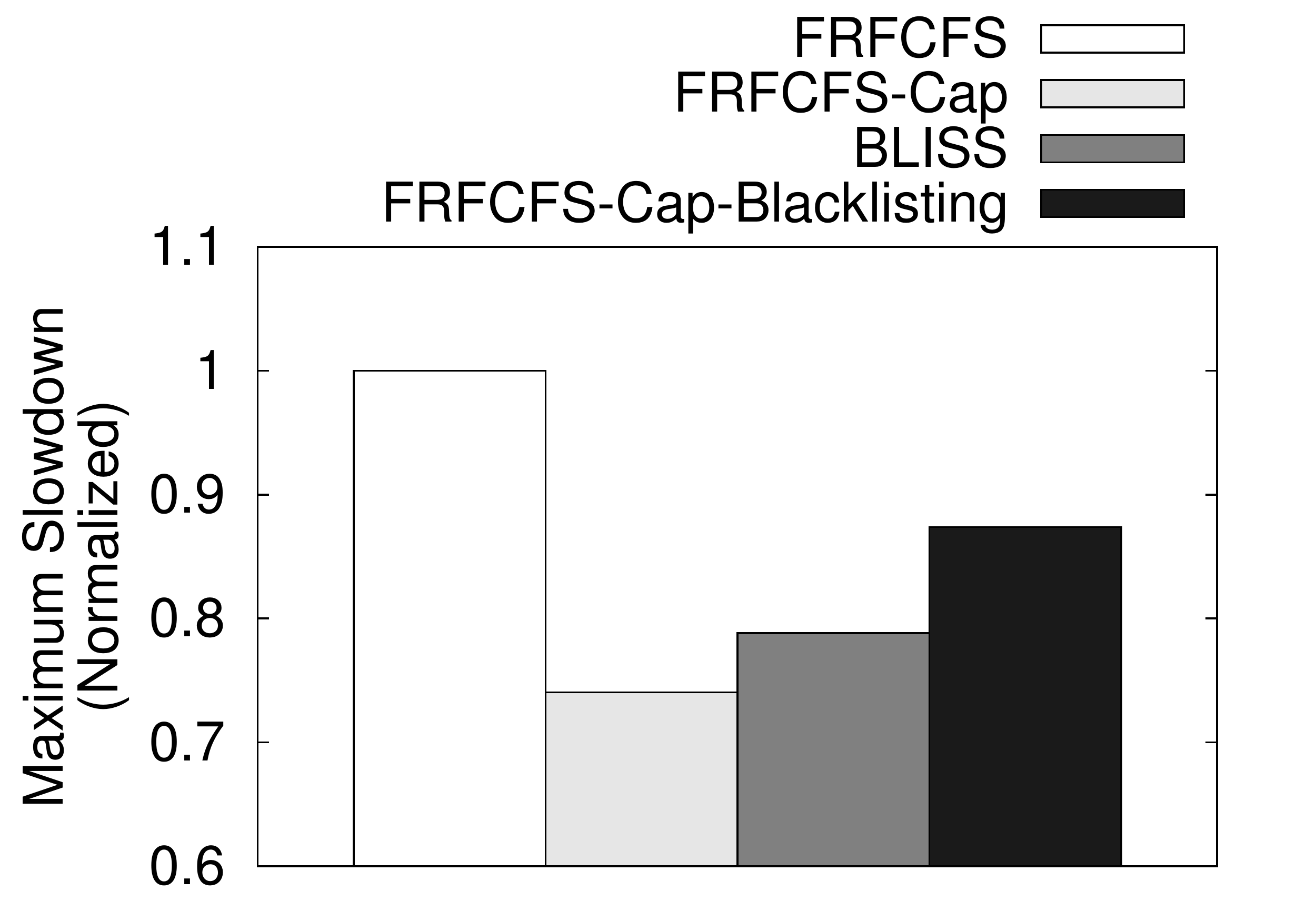}
  \end{minipage}
  \vspace{-1mm}
  \caption{Comparison with FRFCFS-Cap combined with blacklisting}
  \label{fig:frfcfs-cap-blacklisting}
  \vspace{-5mm}
\end{figure*}

Second, ATLAS and TCM, memory schedulers that prioritize requests
of low-memory-intensity applications by employing a full ordered
ranking achieve relatively low average latency. This is because
these schedulers reduce the latency of serving requests from
latency-critical, low-memory-intensity applications significantly.
Furthermore, prioritizing low-memory-intensity applications'
requests does not increase the latency of high-memory-intensity
applications significantly. This is because high-memory-intensity
applications already have high memory access latencies (even when
run alone) due to queueing delays. Hence, average request latency does
not increase much from deprioritizing requests of such
applications. However, always prioritizing such latency-critical
applications results in lower memory throughput for
high-memory-intensity applications, resulting in unfair slowdowns
(as we show in Figure~\ref{fig:main-results}). Third, memory
schedulers that provide the best fairness, PARBS, FRFCFS-Cap and
\bliss have high average memory latencies. This is because these
schedulers, while employing techniques to prevent requests of
vulnerable applications with low memory intensity and low
row-buffer locality from being delayed, also avoid unfairly
delaying requests of high-memory-intensity applications. As a
result, they do not reduce the request service latency of
low-memory-intensity applications significantly, at the cost of
denying memory throughput to high-memory-intensity applications,
unlike ATLAS or TCM. Based on these observations, we conclude that while some
applications benefit from low memory access latencies, other
applications benefit more from higher memory throughput than lower
latency. Hence, average memory latency is {\em not} a suitable
metric to estimate system performance or fairness.

\subsection{Impact of Clearing the Blacklist Asynchronously}
\label{sec:blacklist-asynchronous-clearing}

The Blacklisting scheduler we have presented and evaluated so far
clears the blacklisting information periodically (every 10000
cycles in our evaluations so far), such that \emph{all}
applications are removed from the blacklist at the end of a
\textit{Clearing Interval}. In this section, we evaluate an
alternative design where an individual application is removed from
the blacklist \textit{Clearing Interval} cycles after it has been
blacklisted (independent of the other applications). In order to
implement this alternative design, each application would need an
additional counter to keep track of the number of remaining cycles
until the application would be removed from the blacklist. This
counter is set (to the \textit{Clearing Interval}) when an
application is blacklisted and is decremented every cycle until it
becomes zero. When it becomes zero, the corresponding application
is removed from the blacklist. We use a \textit{Clearing Interval}
of 10000 cycles for this alternative design as well.     

Table~\ref{tab:individual-clearing} shows the system performance
and fairness of the original BLISS design (BLISS) and the
alternative design in which individual applications are removed
from the blacklist asynchronously (BLISS-Individual-Clearing). As
can be seen, the performance and fairness of the two designs are
similar. Furthermore, the first design (BLISS) is simpler since it
does not need to maintain an additional counter for each
application. We conclude that the original BLISS design is
more efficient, in terms of performance, fairness and complexity.

\begin{table}[h!]
  \vspace{-2mm}
  \centering
  \input{tables/individual-clearing}
  \vspace{-1.5mm}
  \caption{Clearing the blacklist asynchronously}
  \label{tab:individual-clearing}
  \vspace{-3mm}
\end{table}

\subsection{Comparison with TCM's Clustering Mechanism}
\label{sec:tcm-clustering}

Figure~\ref{fig:tcm-no-ranking} shows the system performance and
fairness of \bliss, TCM and TCM's clustering mechanism
(TCM-Cluster). TCM-Cluster is a modified version of TCM that
performs clustering, but does not rank applications within each
cluster. We draw two major conclusions. First, TCM-Cluster has
similar system performance as \bliss, since both \bliss and
TCM-Cluster prioritize vulnerable applications by separating them
into a group and prioritizing that group rather than ranking
individual applications. Second, TCM-Cluster has significantly
higher unfairness compared to \bliss. This is because TCM-Cluster
always deprioritizes high-memory-intensity applications,
regardless of whether or not they are causing interference (as
described in Observation 2 in Section~\ref{sec:observations}).
\bliss, on the other hand, observes an application at fine time
granularities, independently at every memory channel and
blacklists an application at a channel {\em only when} it is
generating a number of consecutive requests (i.e., potentially
causing interference to other applications).

\subsection{Evaluation of Row Hit Based Blacklisting}
\label{sec:blacklisting-frfcfs-cap}

\bliss, by virtue of restricting the number of consecutive
requests that are served from an application, attempts to mitigate
the interference caused by both high-memory-intensity and high-
row-buffer-locality applications. In this section, we attempt to
isolate the benefits from restricting consecutive row-buffer
hitting requests vs. non-row-buffer hitting requests. To this end,
we evaluate the performance and fairness benefits of a mechanism
that places an application in the blacklist when a certain number
of row-buffer hitting requests (\textrm{N}) to the same row have
been served for an application (we call this
FRFCFS-Cap-Blacklisting as the scheduler essentially is FRFCFS-Cap
with blacklisting). We use an \textrm{N} value of 4 in our
evaluations.

\begin{sloppypar}
Figure~\ref{fig:frfcfs-cap-blacklisting} compares the system
performance and fairness of \bliss with FRFCFS-Cap-Blacklisting.
We make three major observations. First, FRFCFS-Cap-Blacklisting
has similar system performance as \bliss.  On further analysis of
individual workloads, we find that FRFCFS-Cap-Blacklisting
blacklists only applications with high row-buffer locality,
causing requests of non-blacklisted high-memory-intensity
applications to interfere with requests of low-memory-intensity
applications. However, the performance impact of this interference
is offset by the performance improvement of high-memory-intensity
applications that are not blacklisted. Second,
FRFCFS-Cap-Blacklisting has higher unfairness (higher maximum
slowdown and lower harmonic speedup) than \bliss. This is because
the high-memory-intensity applications that are not blacklisted
are prioritized over the blacklisted high-row-buffer-locality
applications, thereby interfering with and slowing down the high-
row-buffer-locality applications significantly. Third,
FRFCFS-Cap-Blacklisting requires a per-bank counter to count and
cap the number of row-buffer hits, whereas \bliss needs only one
counter per-channel to count the number of consecutive requests
from the same application. Therefore, we conclude that \bliss is
more effective in mitigating unfairness while incurring lower
hardware cost, than the FRFCFS-Cap-Blacklisting scheduler that we
build combining principles from FRFCFS-Cap and BLISS.
\end{sloppypar}

\subsection{Comparison with Criticality-Aware Scheduling}

\begin{figure}[ht!]
  \vspace{-4mm}
  \centering
  \begin{minipage}{0.24\textwidth}
    \centering
    \includegraphics[scale=0.16, angle=270]{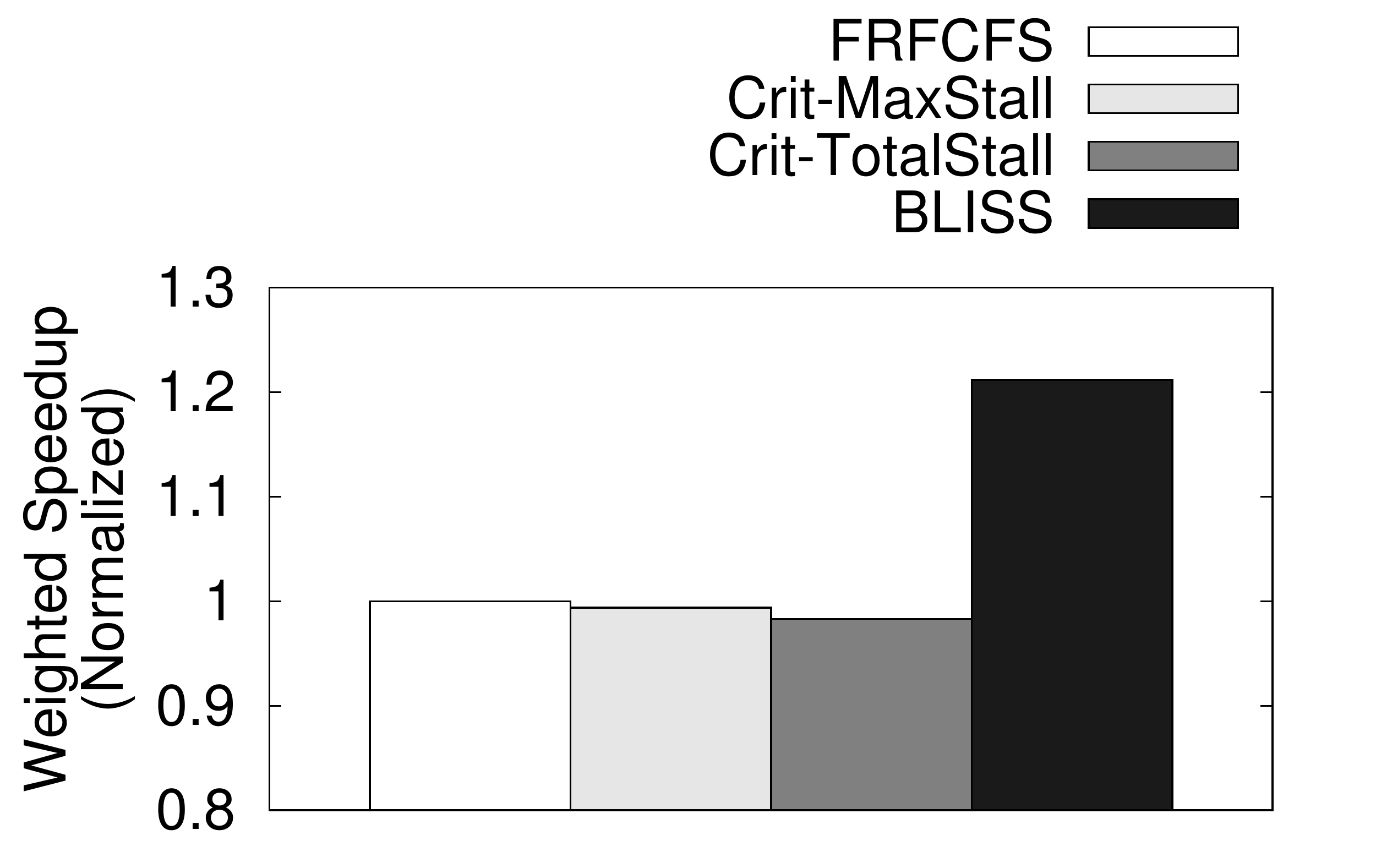}
  \end{minipage}
  \begin{minipage}{0.24\textwidth}
   \centering
   \includegraphics[scale=0.16, angle=270]{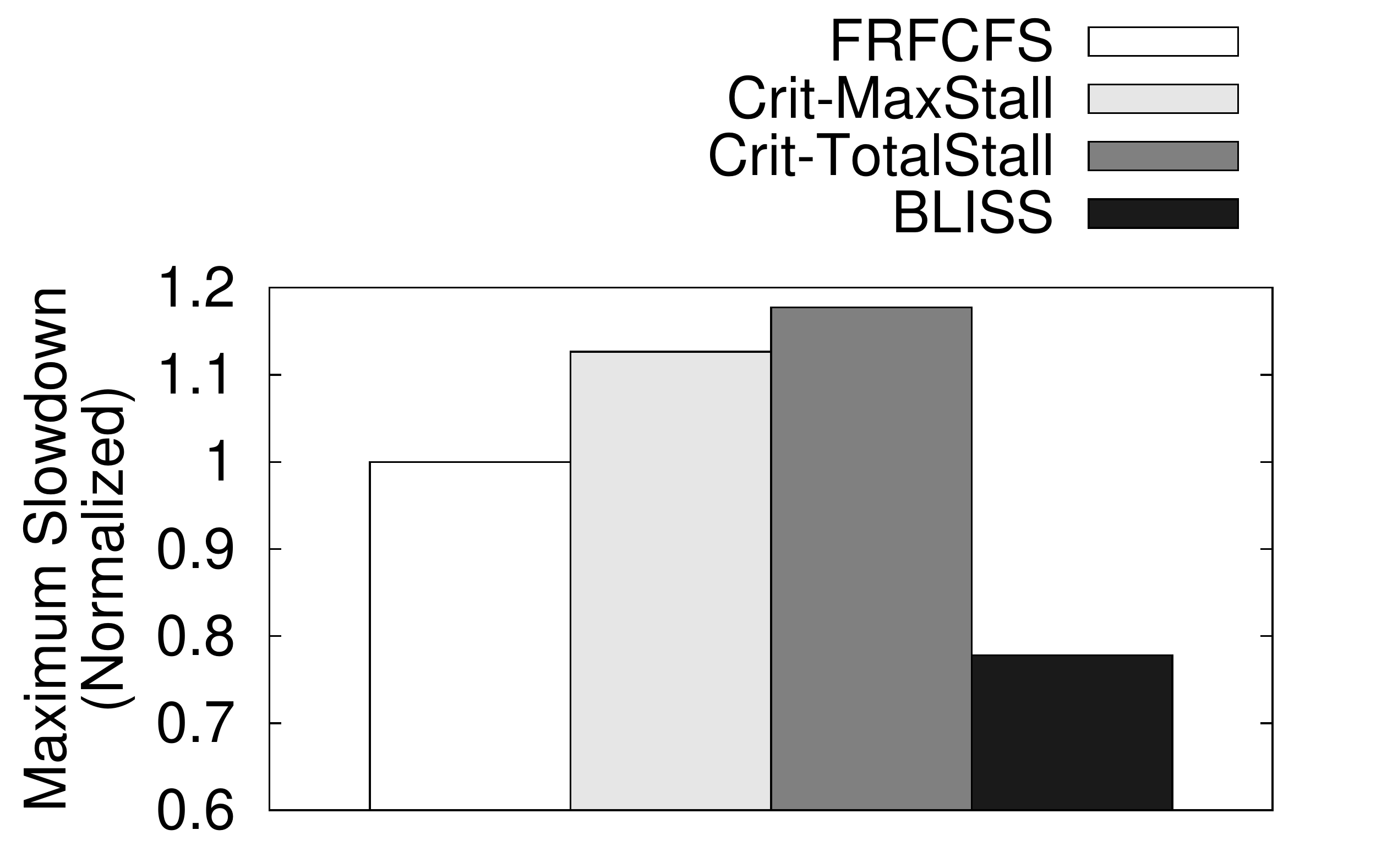}
  \end{minipage}
  \vspace{-2mm}
  \caption{Comparison with criticality-aware scheduling}
  \label{fig:crit-comparison}
  \vspace{-3.5mm}
\end{figure}

We compare the system performance and fairness of \bliss with
those of criticality-aware memory
schedulers~\cite{crit-scheduling-cornell}. The basic idea behind
criticality-aware memory scheduling is to prioritize memory
requests from load instructions that have stalled the instruction
window for long periods of time in the past. Ghose et
al.~\cite{crit-scheduling-cornell} evaluate prioritizing load
requests based on both maximum stall time (Crit-MaxStall) and
total stall time (Crit-TotalStall) caused by load instructions in
the past. Figure~\ref{fig:crit-comparison} shows the system
performance and fairness of \bliss and the criticality-aware
scheduling mechanisms, normalized to FRFCFS, across 40 workloads.
Two observations are in order. First, \bliss significantly
outperforms criticality-aware scheduling mechanisms in terms of
both system performance and fairness. This is because the
criticality-aware scheduling mechanisms unfairly deprioritize and
slow down low-memory-intensity applications that inherently
generate fewer requests, since stall times tend to be low for such
applications. Second, criticality-aware scheduling incurs hardware
cost to prioritize requests with higher stall times. Specifically,
the number of bits to represent stall times is on the order of
12-14, as described in~\cite{crit-scheduling-cornell}. Hence, the
logic for comparing stall times and prioritizing requests with
higher stall times would incur even higher cost than
per-application ranking mechanisms where the number of bits to
represent a core's rank grows only as as $log_2 Number Of Cores$
(e.g. 5 bits for a 32-core system). Therefore, we conclude that
\bliss achieves significantly better system performance and
fairness, while incurring lower hardware cost.

\subsection{Effect of Workload Memory Intensity and Row-buffer Locality}

In this section, we study the impact of workload memory intensity
and row-buffer locality on performance and fairness of \bliss and
five previous schedulers.

\begin{figure}[h!]
  \vspace{-5mm}
  \centering
  \begin{minipage}{0.24\textwidth}
    \centering
    \includegraphics[scale=0.17, angle=270]{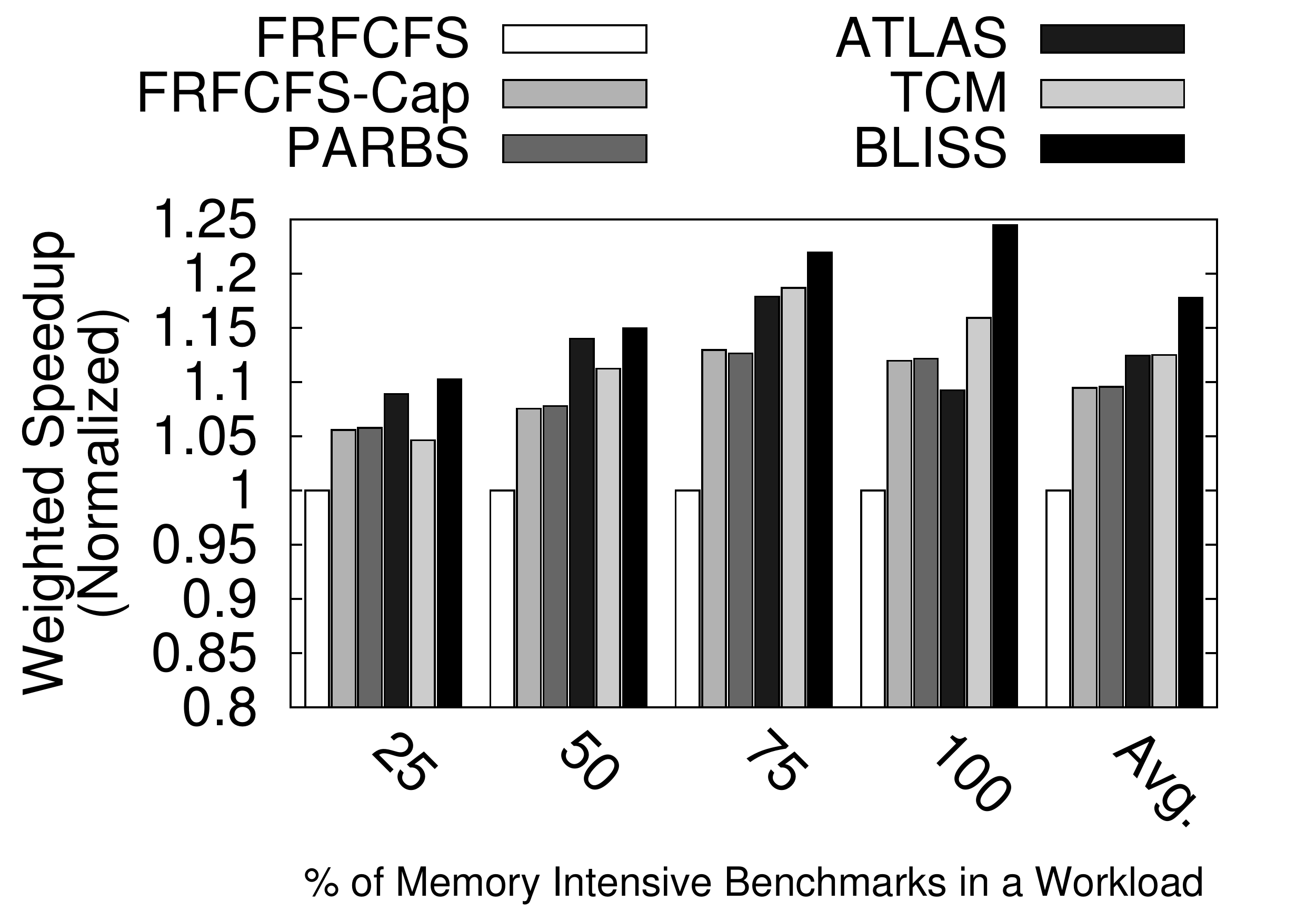}
  \end{minipage}
  \begin{minipage}{0.24\textwidth}  
    \centering
    \includegraphics[scale=0.17, angle=270]{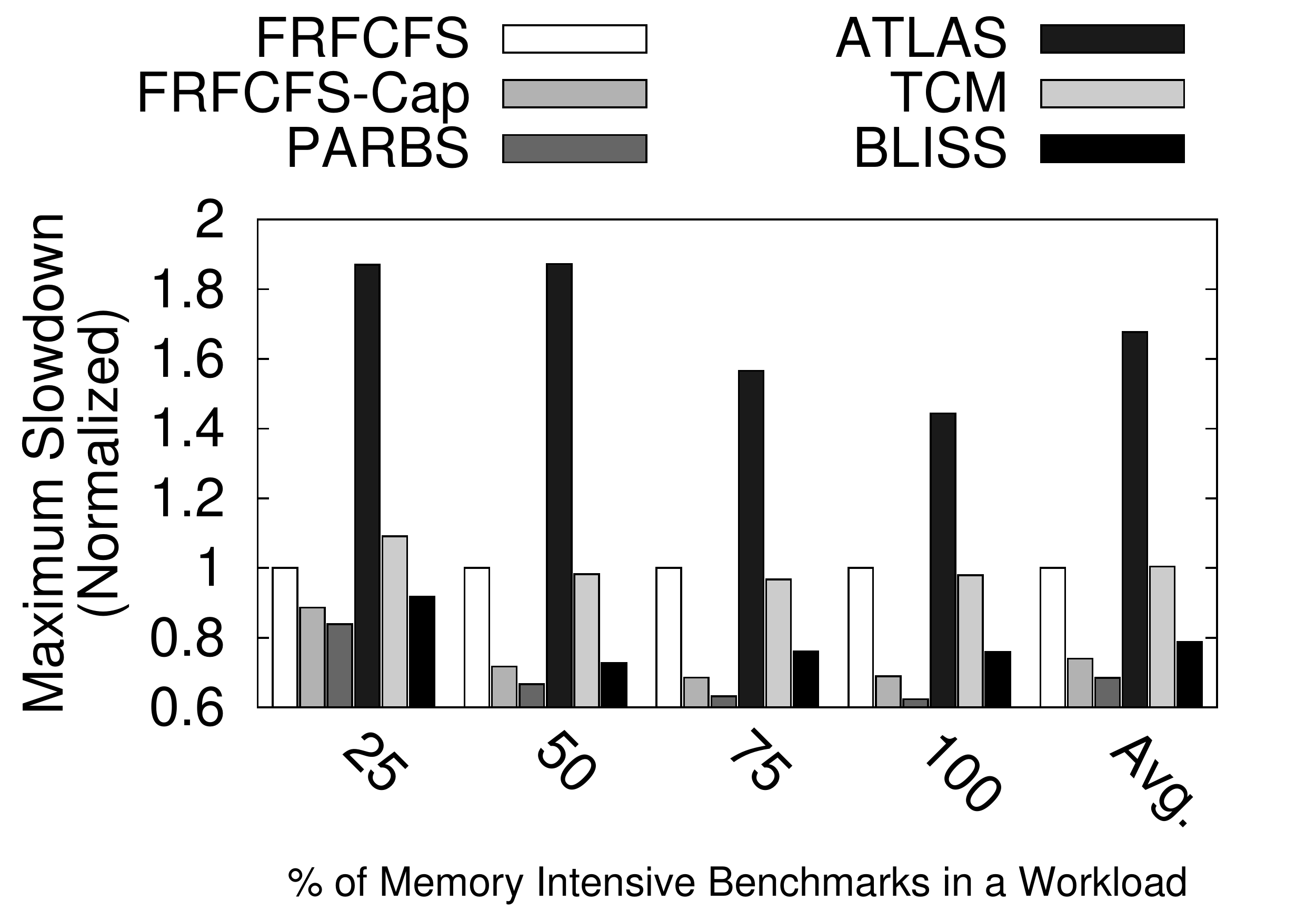}
  \end{minipage}
  \vspace{-1mm}
  \caption{Sensitivity to workload memory intensity}
  \label{fig:intensity-results}
  \vspace{-4mm}
\end{figure}

\noindent\textbf{Workload Memory Intensity.}
Figure~\ref{fig:intensity-results} shows system performance and
fairness for workloads with different memory intensities,
classified into different categories based on the fraction of
high-memory-intensity applications in a workload.\footnote{We
classify applications with MPKI less than 5 as
low-memory-intensity and the rest as high-memory-intensity.}  We
draw three major conclusions.  First, \bliss outperforms previous
memory schedulers in terms of system performance across all
intensity categories. Second, the system performance benefits of
\bliss increase with workload memory intensity. This is because as
the number of high-memory-intensity applications in a workload
increases, ranking individual applications, as done by previous
schedulers, causes more unfairness and degrades system
performance. Third, \bliss achieves significantly lower unfairness
than previous memory schedulers, except FRFCFS-Cap and PARBS,
across all intensity categories. Therefore, we conclude that
\bliss is effective in mitigating interference and improving
system performance and fairness across workloads with different
compositions of high- and low-memory-intensity applications.

\begin{figure}[h!]
  \vspace{-4mm}
  \centering
  \begin{minipage}{0.24\textwidth}
    \centering
    \includegraphics[scale=0.17, angle=270]{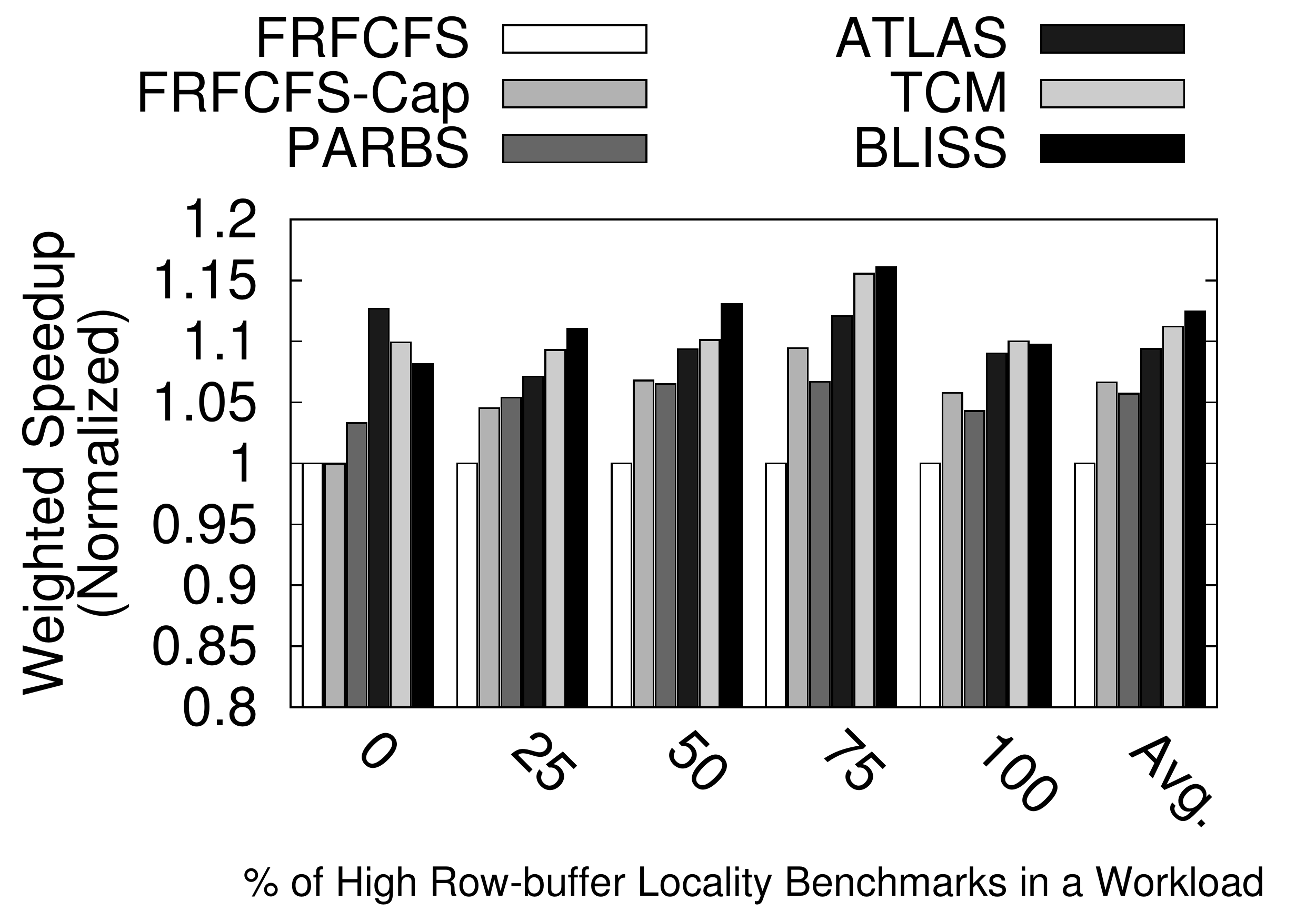}
  \end{minipage}
  \begin{minipage}{0.24\textwidth}  
    \centering
    \includegraphics[scale=0.17, angle=270]{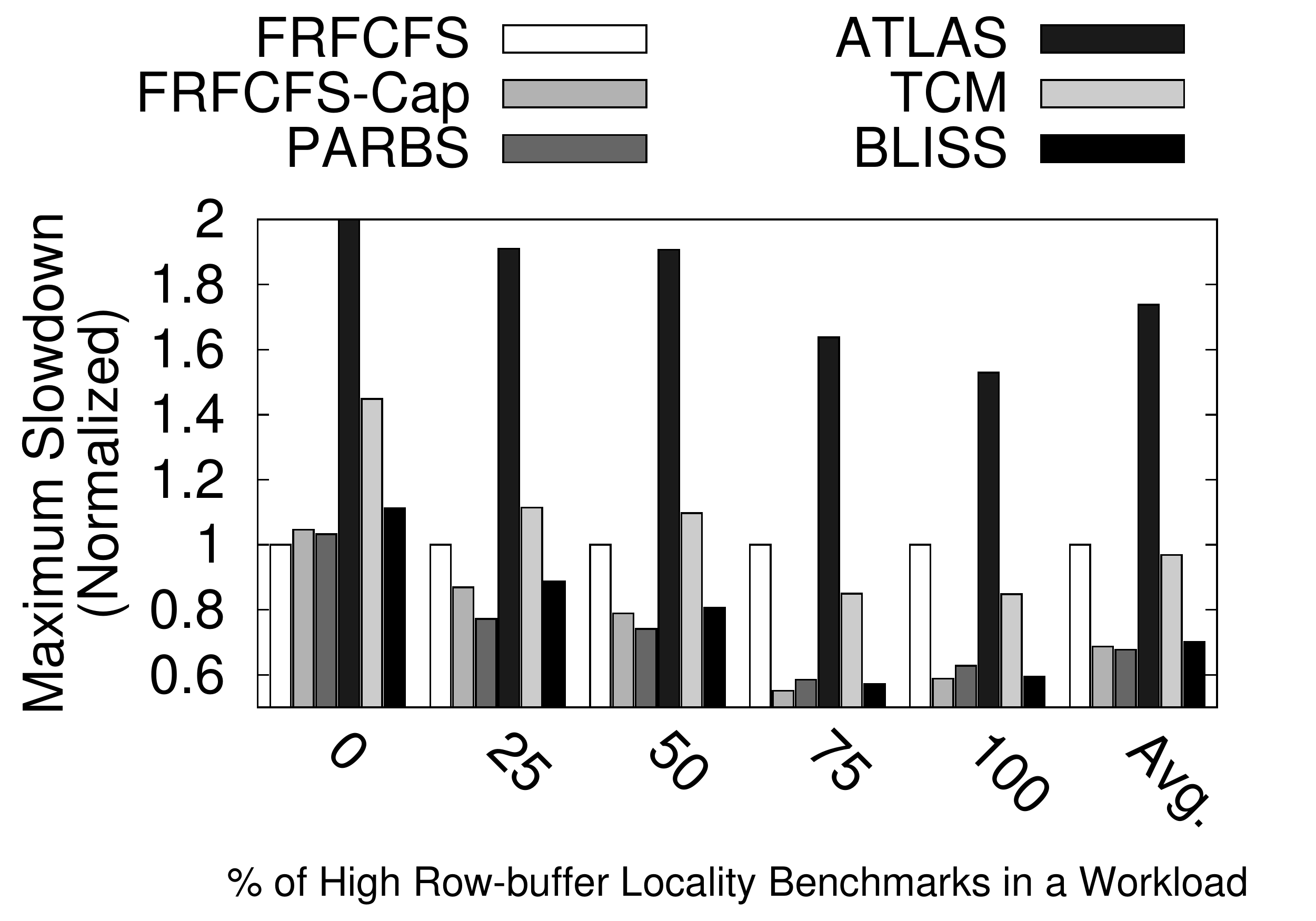}
  \end{minipage}
  \vspace{-1mm}
  \caption{Sensitivity to row-buffer locality}
  \label{fig:row-locality-results}
  \vspace{-4mm}
\end{figure}

\noindent\textbf{Workload Row-buffer Locality.}
Figure~\ref{fig:row-locality-results} shows the system performance
and fairness of five previous schedulers and BLISS when the number
of high row-buffer locality applications in a workload is
varied.\footnote{We classify an application as having high
row-buffer locality if its row-buffer hit rate is greater than
90\%.} We draw three observations. First, BLISS achieves the best
performance and close to the best fairness in most row-buffer
locality categories. Second, BLISS' performance and fairness
benefits over baseline FRFCFS increase as the number of
high-row-buffer-locality applications in a workload increases.
As the number of high-row-buffer-locality
applications in a workload increases, there is more interference to
the low-row-buffer-locality applications that are vulnerable.
Hence, there is more opportunity for BLISS to mitigate this
interference and improve performance and fairness. Third, when all
applications in a workload have high row-buffer locality (100\%),
the performance and fairness improvements of BLISS over baseline
FRFCFS are a bit lower than the other categories. This is because,
when all applications have high row-buffer locality, they each hog
the row-buffer in turn and are not as susceptible to interference
as the other categories in which there are vulnerable
low-row-buffer-locality applications. However, the
performance/fairness benefits of BLISS are still significant since BLISS is
effective in regulating how the row-buffer is shared among 
different applications. Overall, we conclude that BLISS is
effective in achieving high performance and fairness across
workloads with different compositions of high- and low-row-buffer-locality
applications.

\subsection{Sensitivity to System Parameters}
\label{sec:sensitivity-system}

\begin{figure}[h!]
   \vspace{-4mm}
    \centering
    \begin{minipage}{0.24\textwidth}
       \centering
       \includegraphics[scale=0.16, angle=270]{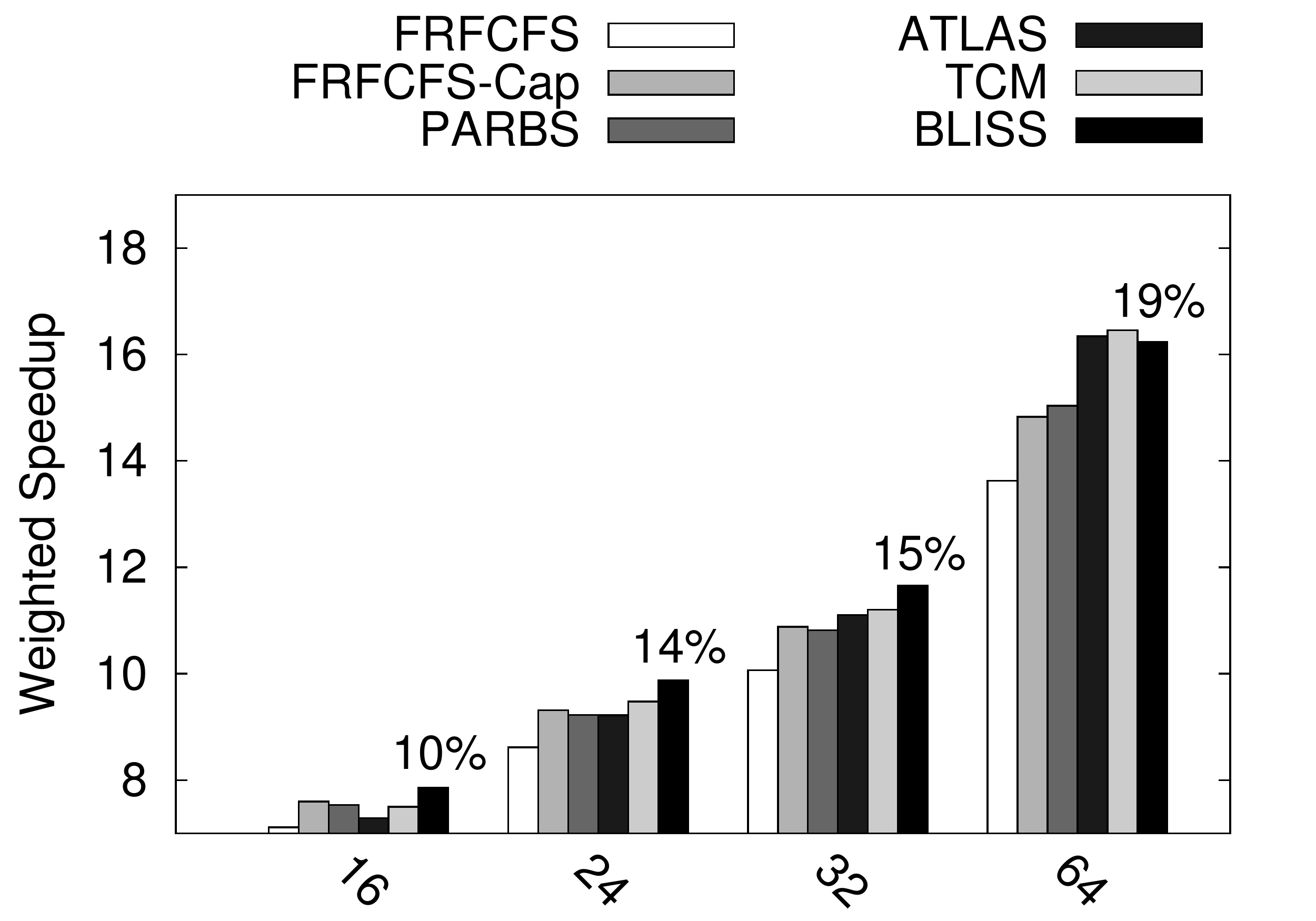}
    \end{minipage}
    \begin{minipage}{0.24\textwidth}
       \centering
        \includegraphics[scale=0.16, angle=270]{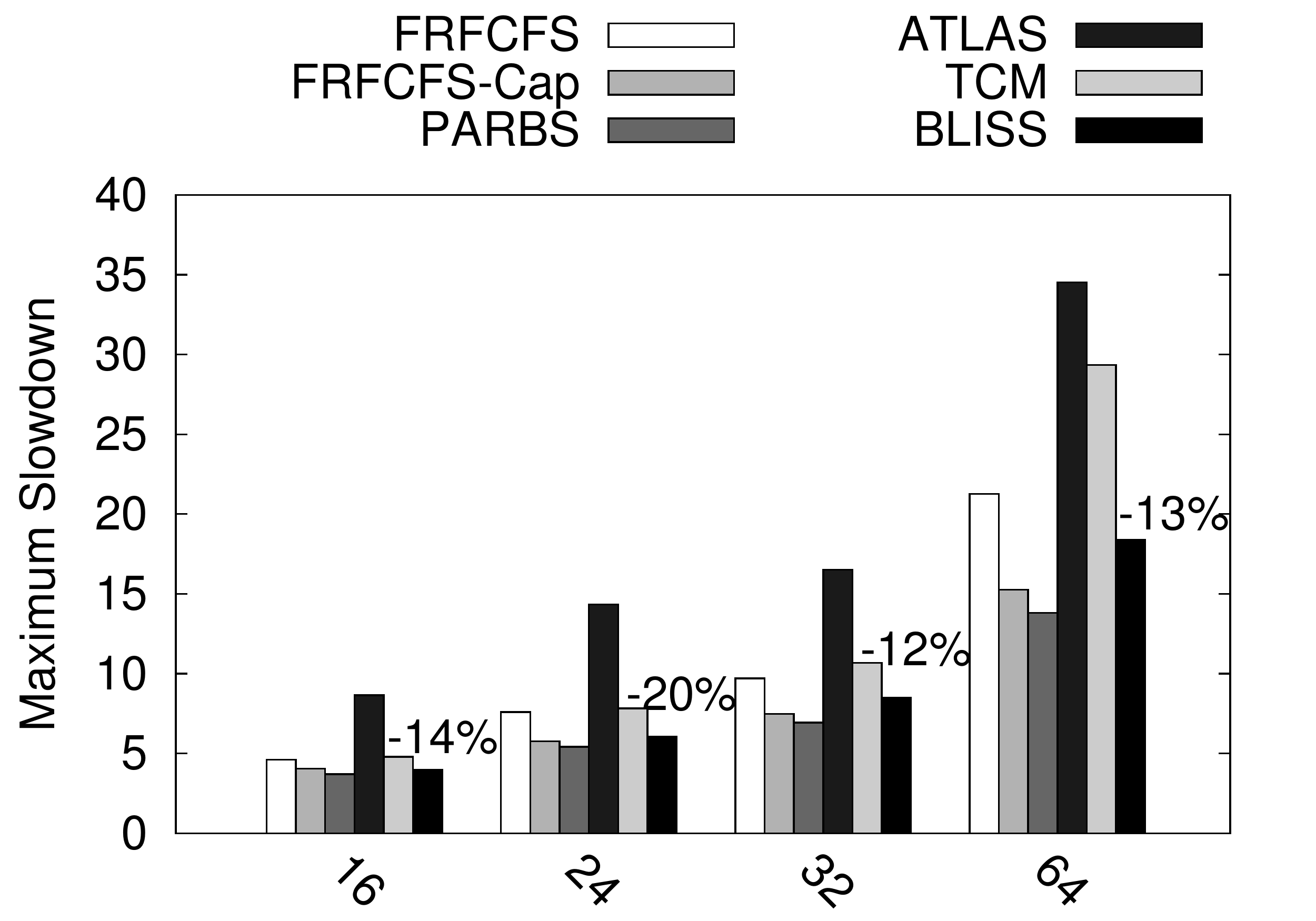}
    \end{minipage}
    \vspace{-2mm}
    \caption{Sensitivity to number of cores}
    \label{fig:sensitivity-num-cores}
   \vspace{-7mm}
\end{figure}

\begin{figure}[h!]
%   \vspace{-3mm}
    \centering
    \begin{minipage}{0.24\textwidth}
       \centering
       \includegraphics[scale=0.16, angle=270]{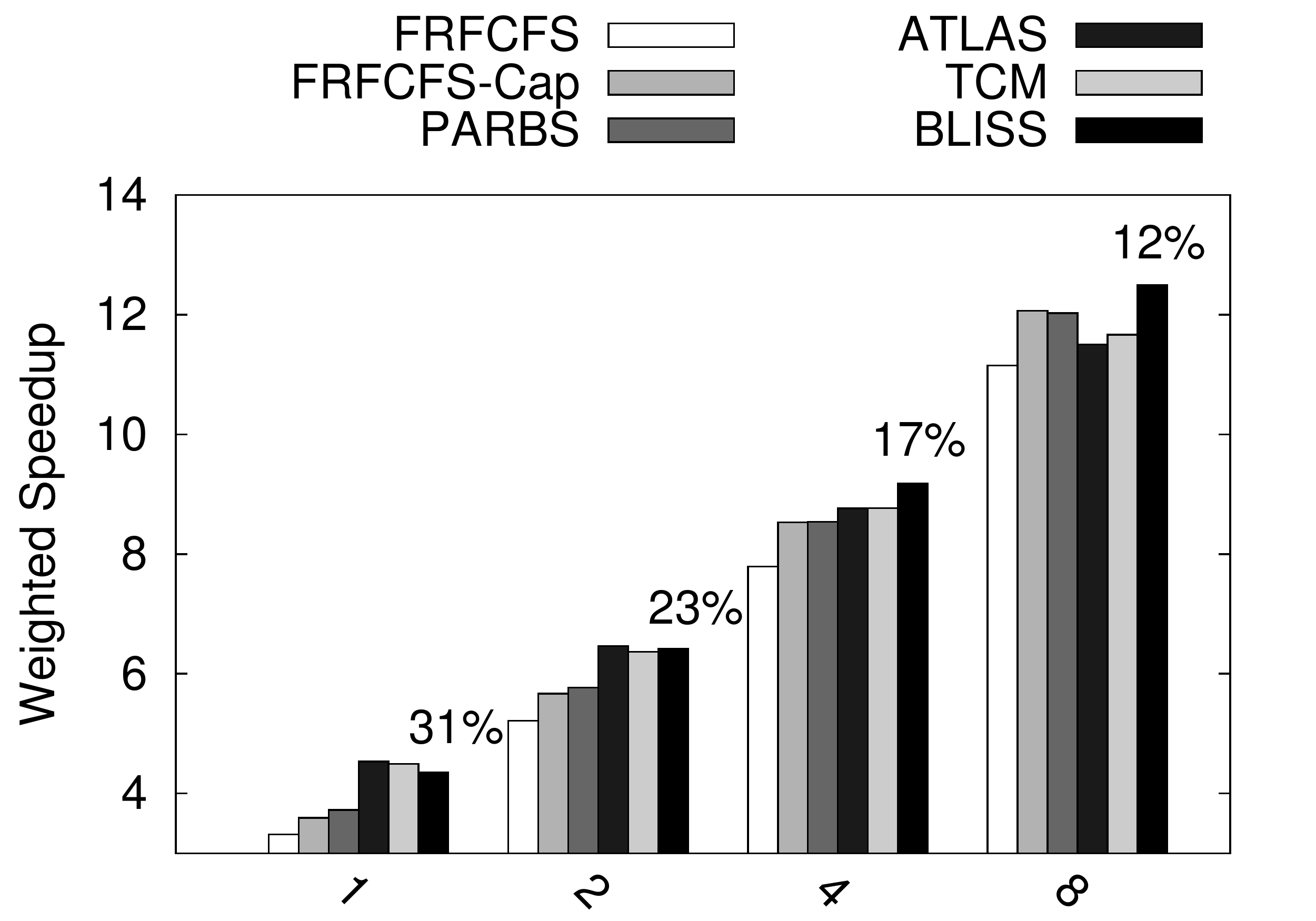}
    \end{minipage}
    \begin{minipage}{0.24\textwidth}
       \centering
        \includegraphics[scale=0.16, angle=270]{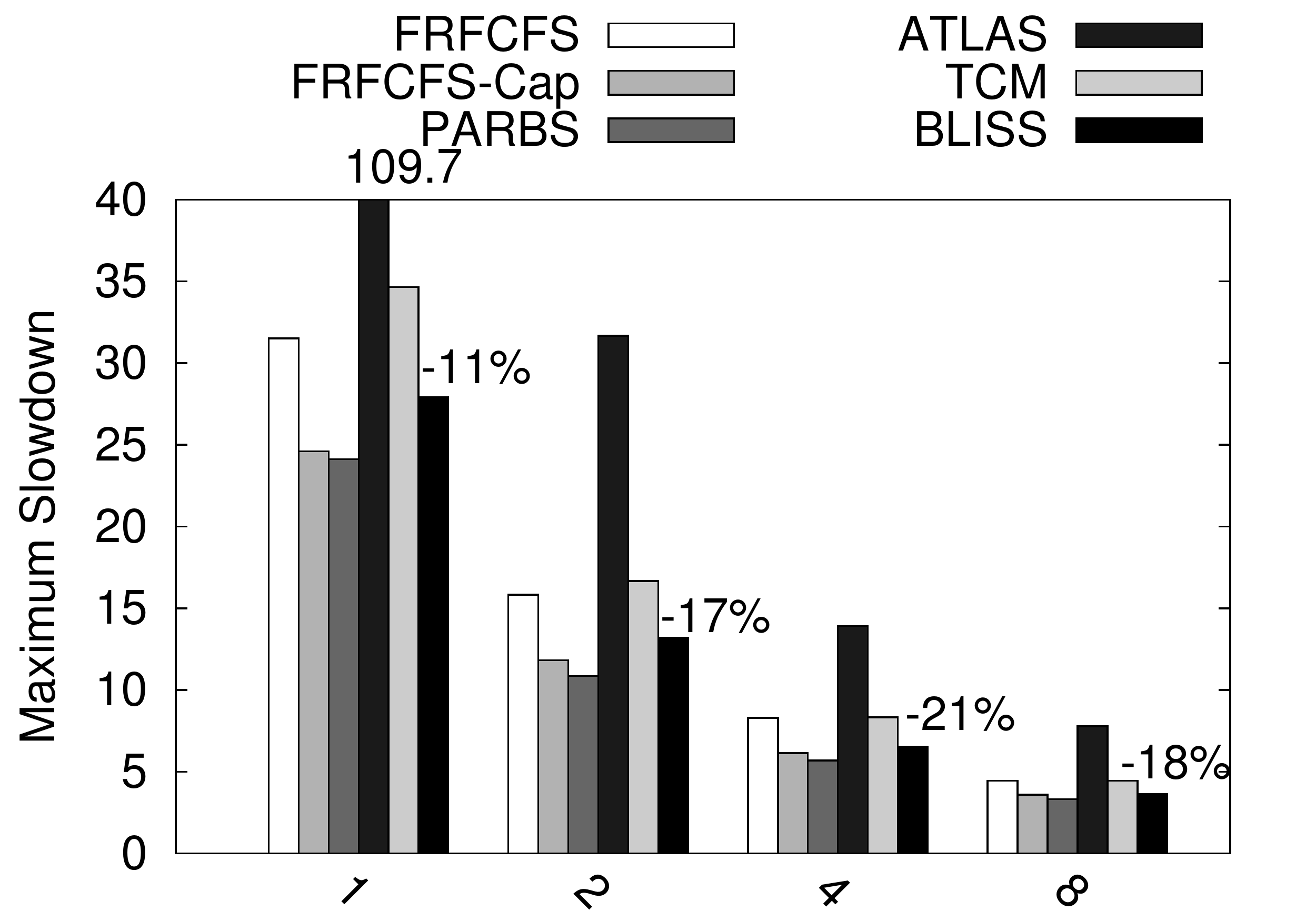}
    \end{minipage}
    \vspace{-2mm}
    \caption{Sensitivity to number of channels}
    \label{fig:sensitivity-num-channels}
   \vspace{-4mm}
\end{figure}

\noindent\textbf{Core and channel count.}
Figures~\ref{fig:sensitivity-num-cores}
and~\ref{fig:sensitivity-num-channels} show the system performance
and fairness of FRFCFS, PARBS, TCM and \bliss for different core
counts (when the channel count is 4) and different channel counts
(when the core count is 24), across 40 workloads for each
core/channel count. The numbers over the bars indicate percentage
increase or decrease compared to FRFCFS. We did not optimize the
parameters of different schedulers for each configuration as this
requires months of simulation time. We draw three major
conclusions. First, the absolute values of weighted speedup
increases with increasing core/channel count, whereas the absolute
values of maximum slowdown increase/decrease with increasing
core/channel count respectively, as expected. 
%\vspace{-2mm}
\noindent
Second, \bliss achieves higher system performance and lower
unfairness than all the other scheduling policies (except PARBS,
in terms of fairness) similar to our results on the 24-core,
4-channel system, by virtue of its effective interference
mitigation. The only anomaly is that TCM has marginally higher
weighted speedup than \bliss for the 64-core system. However, this
increase comes at the cost of significant increase in unfairness.
Third, \bliss' system performance benefit (as indicated by the
percentages on top of bars, over FRFCFS) increases when the system
becomes more bandwidth constrained, i.e., high core counts and low
channel counts. As contention increases in the system, \bliss has
greater opportunity to mitigate it.\footnote{Fairness benefits
reduce at very high core counts and very low channel counts, since
memory bandwidth becomes highly saturated.}

\begin{figure}[ht!]
   \vspace{-4mm}
    \centering
    \begin{minipage}{0.24\textwidth}
       \centering
       \includegraphics[scale=0.16, angle=270]{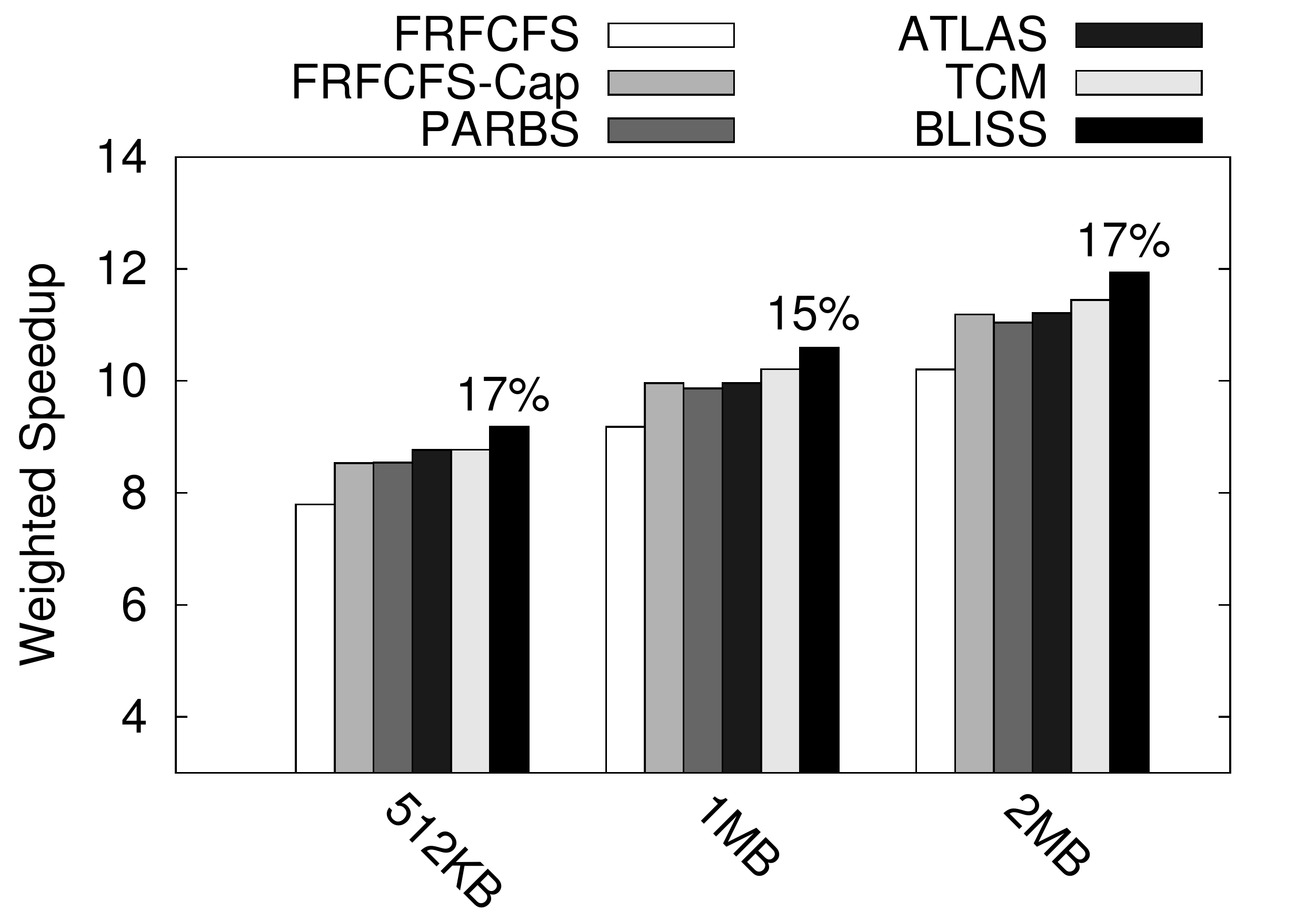}
    \end{minipage}
    \begin{minipage}{0.24\textwidth}
       \centering
        \includegraphics[scale=0.16, angle=270]{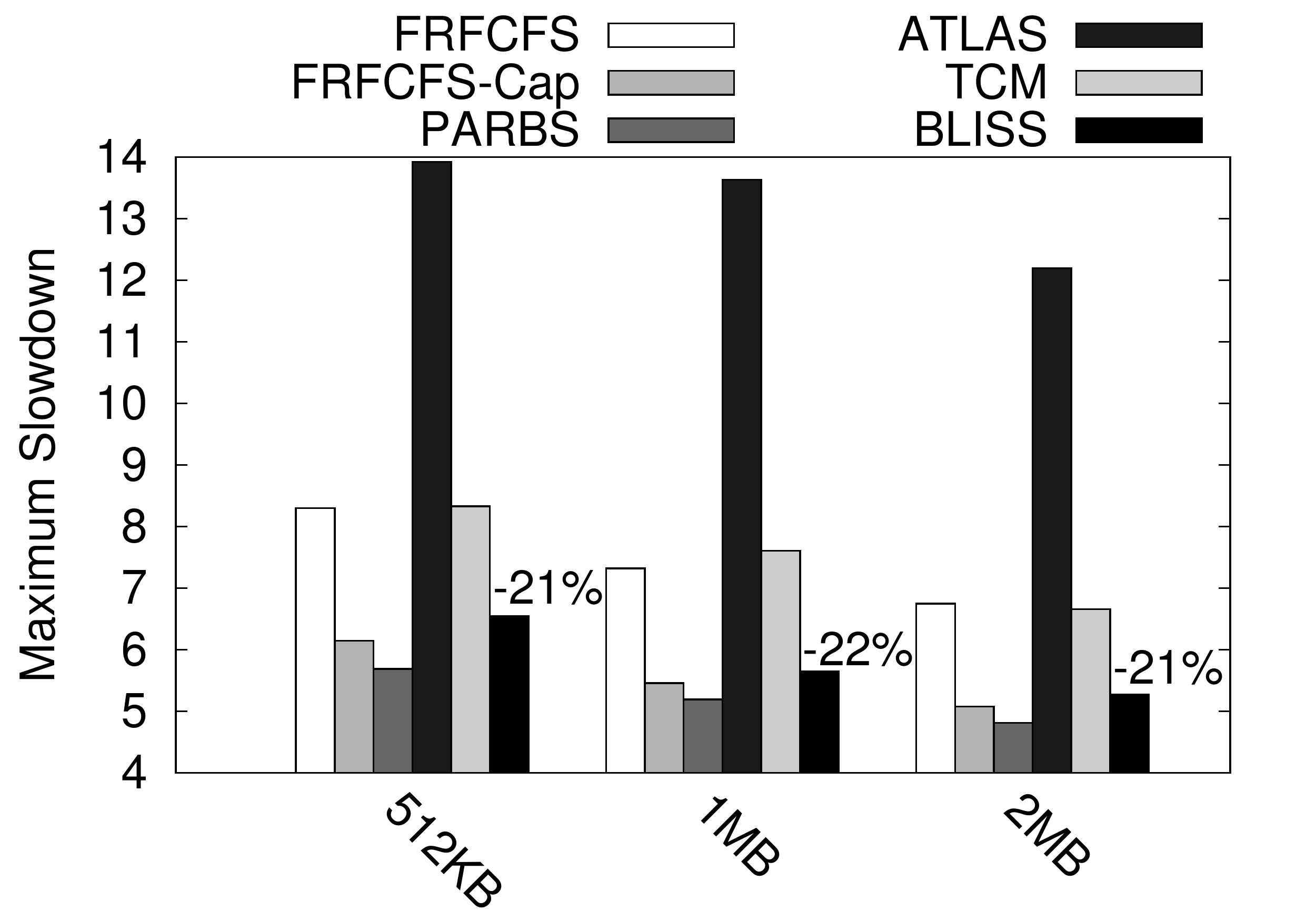}
    \end{minipage}
    \vspace{-2mm}
    \caption{Sensitivity to cache size}
    \label{fig:sensitivity-cache-size}
   \vspace{-4mm}
\end{figure}

\noindent\textbf{Cache size.}
Figure~\ref{fig:sensitivity-cache-size} shows the system
performance and fairness for five previous schedulers and BLISS
with different last level cache sizes (private to each core).  

We make two observations. First, the absolute values of weighted
speedup increase and maximum slowdown decrease, as the cache size
becomes larger for all schedulers, as expected. This is because
contention for memory bandwidth reduces with increasing cache
capacity, improving performance and fairness. Second, across all
the cache capacity points we evaluate, BLISS achieves significant
performance and fairness benefits over the best-performing
previous schedulers, while approaching close to the fairness of
the fairest previous schedulers.

\begin{figure}[ht!]
   \vspace{-5mm}
    \centering
    \begin{minipage}{0.24\textwidth}
       \centering
       \includegraphics[scale=0.16, angle=270]{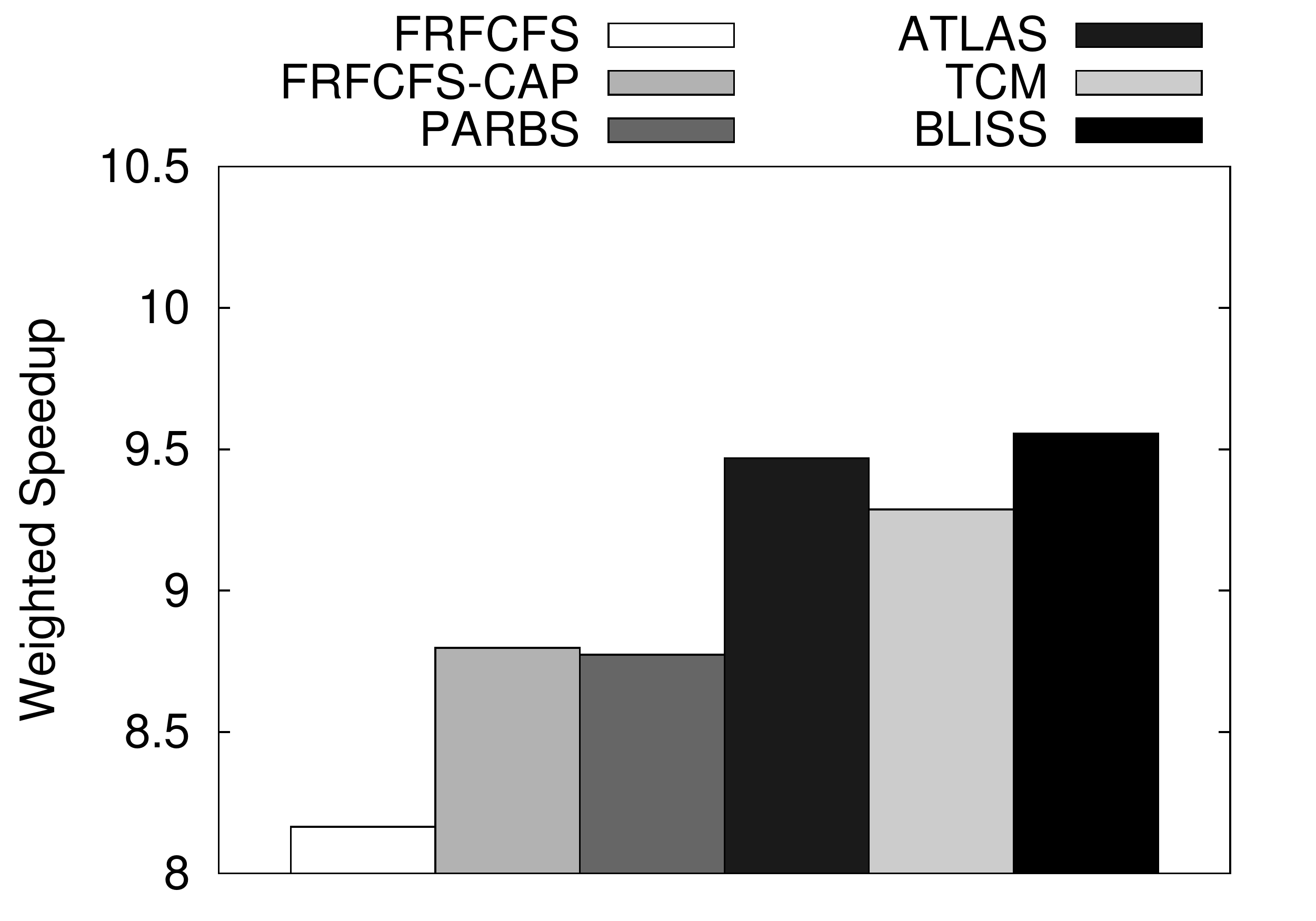}
    \end{minipage}
    \begin{minipage}{0.24\textwidth}
       \centering
        \includegraphics[scale=0.16, angle=270]{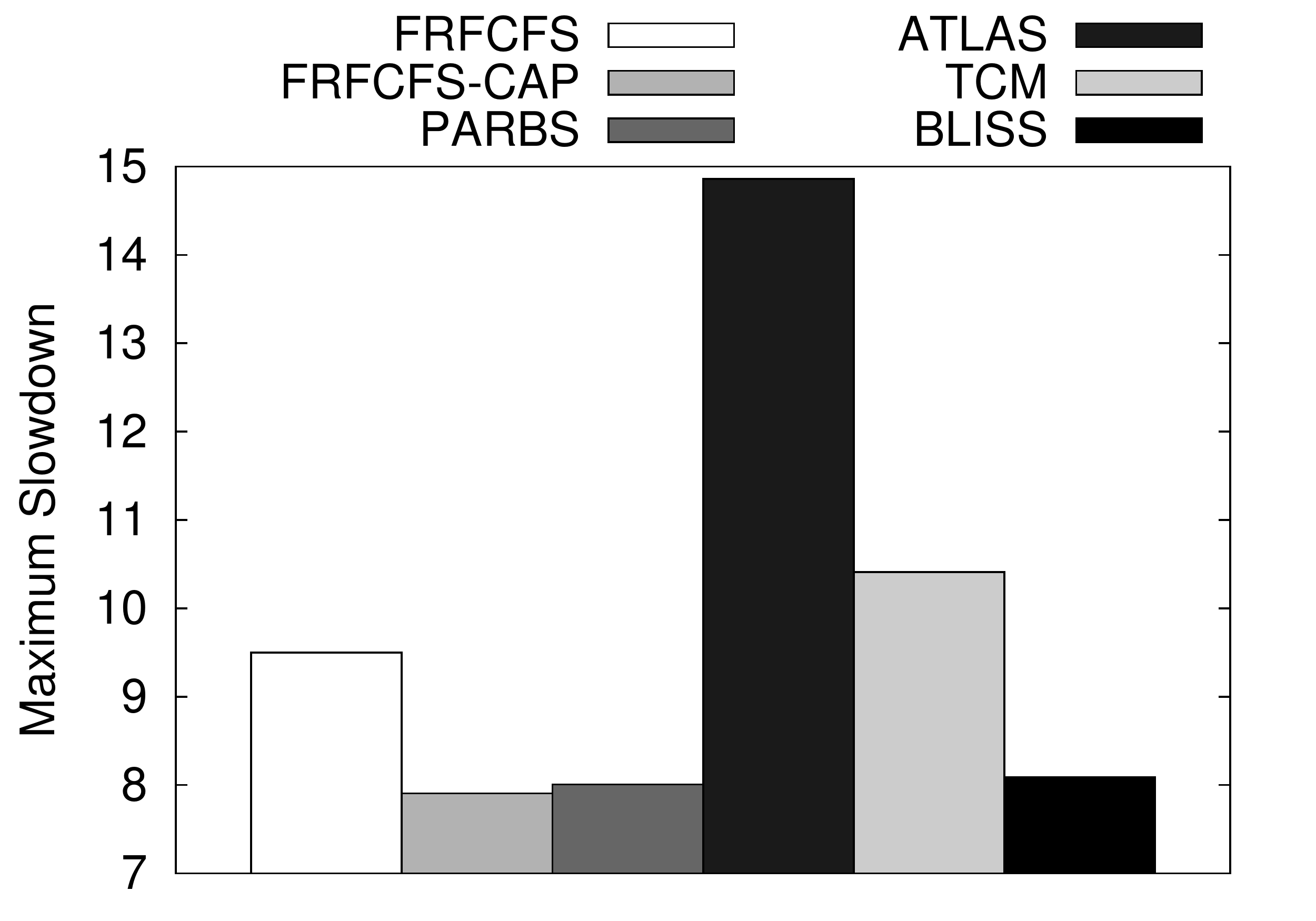}
    \end{minipage}
    \vspace{-2mm}
    \caption{Performance and fairness with a shared cache}
    \label{fig:shared-cache}
   \vspace{-3.5mm}
\end{figure}

\noindent\textbf{Shared Caches.} Figure~\ref{fig:shared-cache}
shows system performance and fairness with a 32 MB shared cache
(instead of the 512 KB per core private caches used in our other
experiments). \bliss achieves 5\%/24\% better performance/fairness
compared to TCM, demonstrating that \bliss is effective in
mitigating memory interference in the presence of large shared caches as
well.

\subsection{Sensitivity to Algorithm Parameters}
\label{sec:sensitivity-algorithm}

Tables~\ref{tab:sensitivity-ws} and \ref{tab:sensitivity-ms} show
the system performance and fairness respectively of \bliss for
different values of the \textit{Blacklisting Threshold} and
\textit{Clearing Interval}. Three major conclusions are in order.
First, a \textit{Clearing Interval} of 10000 cycles provides a
good balance between performance and fairness. If the blacklist is
cleared too frequently (1000 cycles), interference-causing
applications are not deprioritized for long enough, resulting in
low system performance. In contrast, if the blacklist is cleared
too infrequently, interference-causing applications are
deprioritized for too long, resulting in high unfairness. Second,
a \textit{Blacklisting Threshold} of 4 provides a good balance
between performance and fairness. When \textit{Blacklisting
Threshold} is very small, applications are blacklisted as soon as
they have very few requests served, resulting in poor interference
mitigation as too many applications are blacklisted. On the other
hand, when \textit{Blacklisting Threshold} is large, low- and
high-memory-intensity applications are not segregated effectively,
leading to high unfairness.

\begin{table}[ht]
  \vspace{-2mm}
  \centering
  \input{tables/sensitivity-ws}

  \vspace{-1mm}
  \caption{Perf. sensitivity to threshold and interval}
  \label{tab:sensitivity-ws}
  \vspace{-3.5mm}
\end{table}

\begin{table}[ht]
  \vspace{-3mm}
  \centering
  \input{tables/sensitivity-ms}

  \vspace{-1mm}
  \caption{Unfairness sensitivity to threshold and interval}
  \label{tab:sensitivity-ms}
  \vspace{-4mm}
\end{table}

\subsection{Interleaving and Scheduling Interaction}
\label{sec:interleaving-interaction}

In this section, we study the impact of the address interleaving
policy on the performance and fairness of different schedulers.
Our analysis so far has assumed a row-interleaved policy, where
data is distributed across channels, banks and rows at the
granularity of a row. This policy optimizes for row-buffer
locality by mapping a consecutive row of data to the same channel,
bank, rank. In this section, we will consider two other
interleaving policies, cache block interleaving and sub-row
interleaving.

\noindent\textbf{Interaction with cache block interleaving.} In a
cache-block-interleaved system, data is striped across channels,
banks and ranks at the granularity of a cache block.  Such a
policy optimizes for bank level parallelism, by distributing data
at a small (cache block) granularity across channels, banks and
ranks. 

\begin{figure}[ht]
\vspace{-4.5mm}
  \centering
  \begin{minipage}{0.24\textwidth}
    \centering
    \includegraphics[scale=0.17, angle=270]{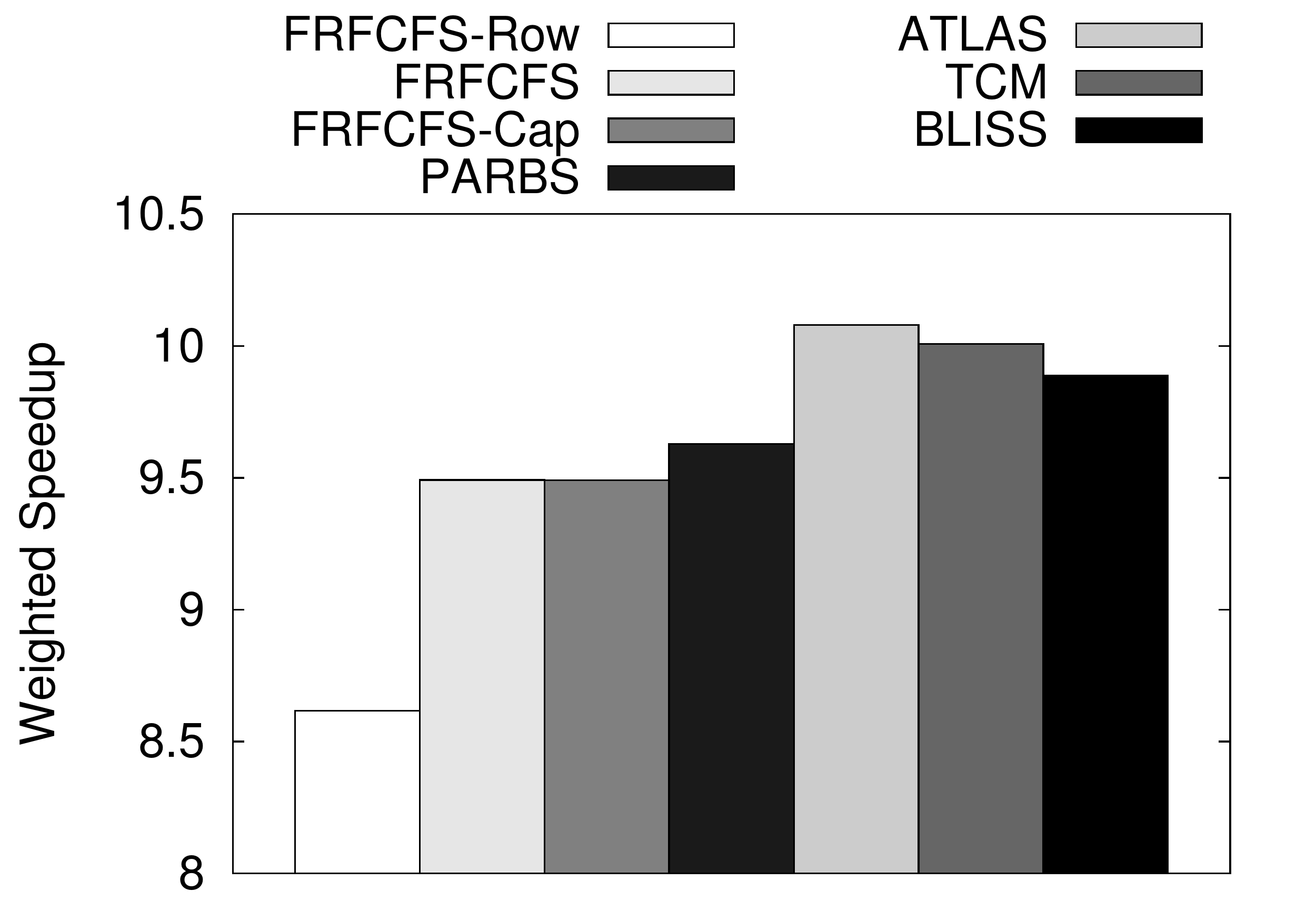}
  \end{minipage}
%  \vspace{-2mm}
  \begin{minipage}{0.24\textwidth}
    \centering
    \includegraphics[scale=0.17, angle=270]{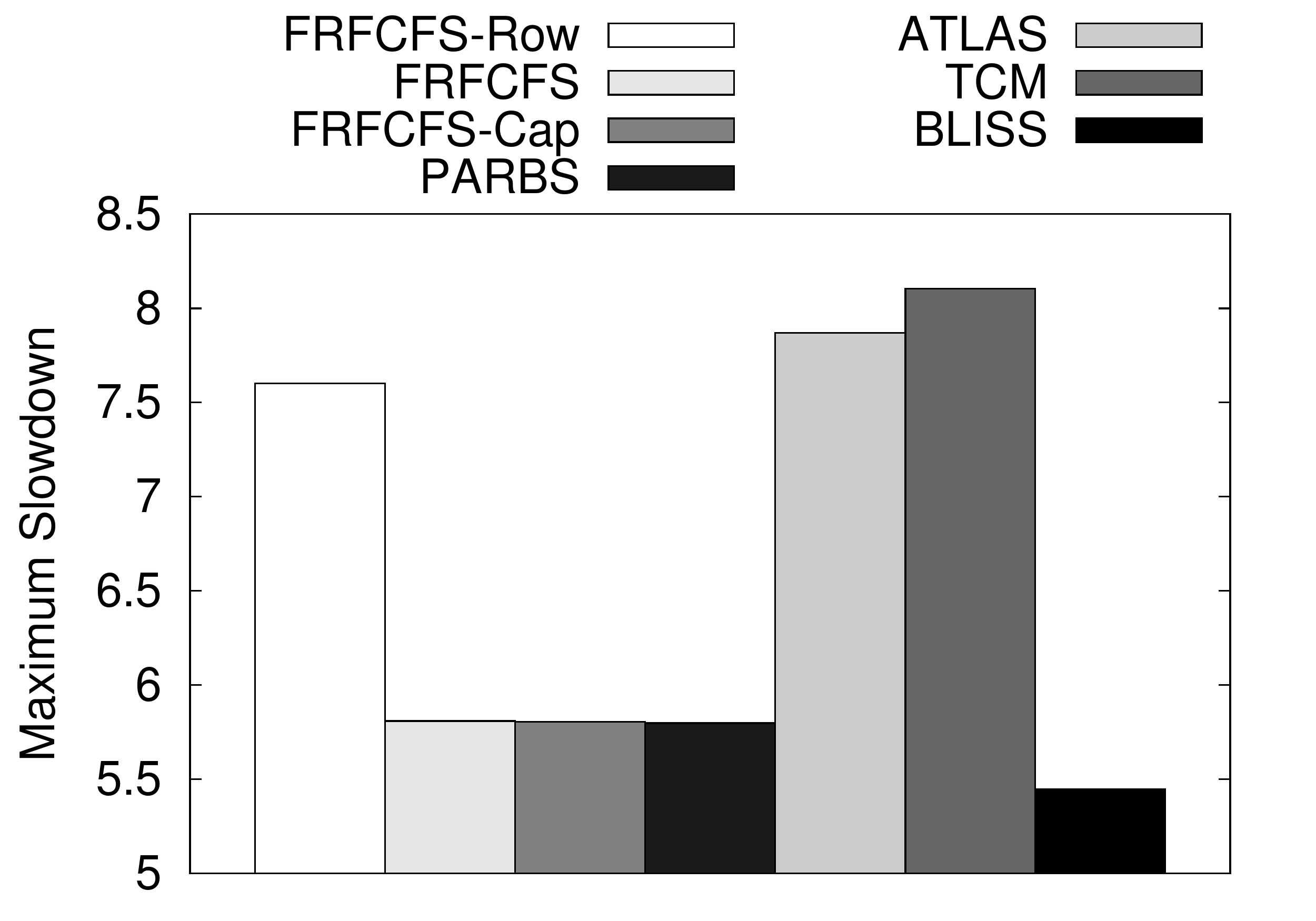}
  \end{minipage}
  \vspace{-2mm}
  \caption{Scheduling and cache block interleaving}
  \label{fig:cacheblockint-interaction}
\vspace{-4mm}
\end{figure}

Figure~\ref{fig:cacheblockint-interaction} shows the system
performance and fairness of FRFCFS with row interleaving
(FRFCFS-Row), as a comparison point, five previous schedulers, and
BLISS with cache block interleaving. We draw three observations.
First, system performance and fairness of the baseline FRFCFS
scheduler improve significantly with cache block interleaving,
compared to with row interleaving. This is because cache block
interleaving enables more requests to be served in parallel at the
different channels and banks, by distributing data across channels
and banks at the small granularity of a cache block. Hence, most
applications, and particularly, applications that do not have very
high row-buffer locality benefit from cache block interleaving.

Second, as expected, application-aware schedulers such as ATLAS
and TCM achieve the best performance among previous schedulers, by
means of prioritizing requests of applications with low memory
intensities. However, PARBS and FRFCFS-Cap do not improve fairness
over the baseline, in contrast to our results with row
interleaving. This is because cache block interleaving already
attempts to provide fairness by increasing the parallelism in the
system and enabling more requests from across different
applications to be served in parallel, thereby reducing unfair
applications slowdowns. More specifically, requests that would
be row-buffer hits to the same bank, with row interleaving,
are now distributed across multiple channels and banks, with cache
block interleaving. Hence, applications' propensity to cause
interference reduces, providing lower scope for request capping
based schedulers such as FRFCFS-Cap and PARBS to mitigate
interference.  Third, BLISS achieves within 1.3\% of the
performance of the best performing previous scheduler (ATLAS),
while achieving 6.2\% better fairness than the fairest previous
scheduler (PARBS). BLISS effectively mitigates interference by
regulating the number of consecutive requests served from
high-memory-intensity applications that generate a large number of
requests, thereby achieving high performance and fairness.\\
\noindent\textbf{Interaction with sub-row interleaving.} While
memory scheduling has been a prevalent approach to mitigate memory
interference, previous work has also proposed other solutions, as
we describe in Section~\ref{sec:related_work}. One such previous
work by Kaseridis et al.~\cite{mop} proposes {\em minimalist open
page}, an interleaving policy that distributes data across
channels, ranks and banks at the granularity of a sub-row (partial
row), rather than an entire row, exploiting both row-buffer
locality and bank-level parallelism. We examine BLISS' interaction
with such a sub-row interleaving policy.

\begin{figure}[h]
  \vspace{-4mm}
  \centering
  \begin{minipage}{0.24\textwidth}
    \centering
    \includegraphics[scale=0.17, angle=270]{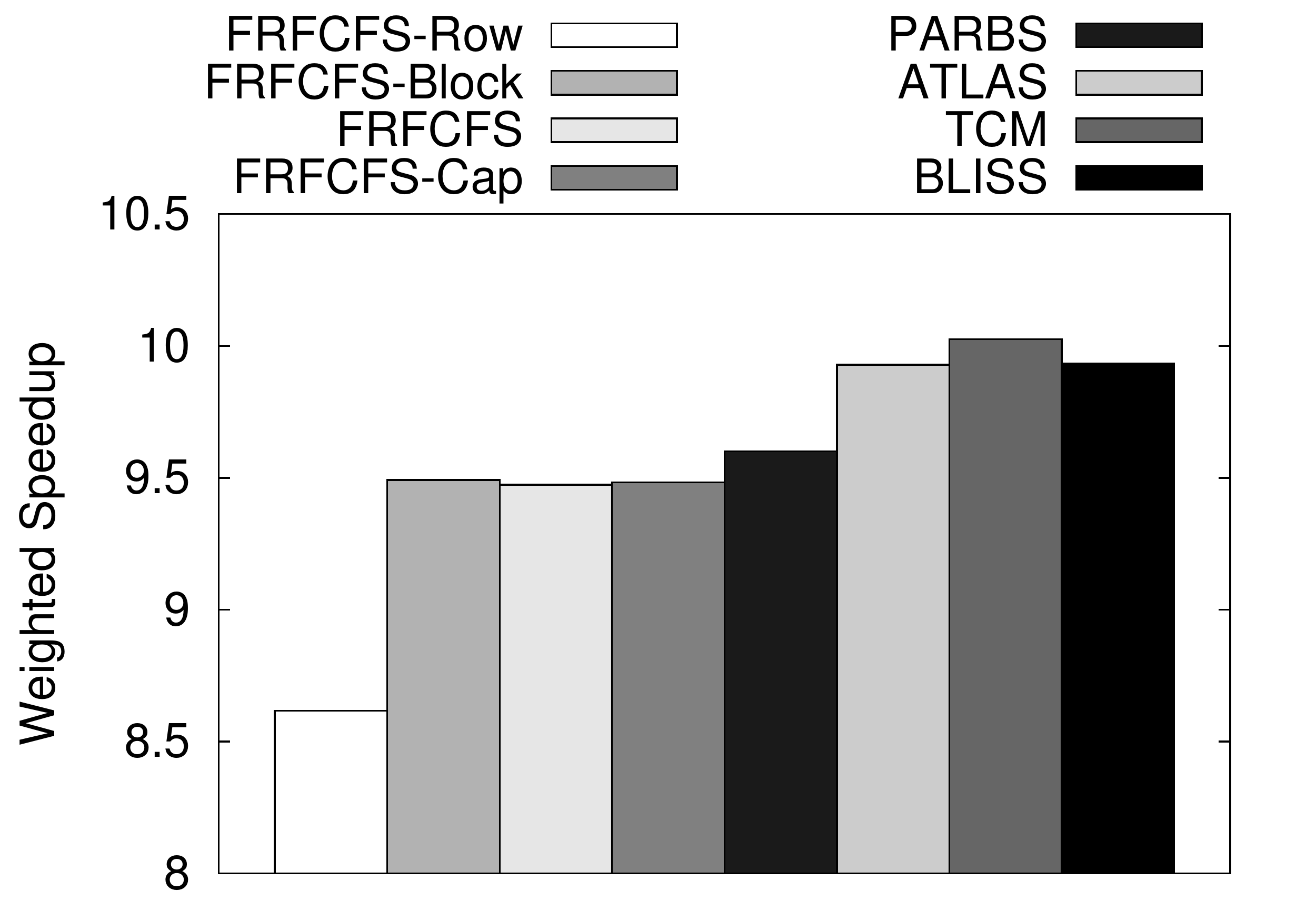}
  \end{minipage}
%  \vspace{-2mm}
  \begin{minipage}{0.24\textwidth}
    \centering
    \includegraphics[scale=0.17, angle=270]{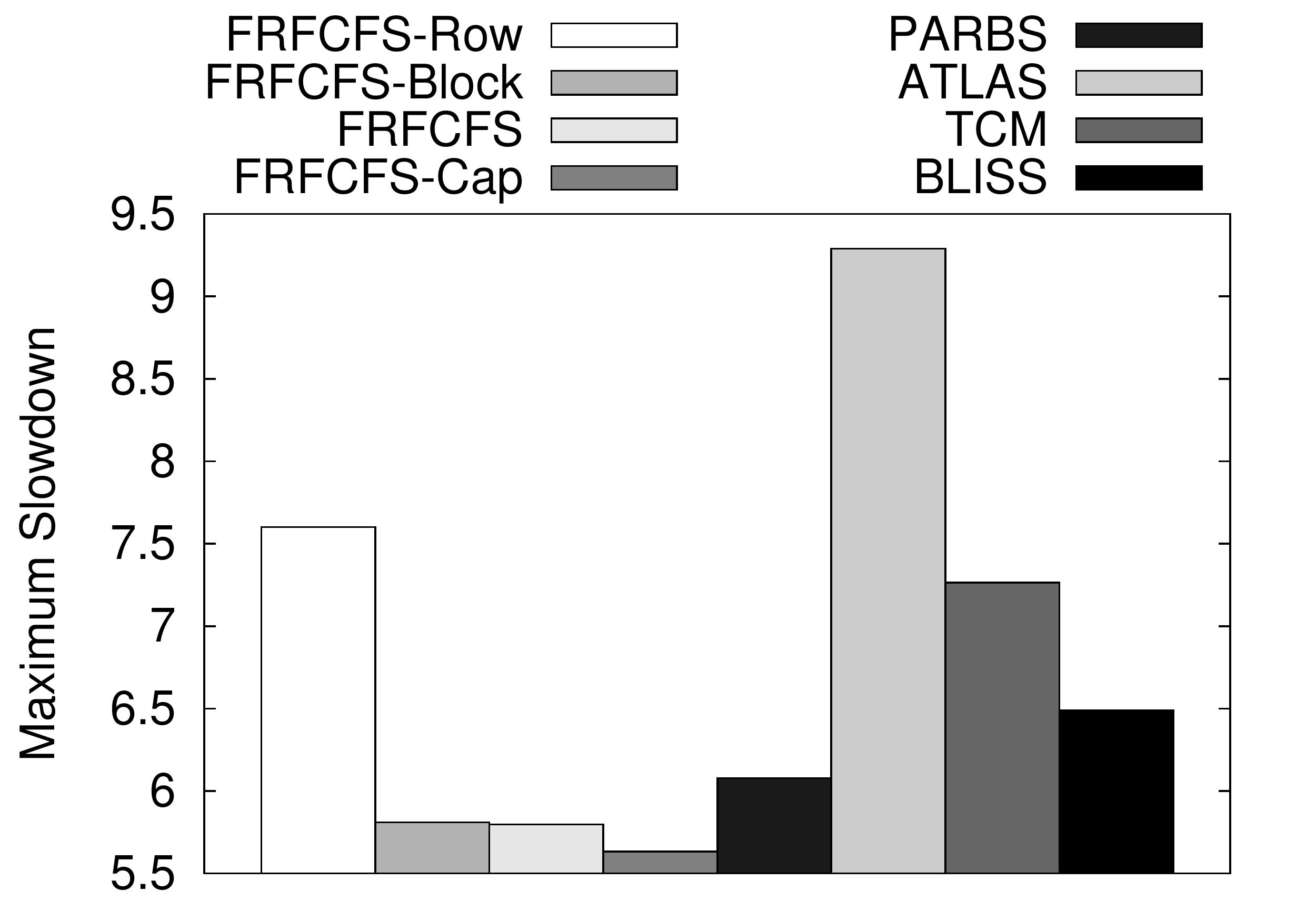}
  \end{minipage}
  \vspace{-2mm}
  \caption{Scheduling and sub-row interleaving}
  \label{fig:subrowint-interaction}
\vspace{-3.5mm}
\end{figure}

Figure~\ref{fig:subrowint-interaction} shows the system
performance and fairness of FRFCFS with row interleaving
(FRFCFS-Row), FRFCFS with cache block interleaving (FRFCFS-Block)
and five previously proposed schedulers and \bliss, with sub-row
interleaving (when data is striped across channels, ranks and
banks at the granularity of four cache blocks). Three observations
are in order. First, sub-row interleaving provides significant
benefits over row interleaving, as can be observed for FRFCFS (and
other scheduling policies by comparing with
Figure~\ref{fig:main-results}). This is because sub-row
interleaving enables applications to exploit both row-buffer
locality and bank-level parallelism, unlike row interleaving that
is mainly focused on exploiting row-buffer locality. Second,
sub-row interleaving achieves similar performance and fairness as
cache block interleaving. We observe that this is because cache
block interleaving enables applications to exploit parallelism
effectively, which makes up for the lost row-buffer locality from
distributing data at the granularity of a cache block across all
channels and banks. Third, BLISS achieves close to the performance
(within 1.5\%) of the best performing previous scheduler (TCM),
while reducing unfairness significantly and approaching the
fairness of the fairest previous schedulers. One thing to note is
that BLISS has higher unfairness than FRFCFS, when a
sub-row-interleaved policy is employed. This is because the
capping decisions from sub-row interleaving and BLISS could
collectively restrict high-row-buffer locality applications to a
large degree, thereby slowing them down and causing higher
unfairness. Co-design of the scheduling and interleaving policies
to achieve different goals such as performance/fairness is an
important area of future research. We conclude that a \bliss-like
scheduler, with its high performance and low complexity is a
significantly better alternative to schedulers such as ATLAS/TCM
in the pursuit of such scheduling-interleaving policy co-design.

%% file: tables/individual-clearing.tex
\begin{tabular}{|c|c|c|c|}
    \hline
    Metric & BLISS & BLISS-Individual-Clearing\\
    \hline
    Weighted Speedup & 9.18 & 9.12\\
    \hline
    Maximum Slowdown & 6.54 & 6.60\\
    \hline
\end{tabular}

%% file: tables/sensitivity-ws.tex
\begin{tabular}{|c|c|c|c|c|c|}
  \hline
  \backslashbox{Threshold}{Interval} & 1000 & 10000 & 100000\\
  \hline
  2 & 8.76 & 8.66 & 7.95\\
  \hline
  4 & 8.61 & 9.18 & 8.60\\
  \hline
  8 & 8.42 & 9.05 & 9.24\\
  \hline
\end{tabular}

%% file: tables/sensitivity-ms.tex
\begin{tabular}{|c|c|c|c|c|c|}
  \hline
  \backslashbox{Threshold}{Interval} & 1000 & 10000 & 100000\\
  \hline
  2 & 6.07 & 6.24 & 7.78\\
  \hline
  4 & 6.03 & 6.54 & 7.01\\
  \hline
  8 & 6.02 & 7.39 & 7.29\\
  \hline
\end{tabular}

%% file: sections/related_work.tex
\section{Related Work}
\label{sec:related_work}

To our knowledge, BLISS is the first memory scheduler design that
attempts to optimize, at the same time, for high performance,
fairness and low complexity, which are three competing yet
important goals. The closest previous works to \bliss are other
memory schedulers. We have already compared \bliss both
qualitatively and quantitatively to previously proposed memory
schedulers, FRFCFS~\cite{frfcfs,frfcfs-patent},
PARBS~\cite{parbs,podc-08}, ATLAS~\cite{atlas}, TCM~\cite{tcm} and
criticality-aware memory
scheduling~\cite{crit-scheduling-cornell}, which have been
designed to mitigate interference in a multicore system. Other
previous schedulers~\cite{mph,stfm,fqm} have been proposed earlier
that PARBS, ATLAS and TCM have been shown to
outperform~\cite{parbs,atlas,tcm}. 

Parallel Application Memory Scheduling (PAMS)~\cite{pams} tackles
the problem of mitigating interference between different threads
of a multithreaded application, while Staged Memory Scheduling
(SMS)~\cite{sms} attempts to mitigate interference between the CPU
and GPU in CPU-GPU systems. Principles from \bliss can be employed
in both of these contexts to identify and deprioritize
interference-causing threads, thereby mitigating interference
experienced by vulnerable threads/applications.
FIRM~\cite{firm-micro14} proposes request scheduling
mechanisms to tackle the problem of heavy write traffic in
persistent memory systems. BLISS can be combined with FIRM's write
handling mechanisms to achieve better fairness in persistent
memory systems. Complexity effective memory access
scheduling~\cite{complexity-effective} attempts to achieve the
performance of FRFCFS using a First Come First Served scheduler in
GPU systems, by preventing row-buffer locality from being
destroyed when data is transmitted over the on-chip network. Their
proposal is complementary to ours. \bliss could be combined with
such a scheduler design to prevent threads from hogging the
row-buffer and banks.

While memory scheduling is a major solution direction towards mitigating
interference, previous works have also explored other approaches such
as address interleaving~\cite{mop}, memory bank/channel
partitioning~\cite{mcp,bank-part,pact-bank-part,bank-part-hpca14}, source
throttling~\cite{fst,selftuned,baydal05,hat,nychis,cc-hotnets10,kayiran-micro14} and thread
scheduling~\cite{zhuravlev-thread-scheduling,tang-thread-scheduling,a2c,adrm} to mitigate interference.\\
\noindent\textbf{Subrow Interleaving:} Kaseridis et al.~\cite{mop}
propose minimalist open page, a data mapping policy that
interleaves data at the granularity of a sub-row across channels
and banks such that applications with high row-buffer locality are
prevented from hogging the row buffer, while still preserving some
amount of row-buffer-locality. We study the interactions of BLISS
with minimalist open page in
Section~\ref{sec:interleaving-interaction} showing BLISS' benefits
on a sub-row interleaved memory system.\\ 
%%% ONUR-8-24: Please keep the tense consistent. Present instead of present perfect. 
\noindent\textbf{Memory Channel/Bank Partitioning:} Previous
works~\cite{mcp,bank-part,pact-bank-part,bank-part-hpca14}
propose techniques to mitigate inter-application interference by
partitioning channels/banks among applications such that the data
of interfering applications are mapped to different
channels/banks. Our approach is complementary to these schemes and
can be used in conjunction with them to achieve more effective
interference mitigation.\\
\noindent\textbf{Source Throttling:} Source throttling techniques
(e.g.,~\cite{fst,selftuned,baydal05,hat,nychis,cc-hotnets10,kayiran-micro14})
propose to throttle the memory request injection rates of
interference-causing applications at the processor core itself
rather than regulating an application's access behavior at the
memory, unlike memory scheduling, partitioning or interleaving.
\bliss is complementary to source throttling and can be combined
with it to achieve better interference mitigation.\\
\noindent\textbf{OS Thread Scheduling:} Previous
works~\cite{zhuravlev-thread-scheduling,tang-thread-scheduling,adrm}
propose to mitigate shared resource contention by co-scheduling
threads that interact well and interfere less at the shared
resources. Such a solution relies on the presence of enough
threads with such symbiotic properties, whereas our proposal can
mitigate memory interference even if interfering threads are
co-scheduled.  Furthermore, such thread scheduling policies and
\bliss can be combined in a synergistic manner to further improve
system performance and fairness. Other techniques to map
applications to cores to mitigate memory interference, such
as~\cite{a2c}, can be combined with BLISS.

%% file: sections/conclusion.tex
\vspace{-0.25mm}
\section{Conclusion}
\vspace{-0.25mm}
\begin{sloppypar}

We introduce the Blacklisting memory scheduler (\bliss), a new and
simple approach to memory scheduling in systems with multiple threads. We
observe that the per-application ranking mechanisms employed by
previously proposed application-aware memory schedulers incur high
hardware cost, cause high unfairness, and lead to high scheduling
latency to the point that the scheduler cannot meet the fast
command scheduling requirements of state-of-the-art DDR protocols.
\bliss overcomes these problems based on the key observation that it is
sufficient to group applications into only two groups, rather than
employing a total rank order among all applications. Our
evaluations across a variety of workloads and systems demonstrate
that \bliss has better system performance and fairness than
previously proposed ranking-based schedulers, while incurring
significantly lower hardware cost and latency in making scheduling
decisions. We conclude that \bliss, with its low complexity, high
system performance and high fairness, can be an efficient and
effective memory scheduling substrate for current and future
multicore and multithreaded systems.
\end{sloppypar}